\documentclass{aa}  

\usepackage{graphicx}
\usepackage{txfonts}
\usepackage{booktabs}

\usepackage{natbib}
\bibpunct{(}{)}{;}{a}{}{,} 

\usepackage{longtable}
\usepackage{ctable}
\usepackage{lscape}

\begin{document}

\title{Quasars as standard candles II: The non linear relation between UV and X-ray emission at high redshifts}

\author{F. Salvestrini\inst{\ref{inst1},\ref{inst2}}
\and
G. Risaliti\inst{\ref{inst3},\ref{inst4}}
\and 
S. Bisogni\inst{\ref{inst3},\ref{inst4}}
\and
E. Lusso\inst{\ref{inst3},\ref{inst4},\ref{inst5}}
\and
C. Vignali\inst{\ref{inst1},\ref{inst2}}}

\institute{
Dipartimento di Astronomia, Universit\`a degli Studi di Bologna, via Gobetti 93/2, 40129 Bologna, Italy\label{inst1}
\and
INAF - Osservatorio di Astrofisica e Scienza dello Spazio di Bologna, via Gobetti 93/3 - 40129 Bologna - Italy\label{inst2}
\and
Dipartimento di Fisica e Astronomia, Universit\`a di Firenze, Via G. Sansone 1, 50019 Sesto Fiorentino (FI) , Italy\label{inst3}
\and
INAF, Osservatorio Astrofisico di Arcetri, Largo E. Fermi 5, I-50125 Firenze, Italy\label{inst4}
\and
Centre for Extragalactic Astronomy, Durham University, South Road, Durham, DH1 3LE, UK\label{inst5}
}

\date{}

\abstract{
A tight non-linear relation between the X-ray and the optical-ultraviolet (UV) emission has been observed in Active Galactic Nuclei (AGN) over a wide range of redshift and several orders of magnitude in luminosity, suggesting the existence of an ubiquitous physical mechanism regulating the energy transfer between the accretion disc and the X-ray emitting corona.
Recently, our group developed a method to use this relation in the observational cosmology, turning quasars into standardizable candles.
This work has the main aim to investigate the potential evolution of this correction at high redshifts. 
We thus studied the $L_{\rm X}-L_{\rm UV}$ relation for a sample of quasars in the redshift range 4<$z$<7, adopting the selection criteria proposed in our previous work regarding their spectral properties.
The resulting sample consists of 53 Type 1 (unobscured) quasars, observed either with \emph{Chandra} or XMM-\emph{Newton}, for which we performed a full spectral analysis, determining the rest-frame 2 keV flux density, as well as more general X-ray properties such as the estimate of photon index, and the soft (0.5-2 keV) and hard (2-10 keV) unabsorbed luminosities.
We find that the relation shows no evidence for evolution with redshift.
The intrinsic dispersion of the L$_X$-L$_{UV}$ for a sample free of systematics/contaminants is of the order of 0.22 dex, which is consistent with previous estimates from our group on quasars at lower redshift. 
}
 
\keywords{quasars - active galactic nuclei - X-ray - high redshift}
 
\maketitle
%
%
\section{Introduction}
An observational non-linear relation between the UV and the X-ray monochromatic luminosities in active galactic nuclei (AGN) has been known for decades ($L_{\rm X}\propto L_{\rm UV}^\gamma$, e.g. \citealt{AvniTananbaum86}).
This relation shows a slope $\gamma$ around 0.6 over several orders of magnitude in luminosity and up to high redshifts irrespective on the sample selection (e.g. X-ray or optically selected samples, \citealt{Vignali03c}; \citealt{Strateva05}; \citealt{Steffen06}; \citealt{Just07}; \citealt{Lusso10}; \citealt{Young10}), suggesting that a universal physical mechanism is driving the non-linear dependence between the X-ray and UV emission.
These properties indicate that the physical mechanism responsible for the observed relation has to be universal.
Indeed, these sources are powered by the accretion of matter onto the central supermassive black hole (SMBH), through an accretion disk \citep{SS73}, where the gravitational energy of the in-falling material is efficiently transformed into UV radiation.
This is the so-called Big Blue Bump (BBB), which is the major contribution to the Spectral Energy Distribution (SED) of a quasar.
The observed emission in the X-ray band (corresponding to $\sim1-10\%$ of the total power, e.g. \citealt{Lusso12}) is due to inverse-Compton reprocessing of seed photons from the disk, by a corona of hot electrons located in the vicinity of the SMBH (e.g., \citealt{HM93}).
In order to maintain a stable emission, the hot coronal gas needs to be continuously reheated, but the physical process responsible for the steady energy transfer from the disk to the corona is not yet well understood.
A fully consistent physical model able to predict the observed relation has yet to be found, despite some toy model have been proposed (e.g., \citealt{HM91}; \citealt{HM93}; \citealt{SZ94}; \citealt{DiMatteo98}; \citealt{Merloni03}; \citealt{LussoRisaliti17}).
A better understanding of the properties of the $L_{\rm X}-L_{\rm UV}$ relation can provide stringent constraints on the unknown physical process which stands behind it. 
Recently, our group developed a technique that uses this non-linear relation in observational cosmology, turning quasars into \emph{standardizable candles} \citep{RisalitiLusso15}.
Thanks to this technique, we can study the evolution of the universe in the redshift range 2<$z$<7.5, which is poorly investigated by other cosmological probes such as Type Ia supernove ($z$<1.4, \citealt{Betoule}) and Baryon Acoustic Oscillations (BAO at $z$$\sim$2, \citealt{Aubourg15}; \citealt{duMasdesBourboux17}), except for gamma-ray burst (e.g., \citealt{Ghirlanda04}).
Since a potential evolution of the relation with redshift could hamper the use of quasars as cosmological tools, in this work we investigate the presence of potential systematics of the $L_{\rm X}-L_{\rm UV}$ relation at high redshifts, using the largest quasar sample available in the redshift range 2<$z$<7 of finely selected objects, and taking advantages of the method developed by our group in previous works.
The paper is organized as follows: in $\S$\ref{sampleselc} we introduce the sample and the selection criteria adopted, in $\S$\ref{xraydata} and in $\S$\ref{opticdata} we outline the procedures performed in order to obtain the X-ray and UV flux estimates, respectively; in $\S$\ref{quasar_properties} we present the properties of the sample; in $\S$\ref{relation-analysis} the analysis of the relation is presented, along with the results, then in $\S$\ref{conclusions} the conclusions of this work are presented.
The luminosity distances were estimated assuming a concordance flat $\Lambda CDM$ model with the matter density parameter $\Omega_{M}=0.30$, the dark energy density parameter $\Omega_{\Lambda}=0.70$ and the Hubble constant $H_0=70$ km s$^{-1}$ Mpc$^{-1}$ \citep{komatsu09}.
%
%
\section{Sample selection}
\label{sampleselc}
\begin{figure*}
     \includegraphics[width=17cm]{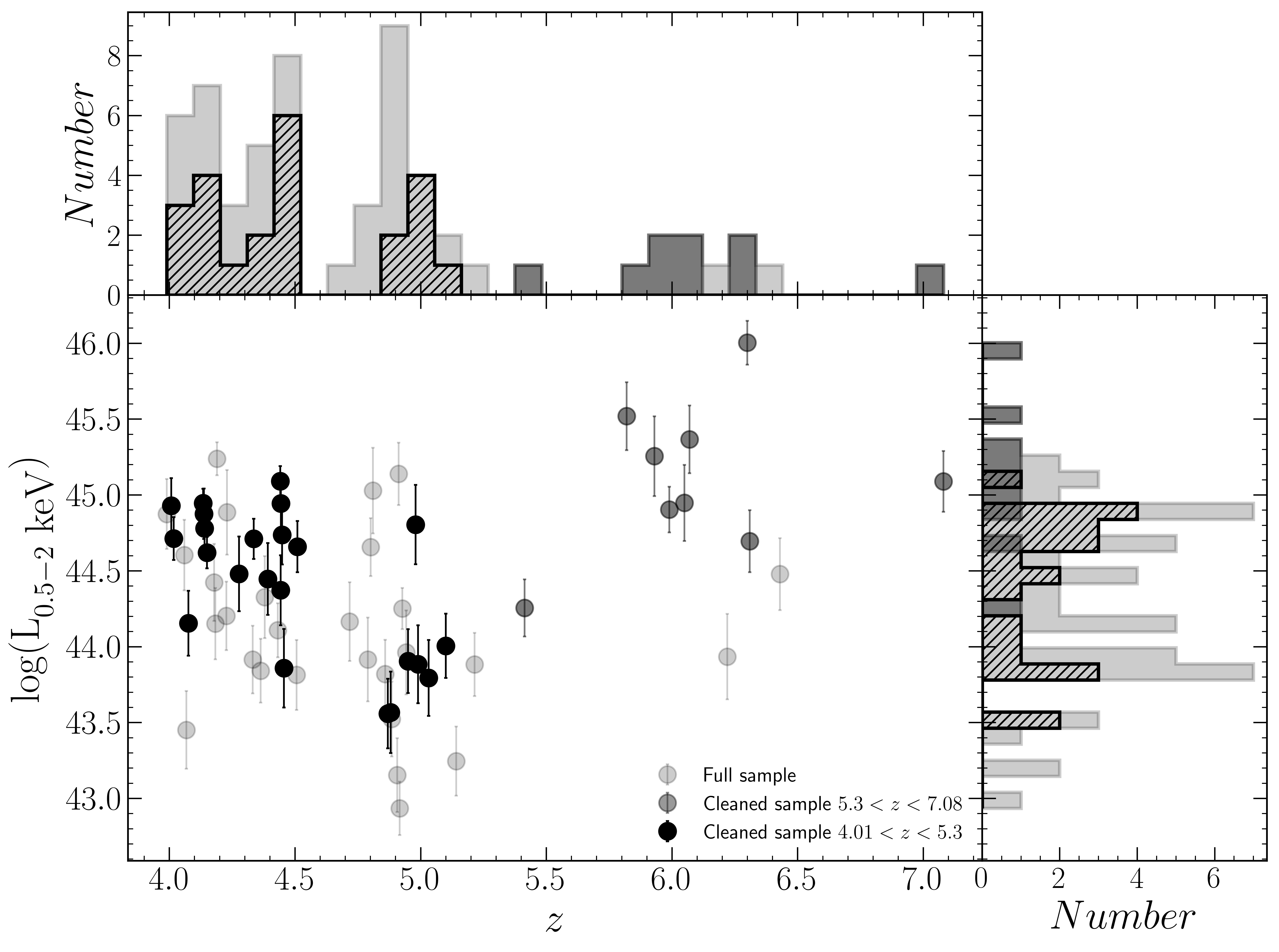}
     \caption{\emph{Central panel:} the distribution of the \textbf{estimated} rest-frame soft X-ray luminosity (0.5-2 keV) vs. redshift for the final sample, colour-coded as follows: the cleaned sample of 22 sources with $z$<5.3 are in solid black; in dark greys we report those selected at $z$>5.3 (9 out of 53); the remaining (22/53) in light grey are those not included in the analysis due to the selection criteria applied (e.g., X-ray flux upper limits, too steep or flat X-ray spectral slope). \newline \emph{Upper panel and right panel:} the redshift and soft X-ray luminosity distribution of the final quasar sample, respectively. The colour coding adopted is the same used in the central panel except for hatched black filling instead of the solid one.}
     \label{fig:sample}
\end{figure*}

In the last few years, our group proved that the $L_{\rm X}-L_{\rm UV}$ relation in quasars is actually tight ($\sim$0.2 dex), once accurate selection criteria are applied and systematic effects are properly taken into account (e.g. non-simultaneity/variability of the observations, gas absorption, dust reddening, host galaxy contamination). 
In this paper, we selected a sample of high-redshift quasars spectrally classified as Type 1 (i.e., unobscured) and possibly observed with the same facility in order to maintain the sample as homogeneous as possible and to avoid potential systematic effects, which could lead to a larger observed dispersion.\\
As discussed before, we are interested in the study of the relation at the highest redshift, so we considered the updated catalogue by \citeauthor{Brandt_Vignali_list}\footnote{\emph{http://personal.psu.edu/wnb3/papers/highz-xray-detected.txt}}, which consist of 158 quasars with redshift in the interval 3.96<$z$<7.08, detected in the X-rays.
We selected the 138 optically-selected quasars out of the original 158, which have been observed either with \emph{Chandra} or XMM-\emph{Newton}.
We then included \textsc{SDSS J114816.7+525150.4} at $z$=6.43 and \textsc{SDSS J010013.0+280225.9} at $z$=6.30, both observed with XMM-\emph{Newton}, from the catalogue of high redshift quasars by \cite{Nanni17}.
Given the 140 sources with at least an X-ray observation, we searched for the UV coverage following this approach: i) we cross matched our sample with the catalogue by \cite{Shen}, which provides the rest-frame $2500\AA$ flux density for 36 out of the 138 quasars in the redshift range 4.01<$z$<4.99; ii) we then cross matched the remaining 104 objects with the \emph{Sloan Digital Sky Survey} (SDSS) Data Release 7 (DR7; \citealt{sdssdr7}) and Data Release 12 (DR12; \citealt{Paris17}) catalogues, providing the optical/UV spectra for 6 and 4 additional quasars, respectively, in the redshift range 5.0<$z$<5.4; iii) for sources with a redshift $z$>5.4, we searched in the literature for their UV spectra, and we found data for 12 of them, which have been observed with a number of different facilities from the SDSS (the references for each source are provided at the bottom of Table \ref{table:data}).
Further discussion on the adopted rest-frame $2500\AA$ monochromatic flux estimates for each group of sources is provided in $\S$\ref{opticdata}.\\
The resulting sample consists of 58 quasars, in the redshift range 4.01<$z$<7.08, which benefit from a \textbf{moderate}-quality coverage in both UV and X-ray bands.\\
We then applied to this sample a series of selection criteria following the procedure presented by \cite{lusso-risaliti16}.
Specifically, we chose unobscured optically-selected quasars, classified as radio quiet sources (i.e. with radio-loudness parameter $R\ =\ F_{\nu , 6cm}\ /\ F_{\nu , 4400\AA}$ lower than 10, here 57/58), showing no Broad Absorption Line features (identified as BAL in the literature, 4 in the sample).\\
Our final clean sample of high redshift objects is thus composed by 53 objects spanning the redshift range 4.01<$z$<7.08.
Taking advantage of the spectral and spatial resolution of the X-ray observations from \emph{Chandra} and \emph{XMM Newton}, we performed a full spectral analysis on the archival data of the quasars in the sample. We catalogued the X-ray properties (i.e. spectral index, rest frame 2 keV monochromatic flux, rest-frame 0.5-2 keV and 2-10 keV X-ray luminosities) for the sample in Table \ref{table:data}.
The distribution of the 53 sources in terms of soft X-ray luminosity and redshift is presented in Fig. \ref{fig:sample}. 
%
%
\section{X-ray data}
\label{xraydata}
\subsection{X-ray data reduction}
Of the 53 quasars, 47 objects have been observed with \emph{Chandra} and 9 with XMM-\emph{Newton} (\emph{ULAS J1120+0641}, \emph{SDSS J114816.7+525150.4} and \emph{SDSS 1030+0524} have been observed with both).
For each observation, we followed the standard data reduction procedures (which depend on the telescope), obtaining a background-subtracted spectrum in the $\sim$0.1-10 keV band. 
We reprocessed \emph{Chandra} data using the dedicated software CIAO v. 4.9.
For on-axis observations (i.e. with source off-axis angle $\theta\ $<\ 1'), we extracted the source and the background counts from a circular radius of 2'', centred on the source optical position, corresponding to 95\% of the \emph{Encircled Energy Fraction} (EEF) at 1.5 keV.
Counts from off-axis sources ($\theta\ $>\ 1') were selected using 10" radius circular regions, corresponding to at least 90\% of the EEF.
Background counts were extracted from contiguous source-free circular regions, having $\sim15''$ radii.\\
In the case of XMM EPIC data we performed a step by step procedure using the Science Analysis Software (SAS) v16.
For each observation, we filtered for time intervals of high-background.
In order to increase the signal-to-noise ratio, we merged the two EPIC-MOS observations, while the pn observation was reduced independently. \\
Source and background counts were extracted from a circular region centred at the optical position of QSOs with radius of 15" for on-axis positions, corresponding to 70\% of the EEF ad 1.5 keV, 30" for off-axis observations, equivalent to at least 40\% of EEF at 1.5 keV. \\
Background counts were extracted from contiguous source-free circular regions, having $\sim60''$ radii.
%
%
\subsection{X-ray analysis}
\label{xrayanalysis}
For the X-ray spectral analysis we used the software XSPEC v. 12.9 \citep{arnaud+96}.
We assumed a \emph{cstat} statistic (Poisson data) for the majority of the spectra, and a $\chi^2$ statistic (Gaussian data) in the case of XMM-\emph{Newton} observations having a number of counts >100.
Galactic absorption is included in all the spectral models, and the fluxes presented in Table \ref{table:data} are corrected for this effect.\\
The sample is a collection of unobscured (type-I) quasars; their spectra are typically dominated by the continuum emission.
Additional features have been observed in the X-ray spectra of type-I AGN: fluorescence emission lines from the neutral iron (e.g., the Fe K$\alpha$ and K$\beta$ lines at rest-frame 6.40 and 7.06 keV, respectively), emission lines from the ionized iron (e.g., the Fe XXV and XXVI at rest-frame 6.70 and 6.97 keV, respectively), and a potential reflection component by the torus or the accretion disk.
The inclusion of model components either for the potential emission lines or reflection hump (which has been found to be weak in luminous type-I quasars at high redshift; e.g., \citealt{Shemmer05}) was not possible in the vast majority of the observations collected in this work due to the relatively low photon counts statistics of the spectra.
In the few observations with a relatively higher statistics, we tried to include additional Gaussian components among those discussed above, but they did not improve significantly the quality of the fit and the parameters were not constrained (considering the 90 per cent confidence level).
Therefore, the adopted model consists of a single power-law for the primary emission, where the slope and the normalization are free parameters.
When the number of counts was not sufficient to perform a full spectral analysis, we evaluated an upper-limit to flux density, freezing the power-law slope at $\Gamma=1.9$, which is the average value for unobscured quasars (e.g., \citealt{Vignali03d}, \citealt{Nanni17}).
This occurred for 9 sources, properly flagged in Table \ref{table:data}. 
For XMM-\emph{Newton} observations we fitted together the EPIC-pn and the merged EPIC-MOS spectra, introducing a cross-calibration constant between the two datasets, in order to account for the different camera responses.
The values obtained for this constant are fully consistent within 8\% (e.g., \citealt{read-guainazzi}).
In the case of multiple observations from the same telescope, we adopted the following approach: i) we chose the observation with the longest exposure if the difference between two observations is significant (e.g. in the case of \emph{PSS0133+0400} we chose the 64 ks observation instead of the 6 ks one). 
ii) Otherwise, we checked for any potential variability, both in the slope and in the flux and, if the results from the fitting procedures were consistent within the uncertainties, we fitted jointly all the observations with the same model, with a free cross-normalization constant for any potential minor variability in the flux or calibration within different observations.
If the sources were observed with both \emph{Chandra} and XMM-\emph{Newton}, we compared the two best fit models in order to test for X-ray variability.
We then considered the result from the model with the longest exposure time or, in case of similar ones, we used the best fit parameters from the model with the lowest $\chi^{2}$.\\
The fluxes have been estimated integrating the continuum emission over a given energy band (e.g., rest-frame 0.5-2keV).
To this purpose, we used the \emph{cflux} convolution model in XSPEC, which provides the integrated emission and the associated uncertainty.
\textbf{Given the redshift of the sample (4<$z$<7), the soft band (rest-frame 0.5-2 keV) has been marginally detected by \emph{Chandra} and XMM-\emph{Newton} only for the sources with the lowest redshift ($z$<5.5).
Then, the soft band fluxes (and luminosities) have been estimated from the extrapolation of the power law fitted to the hard band (rest-frame 2-10 keV) spectrum, assuming that it follows the same law.}
The rest-frame 2 keV flux density has been estimated dividing the flux of the continuum emission over a narrow energy band (corresponding to rest-frame 0.01 keV) centred on the rest-frame 2 keV by the width of the energy band itself.
In the case of sources with a redshift for which the rest-frame 2 keV fell outside the observable energy bands of \emph{Chandra} and XMM-\emph{Newton}, we estimated the fluxes from the extrapolation of the power law modelling the continuum emission \textbf{in the rest-frame hard band}.\\
Apart from one case that we will briefly discuss below, we did not find evidence for significant X-ray variability.
This is in agreement with the results from recent works on the monitoring of samples of high-redshift quasars, which showed that these objects appeared to be less variable in the X-rays with respect to active nuclei in the local Universe (e.g., \citealt{Shemmer17}).
The only source showing evidence for a strong X-ray variability is \emph{SDSS 1030+0524}, which has been the target of three different observations over fifteen years.
The recent 500ks monitoring by \emph{Chandra}, presented by \cite{Nanni18}, confirmed the source properties obtained from the previous one (2002) with the same telescope, but are in significant disagreement with those from the 2003 \emph{XMM-Netwon} observation.
The observed variability, affecting both the flux (up to a factor $\sim2.5$) and the spectral shape, raises questions about the potential variability in quasars observed when the Universe had less than 1 Gyr, when these sources could still be going through the early stages of their evolution. 
In this particular case, the low monitoring and the poor counting statistics do not allow us to disentangle between different scenarios, e.g., the variation in the obscuration level along the line of sight (as observed in local AGN, e.g. \citealt{Risaliti07}) or variation in the accretion process onto the central supermassive black hole. 
Further investigation on this issue are needed, but they require long-term monitoring and higher statistics observations for these high-redshift sources.\\
Except for this particular case, there are no hints for any significant evolution in the X-ray properties of our properly selected samples of quasars.\\
In Table \ref{table:data} we present their X-ray properties, i.e. the spectral slope and the estimates of the fluxes.
We found results which are in agreement, within the uncertainties, with literature works on similar collection of high-redshift quasars (e.g., \citealt{Nanni17}, \citealt{Shemmer17}).
We observe a mean spectral slope $\Gamma=1.9$ with a dispersion of $\sim0.5$, in agreement with the spectral properties of quasars at similar and lower redshifts (1<$z$<5.5, e.g. \citealt{Shemmer06}, \citealt{Just07}, \citealt{Vignali05}).
\begin{figure}[t!]
  \resizebox{\hsize}{!}{\includegraphics{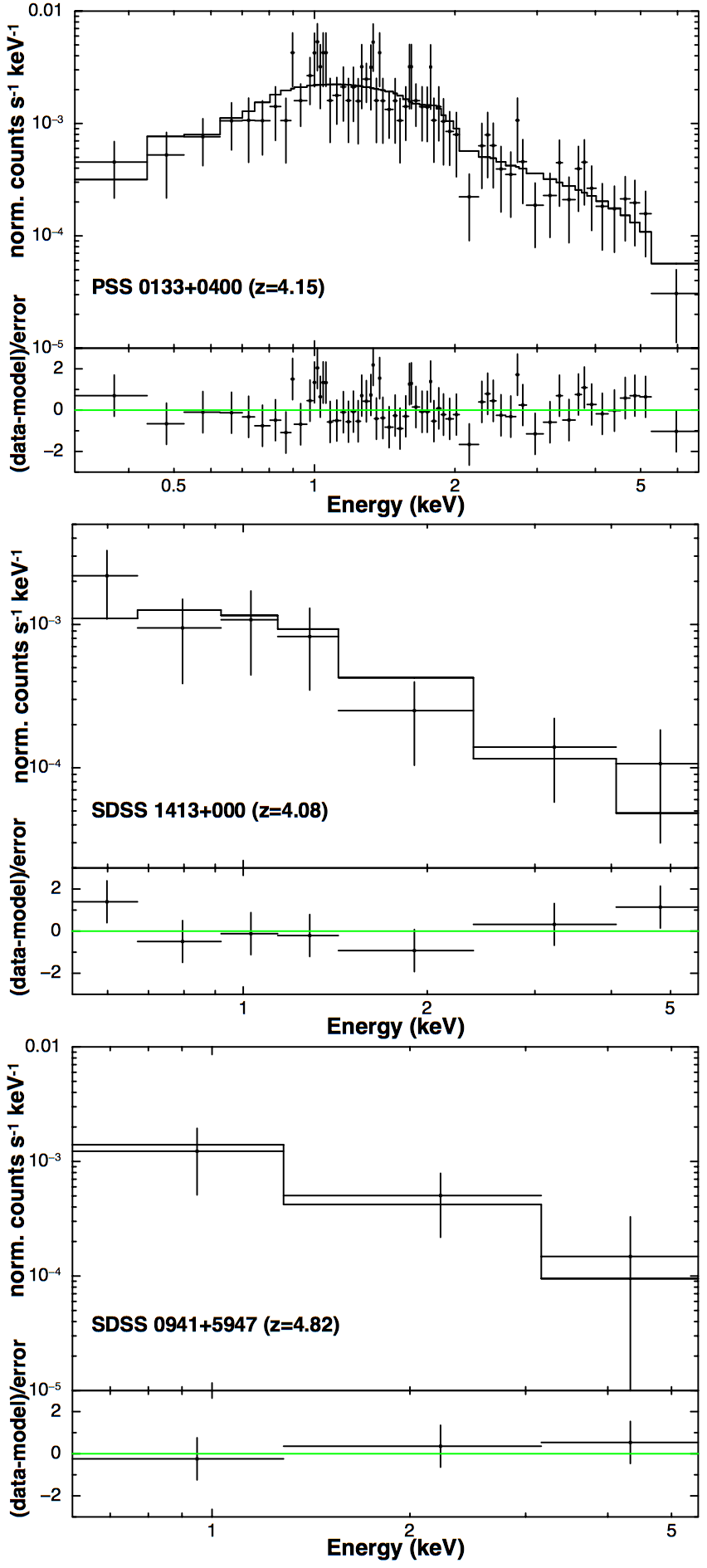}}
  \caption{Examples of spectra of three quasars with decreasing statistics, from top to bottom. The best fit models (consisting of a power law modified by Galactic absorption) and the data are plotted in each of the upper panels, \textbf{as a function of the observed-frame energies, in units of keV}. The residuals (data-model, in units of sigma) are shown in the lower panels.}
  \label{fig:xray_analysis}
\end{figure}
%
%
\section{Optical Data}
\label{opticdata}
The proxy of the UV luminosity is usually obtained from the rest-frame $2500\AA$ flux density. 
For a large part of the sources in our sample (33 out of 53), which were included in the SDSS DR7, we adopted the values compiled by \cite{Shen}. 
The authors provide the $2500\AA$ flux density only for the sources having a redshift $z\lesssim5$, since there are no spectral windows free of emission lines at higher redshift, where they can accurately fit the continuum within the SDSS wavelength coverage (3800-9200$\AA$).
In particular, the authors performed a spectral fit on the available spectral windows (i.e., up to the observed-frame 9200$\AA$) with a power law for the continuum emission and a template including FeII and many emission lines (see \citealt{Shen} for further details).
Then, they extrapolate the rest-frame 2500$\AA$ flux density using the slope of the continuum obtained from the fit.\\
For the remaining sources in our sample observed within the SDSS DR7 (6/53), for which \cite{Shen} do not provide a 2500$\AA$ flux density measurement (i.e. $z$>5), or observed within the SDSS DR12 (4/53), we obtain the monochromatic UV flux assuming the continuum spectrum to be a power law $S\propto \nu^{-\alpha}$ with $\alpha$=0.50 (e.g., \citealt{VandenBerk2001}; \citealt{Lusso15}).
We then use this slope to extrapolate the value at 2500$\AA$, starting from the median flux value in the rest-frame 1430-1470$\AA$ waveband, the last (shortest wavelength) continuum window free from emission lines and for which the Intergalactic Medium absorption is not relevant \citep{Lusso15}.
The reliability of this method (i.e., the extrapolation of the flux density with a fixed slope) is discussed in Appendix \ref{appendix:a}.\\
In the case of quasar at redshift higher than 5.5, the SDSS wavelength coverage is probing the far ultraviolet (i.e. $\lambda<1450\AA$ at the rest-frame), so we searched in the literature for any optical/NIR observation from which we can estimate the rest-frame 2500$\AA$ flux density (10 sources out of 53 having $z$>5.5).
Given the number of different telescopes used to observe this set of sources (references in Table \ref{table:data}), the UV flux estimates can be affected by a larger dispersion due to potential cross-calibration uncertainties among different cameras.
In order to assess the potential contribution of this effect on the shape of the L$_{\rm X}$-L$_{\rm UV}$ relation, in $\S$\ref{relation-analysis} we presented the results obtained with and without the inclusion of this high-redshift subsample.\\
Since the catalogue published by \cite{Shen} does not list any uncertainty on the UV flux density, we calculate the standard deviation of the fluxes in the rest-frame 1430-1470$\AA$ waveband on each of the available SDSS DR7 spectra.
We assume this as the uncertainty on the rest-frame 2500$\AA$ flux density.
We adopt this method also for the 10 sources from the SDSS DR7 (having $z$>5, hence not included in the catalogue by \citealt{Shen}) and DR12 for which we extrapolated the UV flux density with a fixed slope.
This method provides an average uncertainty of 22\% over the entire sample of 43 quasars observed within the SDSS releases.\\
For the quasars at $z$>5.5, we do not have an estimate of the uncertainty on the rest-frame 2500$\AA$ flux density.
The optical spectra we found in the literature for these objects (references in Table \ref{table:data}) have a comparable, if not higher, signal-to-noise ratio with respect to SDSS.
We thus considered a 22\% uncertainty on the rest-frame 2500$\AA$ flux density similarly to what we assumed for the entire sample.
%
%
\section{Quasar properties}
\label{quasar_properties}
The properties of the quasar of our sample are shown in Table 1, where the content of each columns is: \\
\emph{Column 1:} Source name.\\
\emph{Column 2:} Right Ascension (\emph{RA}) in the \emph{J2000} frame.\\
\emph{Column 3:} Declination (\emph{DEC}) in the \emph{J2000} frame.\\
\emph{Column 4:} Archival source redshift.\\
\emph{Column 5:} Rest frame $2500\AA$ flux density, in unity of $10^{-28}$ erg s$^{-1}$ cm$^{-2}$ Hz$^{-1}$. Uncertainties are assumed being about the $22\%$ of the flux. See $\S$\ref{opticdata} for details.\\
\emph{Column 6:} The rest-frame 0.5-2 keV luminosity in units of $10^{44}$ erg s$^{-1}$.\\
\emph{Column 7:} The rest-frame 2-10 keV luminosity in units of $10^{44}$ erg s$^{-1}$.\\
\emph{Column 8:} X-ray spectral photon index, evaluated through the fitting procedure (see $\S$\ref{xraydata}).\\
\emph{Column 9:} Rest frame $2$ keV flux density, in unity of $10^{-32}$ erg s$^{-1}$ cm$^{-2}$ Hz$^{-1}$. See $\S$\ref{xraydata} for details.\\ 
\emph{Column 10:} Reduced $\chi^{2}$ of the X-ray spectral fit procedure.\\
\emph{Column 11:} Number of degrees of freedom in the X-ray spectral fit, i.e. the number of spectral bins minus the number of the free parameters (which are 2 for non-upper limit estimates, i.e. the power-law slope and normalization).\\
\emph{Column 12:} Exposure time of the X-ray observation in units of ks.\\
\emph{Column 13:} Telescope used for this observation: \emph{C} for \emph{Chandra}, \emph{X} for XMM-\emph{Newton}.
%
%
\section{The L$_{\rm 2 keV}$-L$_{\rm 2500\AA}$ relation}
\label{relation-analysis}
The non-linear relation between the X-rays (L$_{\rm X}$) and UV (L$_{\rm UV}$) luminosities can be parametrized as 
\begin{equation}
\log(L_{\rm X})=\beta + \gamma \log(L_{\rm UV}), 
\end{equation}
where $\beta$ is a normalization constant and $\gamma$ is the observed slope.
Expressing the luminosities in terms of fluxes and distances, i.e. $L=F~4~\pi~D_L^2$, we obtain:
\begin{equation}
\log(F_{\rm X})=\beta + \gamma \log(F_{\rm UV}) - (\gamma-1) \log(4\pi D_L^2)
\label{Xflux}
\end{equation}
where $F_{\rm X}$, $F_{\rm UV}$ and $D_{L}$ are the X-ray and UV flux densities, and the luminosity distance, respectively.
In order to perform the analysis of the relation, we considered the following likelihood function:
\begin{equation}
p_{i}(F_{\rm X}|F_{\rm X,est})=\prod_{i} \frac{1}{\sqrt{2\pi s_{i}^{2}}}exp\left[-\frac{(F_{\rm X}-F_{\rm X,est})^2}{2 s_{i}^{2}}\right],
\label{eq:likelihood}
\end{equation}
where $F_{\rm X}$ is the rest-frame 2 keV flux estimated as described in $\S$\ref{xraydata}, while $F_{\rm X,est}$ is the one estimated using Eq. \ref{Xflux} for a given UV flux, and $s_{i}^2=\sigma_{UV,i}^2+\sigma_{X,i}^2+\delta_{\rm intr}^2$, where $\delta_{\rm intr}$ is the intrinsic dispersion, $\sigma_{UV,i}$ and $\sigma_{X,i}$ are the uncertainties on the UV and X-ray fluxes, respectively.
To perform the fitting procedure, we used the \textsc{Python} package \verb#emcee# \citep{emcee}, which is an implementation of the Goodman \& Weare's Affine Invariant Markov chain Monte Carlo (MCMC) Ensemble sampler.
%
%
\subsection{Results}
\begin{figure*}
\centering
   \includegraphics[width=\textwidth]{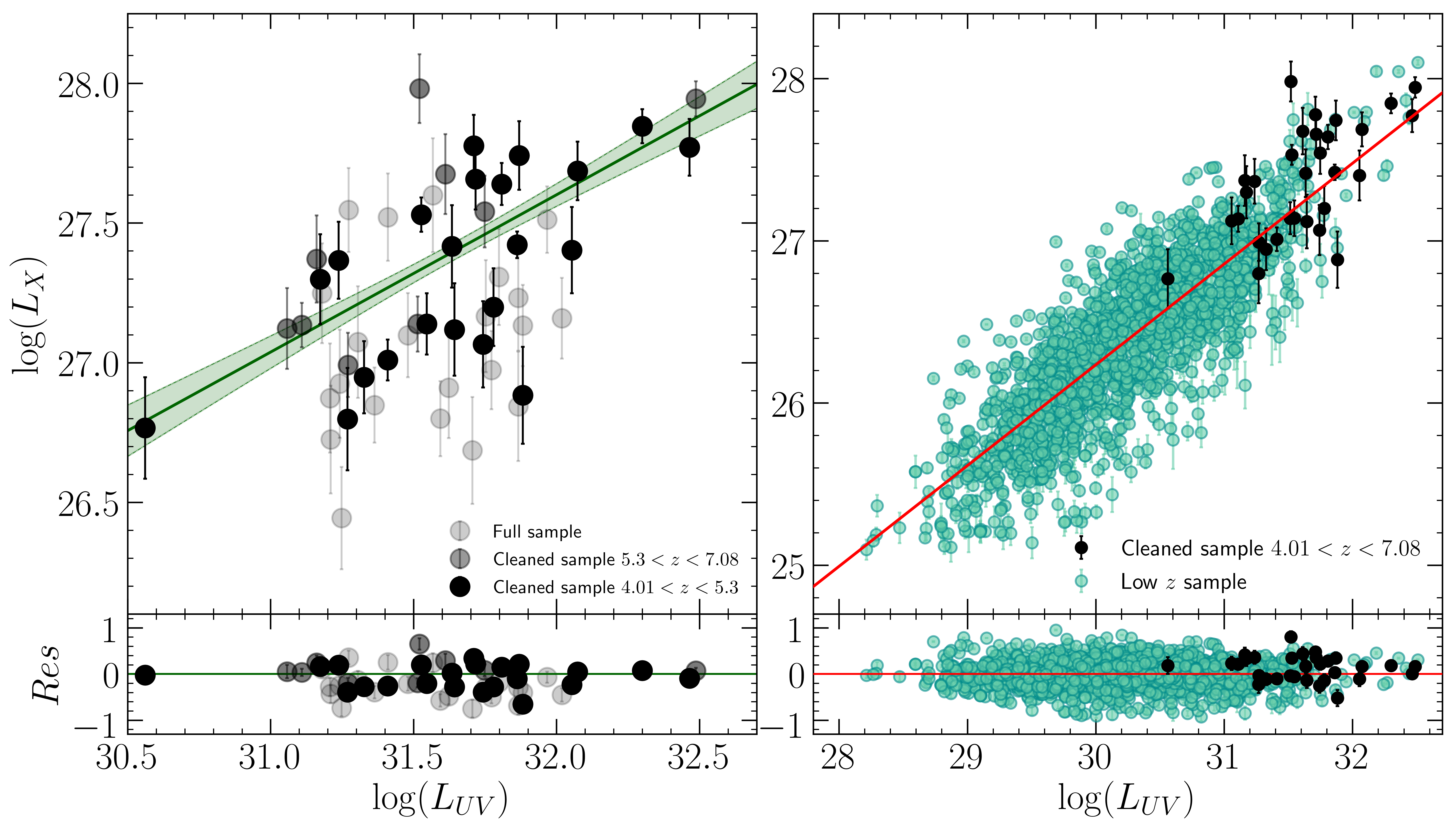}
     \caption{\emph{Left panel:} $\log(L_{\rm X})$ vs. $\log(L_{\rm UV})$ plot for the high-redshift quasars sample ($z$>4), where $L_{\rm X}$ and $L_{\rm UV}$ are the monochromatic luminosities at rest-frame 2 keV and $2500\AA$ in units of in erg s$^{\rm -1}$ Hz$^{\rm -1}$, respectively.
     Full black circles are the quasars (22 sources) of the cleaned sample having $z$<5.3, dark grey circles are those (9) of the cleaned sample having 5.3<$z$<7.08. The remaining sources, represented with light grey circles, are those excluded due to the selection criteria adopted, as explained above (e.g., X-ray flux upper limit, X-ray power-law slope $\Gamma_{\rm X}$ too steep or too flat).
     The best-fit parameters obtained from the MCMC regression analysis on the full cleaned sample are represented by the green solid line, while the light green area covers the parameters space between the 16th and 84th percentile ($\gamma=0.53\pm0.11$, $\beta=27.43\pm0.05$, $\delta_{\rm intr}=0.20\pm0.04$, obtained with the sample of 31 objects). 
    On the bottom, the residual between the data and the best fit model.
     \newline
     \emph{Right panel:} $\log(L_{\rm X})$ vs. $\log(L_{\rm UV})$ plot, in black the full cleaned high-redshift sample ($z$>4), in cyan the sample at lower redshift by \cite{lusso-risaliti16}.
     It is clear that the quasars with $z$>4 follow the same relation of those at the lowest redshift, suggesting that there is no evolution with cosmic time for the relation.
     On the bottom, the residual between the data and the model, obtained with the same MCMC regression analysis, on the collection of sources containing both samples here represented ($\gamma=0.62\pm0.09$, $\beta=27.32\pm0.16$, $\delta_{\rm intr}=0.27\pm0.04$).
     }
     \label{fig:plotlxluv}
\end{figure*}

In order to follow the approach developed by our group in previous works, we further cleaned the sample on the basis of the X-ray properties and tested the effects of these cuts on the shape of the L$_{\rm X}$-L$_{\rm UV}$ relation.
We excluded the sources for which the rest-frame 2 keV flux estimate was an upper limit (9 out of 53 objects), and the quasars showing a too steep or flat X-ray spectra. 
Regarding the latter criterion, we excluded the 13 quasars showing an X-ray spectral slope which differs significantly from the peak values of the distribution of photon index for bright, unobscured quasars, i.e. $\Gamma_{\rm X}$=1.9-2.0 (e.g., \citealt{Bianchi09}). 
The peculiar slope could suggest either intrinsic obscuration (i.e. $\Gamma_{\rm X}<1.5$, which is rare to observe as intrinsic photon index) or an extreme object (i.e. $\Gamma_{\rm X}>2.8$).
These criteria were applied in order to avoid the presence of possible contaminants (e.g., absorbed quasars) and to maintain the sample as homogeneous as possible.\\
The results of the analysis of the L$_{\rm X}$-L$_{\rm UV}$ relation on the cleaned sample, consisting of 31 sources with redshift 4.01<$z$<7.08, are: $\gamma=0.53^{+0.11}_{-0.11}$, $\beta=27.46^{+0.05}_{-0.05}$ and $\delta_{\rm intr}=0.20^{+0.04}_{-0.04}$ dex.
These results are fully consistent with the observed slope in other literature works at various redshifts (e.g., \citealt{Vagnetti13}, \citealt{Vagnetti10}, \citealt{lusso-risaliti16} and \citealt{Nanni17}) and with samples selected upon different criteria.
As said in $\S$\ref{opticdata}, the rest-frame 2500$\AA$ flux density for objects having redshifts $z$>5.5 have been obtained from different telescopes, and that could be a further source of dispersion due to potential cross-normalization uncertainties within different facilities.
In this regard, we excluded \emph{SDSS 0231-0728}, that has been observed within the SDSS DR12, but due to the combination of its redshift ($z=5.41$) and the limited coverage of the SDSS $z$ band (which covers up to $\sim9200\AA$), has an unreliable estimate of the optical flux.
Due to the limited SDSS coverage for quasars at redshift higher than about 5, we decided to perform the whole analysis by excluding these sources (9 objects), leading to a subsample of 22 quasars covering the range 4.01<$z$<5.3. 
The results of the analysis, considering this subsample, are: $\gamma=0.55^{+0.14}_{-0.14}$, $\beta=27.43^{+0.05}_{-0.05}$ and $\delta_{\rm intr}=0.19^{+0.05}_{-0.05}$ dex.
Since the results are fully consistent with those obtained with the sample extending up to the highest redshift, we can conclude that the potential dispersion introduced by including optical flux estimates from different instruments is negligible with respect to others (the large uncertainties affecting the X-ray fluxes, for instance).
It is important to note that the intrinsic dispersion obtained using both the cleaned sample with and without the cut at redshift $z$=5.3, is consistent with that found in similarly cleaned quasar samples at lower redshift (e.g., \citealt{lusso-risaliti16}) and lower with respect to what previously reported in literature (e.g., \citealt{Lusso10}; \citealt{Young10}).
This supports the hypothesis that the real intrinsic dispersion can be reduced once an accurate source selection has been applied and the sample has been made as homogeneous as possible.\\
Then, we tested the effects of the inclusion of different subsamples on the slope on the L$_{\rm X}$-L$_{\rm UV}$ relation.
First, we extended the sample to the sources with a flatter/steeper X-ray photon index, where the former class are those that can be affected by intrinsic obscuration, which have been observed in a significant fraction of quasars optically classified as unobscured (e.g., \citealt{Merloni14}).
Including the sources having 1.3<$\Gamma_X$<2.8, the observed slopes are: $\gamma=0.55^{+0.15}_{-0.15}$ for the quasars with $z$<5.3 (28/53) and $\gamma=0.51^{+0.12}_{-0.12}$ without any cut in redshift (38/53), respectively.
In both cases, the intrinsic dispersion increases (0.22 and 0.23 respectively) with respect to the analysis performed on the cleaned sample.
Similarly, including in the cleaned sample the quasars for which we can estimate at most an upper limit to the X-ray flux density, we get $\gamma=0.62^{+0.15}_{-0.15}$ and $\delta_{\rm intr}$$\sim$0.19 for 31 sources with a redshift cut at $z$=5.3. 
Then, considering the 41 quasars covering the entire redshift range 4.01<$z$<7.08, we get $\gamma=0.58^{+0.12}_{-0.12}$ and $\delta_{\rm intr}$$\sim$0.23.
We can conclude that the inclusion of these two subsamples does not affect significantly the shape of the L$_{\rm X}$-L$_{\rm UV}$ relation, but only the intrinsic dispersion, which increases.
This result is in agreement with those obtained from our group in the previous works on quasar samples at lower redshift, i.e. stricter selection criteria lead to a smaller dispersion due to the exclusion of potential contaminants.\\ 
The results of the analysis of the L$_{\rm X}$-L$_{\rm UV}$ relation, performed on the cleaned sample of 31 sources (4.01<$z$<7.08), are presented in the left panel of Fig. \ref{fig:plotlxluv}.
In the right panel, the same sources are plotted in the $\log(L_{\rm X})-\log(L_{\rm UV})$ plane, along with a sample of similarly selected quasars at lower redshift ($\sim750$ objects), presented by \citealt{lusso-risaliti16}: the high-redshift quasars here appear to follow the same relation as those at lower redshift.
%
%
%
%
\section{Summary \& Conclusions}
\label{conclusions}
We presented the results from our study of the non-linear $L_{\rm X}-L_{\rm UV}$ relation using a sample of 53 high-redshift (4<$z$<7.08) unobscured quasars.
The observed relation between the X-ray and the UV luminosities indicates the presence of an unknown physical mechanism that links the emission from the accretion disk with that one from the X-ray emitting corona.
The study of this relation, that has been observed over several orders of magnitude in luminosities and up to high-redshift, can provide hints on the nature of this mechanism, placing constraints on the energy generation and transfer in the accretion disk and the surrounding environment in AGN.
The main contribution of this work stands in the use of a carefully selected quasars sample in the highest possible redshift range, the X-ray spectral properties of which have been determined through a full spectral analysis, presented in Table \ref{table:data} and in Fig. \ref{fig:app_b} in Appendix \ref{appendix:b}.
The main results obtained in this work are:
\begin{itemize}
\item The observed X-ray spectral properties are consistent with those at lower redshift, e.g., a mean spectral slope $\Gamma_{\rm X}=1.9\pm0.5$.  
\item For the analysis of the relation, we considered only 31 sources, excluding X-ray upper limits, sources with too steep/too flat spectra.
We observed a dependence ($\gamma=0.53^{+0.11}_{-0.11}$) which is consistent with that observed at lower redshift ($\gamma \sim 0.6$), hence no evolution with the cosmic time has been found.
\item The intrinsic dispersion ($\delta_{\rm intr}=0.20^{+0.04}_{-0.04}$ dex), appears to be lower than in archival works. This is due to the adopted selection criteria, reducing some of the potential systematic effects induced by contaminants, and to the accurate flux estimates, as found in a previous work from our group \citep{lusso-risaliti16}.
\item Releasing any of the selection criteria (e.g., including the sources with too steep/too flat spectra) lead to a larger intrinsic dispersion.
\end{itemize}
Our study further supports a non-evolution of the relation between the X-ray and UV luminosities with redshift, suggesting a universal mechanism linking the emission from the hot corona to the one from the accretion disk.
Moreover, the non-linearity of the relation provides a new, powerful way to estimate the absolute luminosity, turning quasars into a new class of \emph{standard candles} that can provide an important contribution in the determination of the cosmological parameters, in particular probing its evolution and expansion in a redshift range that can not be explored using other known observational methods.
Indeed, the preliminary results of the work presented here have been already used in a Hubble diagram of quasars, recently published by our group.
As shown in recent works from our group (e.g., \citealt{RisalitiLusso18}, \citealt{Lusso19}), cosmological investigation with quasars can be pursued even now.
%
%
\bibliographystyle{aa} 
\bibliography{biblio_high_z.bib} 

\begin{thebibliography}{52}
\expandafter\ifx\csname natexlab\endcsname\relax\def\natexlab#1{#1}\fi

\bibitem[{{Abazajian} {et~al.}(2009){Abazajian}, {Adelman-McCarthy},
  {Ag{\"u}eros}, {Allam}, {Allende Prieto}, {An}, {Anderson}, {Anderson},
  {Annis}, {Bahcall}, \& et~al.}]{sdssdr7}
{Abazajian}, K.~N., {Adelman-McCarthy}, J.~K., {Ag{\"u}eros}, M.~A., {et~al.}
  2009, \apjs, 182, 543

\bibitem[{{Arnaud}(1996)}]{arnaud+96}
{Arnaud}, K.~A. 1996, in Astronomical Society of the Pacific Conference Series,
  Vol. 101, Astronomical Data Analysis Software and Systems V, ed. G.~H.
  {Jacoby} \& J.~{Barnes}, 17

\bibitem[{{Aubourg} {et~al.}(2015){Aubourg}, {Bailey}, {Bautista}, {Beutler},
  {Bhardwaj}, {Bizyaev}, {Blanton}, {Blomqvist}, {Bolton}, {Bovy},
  {Brewington}, {Brinkmann}, {Brownstein}, {Burden}, {Busca}, {Carithers},
  {Chuang}, {Comparat}, {Croft}, {Cuesta}, {Dawson}, {Delubac}, {Eisenstein},
  {Font-Ribera}, {Ge}, {Le Goff}, {Gontcho}, {Gott}, {Gunn}, {Guo}, {Guy},
  {Hamilton}, {Ho}, {Honscheid}, {Howlett}, {Kirkby}, {Kitaura}, {Kneib},
  {Lee}, {Long}, {Lupton}, {Maga{\~n}a}, {Malanushenko}, {Malanushenko},
  {Manera}, {Maraston}, {Margala}, {McBride}, {Miralda-Escud{\'e}}, {Myers},
  {Nichol}, {Noterdaeme}, {Nuza}, {Olmstead}, {Oravetz}, {P{\^a}ris},
  {Padmanabhan}, {Palanque-Delabrouille}, {Pan}, {Pellejero-Ibanez},
  {Percival}, {Petitjean}, {Pieri}, {Prada}, {Reid}, {Rich}, {Roe}, {Ross},
  {Ross}, {Rossi}, {Rubi{\~n}o-Mart{\'{\i}}n}, {S{\'a}nchez}, {Samushia},
  {G{\'e}nova-Santos}, {Sc{\'o}ccola}, {Schlegel}, {Schneider}, {Seo},
  {Sheldon}, {Simmons}, {Skibba}, {Slosar}, {Strauss}, {Thomas}, {Tinker},
  {Tojeiro}, {Vazquez}, {Viel}, {Wake}, {Weaver}, {Weinberg}, {Wood-Vasey},
  {Y{\`e}che}, {Zehavi}, {Zhao}, \& {BOSS Collaboration}}]{Aubourg15}
{Aubourg}, {\'E}., {Bailey}, S., {Bautista}, J.~E., {et~al.} 2015, \prd, 92,
  123516

\bibitem[{{Avni} \& {Tananbaum}(1986)}]{AvniTananbaum86}
{Avni}, Y. \& {Tananbaum}, H. 1986, \apj, 305, 83

\bibitem[{{Betoule} {et~al.}(2014){Betoule}, {Kessler}, {Guy}, {Mosher},
  {Hardin}, {Biswas}, {Astier}, {El-Hage}, {Konig}, {Kuhlmann}, {Marriner},
  {Pain}, {Regnault}, {Balland}, {Bassett}, {Brown}, {Campbell}, {Carlberg},
  {Cellier-Holzem}, {Cinabro}, {Conley}, {D'Andrea}, {DePoy}, {Doi}, {Ellis},
  {Fabbro}, {Filippenko}, {Foley}, {Frieman}, {Fouchez}, {Galbany}, {Goobar},
  {Gupta}, {Hill}, {Hlozek}, {Hogan}, {Hook}, {Howell}, {Jha}, {Le Guillou},
  {Leloudas}, {Lidman}, {Marshall}, {M{\"o}ller}, {Mour{\~a}o}, {Neveu},
  {Nichol}, {Olmstead}, {Palanque-Delabrouille}, {Perlmutter}, {Prieto},
  {Pritchet}, {Richmond}, {Riess}, {Ruhlmann-Kleider}, {Sako}, {Schahmaneche},
  {Schneider}, {Smith}, {Sollerman}, {Sullivan}, {Walton}, \&
  {Wheeler}}]{Betoule}
{Betoule}, M., {Kessler}, R., {Guy}, J., {et~al.} 2014, \aap, 568, A22

\bibitem[{{Bianchi} {et~al.}(2009){Bianchi}, {Guainazzi}, {Matt}, {Fonseca
  Bonilla}, \& {Ponti}}]{Bianchi09}
{Bianchi}, S., {Guainazzi}, M., {Matt}, G., {Fonseca Bonilla}, N., \& {Ponti},
  G. 2009, A\&A, 495, 421

\bibitem[{{Brandt} {et~al.}(2004){Brandt}, {Vignali}, {Schneider}, {Alexander},
  {Anderson}, {Bauer}, {Fan}, {Garmire}, {Kaspi}, \&
  {Richards}}]{Brandt_Vignali_list}
{Brandt}, W.~N., {Vignali}, C., {Schneider}, D.~P., {et~al.} 2004, Advances in
  Space Research, 34, 2478

\bibitem[{{Di Matteo}(1998)}]{DiMatteo98}
{Di Matteo}, T. 1998, \mnras, 299, L15

\bibitem[{{du Mas des Bourboux} {et~al.}(2017){du Mas des Bourboux}, {Le Goff},
  {Blomqvist}, {Busca}, {Guy}, {Rich}, {Y{\`e}che}, {Bautista}, {Burtin},
  {Dawson}, {Eisenstein}, {Font-Ribera}, {Kirkby}, {Miralda-Escud{\'e}},
  {Noterdaeme}, {Palanque-Delabrouille}, {P{\^a}ris}, {Petitjean},
  {P{\'e}rez-R{\`a}fols}, {Pieri}, {Ross}, {Schlegel}, {Schneider}, {Slosar},
  {Weinberg}, \& {Zarrouk}}]{duMasdesBourboux17}
{du Mas des Bourboux}, H., {Le Goff}, J.-M., {Blomqvist}, M., {et~al.} 2017,
  \aap, 608, A130

\bibitem[{{Foreman-Mackey} {et~al.}(2013){Foreman-Mackey}, {Hogg}, {Lang}, \&
  {Goodman}}]{emcee}
{Foreman-Mackey}, D., {Hogg}, D.~W., {Lang}, D., \& {Goodman}, J. 2013, \pasp,
  125, 306

\bibitem[{{Gallerani} {et~al.}(2010){Gallerani}, {Maiolino}, {Juarez}, {Nagao},
  {Marconi}, {Bianchi}, {Schneider}, {Mannucci}, {Oliva}, {Willott}, {Jiang},
  \& {Fan}}]{Gallerani10}
{Gallerani}, S., {Maiolino}, R., {Juarez}, Y., {et~al.} 2010, \aap, 523, A85

\bibitem[{{Ghirlanda} {et~al.}(2004){Ghirlanda}, {Ghisellini}, {Lazzati}, \&
  {Firmani}}]{Ghirlanda04}
{Ghirlanda}, G., {Ghisellini}, G., {Lazzati}, D., \& {Firmani}, C. 2004, \apjl,
  613, L13

\bibitem[{{Haardt} \& {Maraschi}(1991)}]{HM91}
{Haardt}, F. \& {Maraschi}, L. 1991, \apjl, 380, L51

\bibitem[{{Haardt} \& {Maraschi}(1993)}]{HM93}
{Haardt}, F. \& {Maraschi}, L. 1993, \apj, 413, 507

\bibitem[{{Hjorth} {et~al.}(2013){Hjorth}, {Vreeswijk}, {Gall}, \&
  {Watson}}]{Hjiort13}
{Hjorth}, J., {Vreeswijk}, P.~M., {Gall}, C., \& {Watson}, D. 2013, \apj, 768,
  173

\bibitem[{{Iwamuro} {et~al.}(2004){Iwamuro}, {Kimura}, {Eto}, {Maihara},
  {Motohara}, {Yoshii}, \& {Doi}}]{iwamuro}
{Iwamuro}, F., {Kimura}, M., {Eto}, S., {et~al.} 2004, \apj, 614, 69

\bibitem[{{Jiang} {et~al.}(2007){Jiang}, {Fan}, {Vestergaard}, {Kurk},
  {Walter}, {Kelly}, \& {Strauss}}]{jiang07}
{Jiang}, L., {Fan}, X., {Vestergaard}, M., {et~al.} 2007, \aj, 134, 1150

\bibitem[{{Just} {et~al.}(2007){Just}, {Brandt}, {Shemmer}, {Steffen},
  {Schneider}, {Chartas}, \& {Garmire}}]{Just07}
{Just}, D.~W., {Brandt}, W.~N., {Shemmer}, O., {et~al.} 2007, \apj, 665, 1004

\bibitem[{{Komatsu} {et~al.}(2009){Komatsu}, {Dunkley}, {Nolta}, {Bennett},
  {Gold}, {Hinshaw}, {Jarosik}, {Larson}, {Limon}, {Page}, {Spergel},
  {Halpern}, {Hill}, {Kogut}, {Meyer}, {Tucker}, {Weiland}, {Wollack}, \&
  {Wright}}]{komatsu09}
{Komatsu}, E., {Dunkley}, J., {Nolta}, M.~R., {et~al.} 2009, \apjs, 180, 330

\bibitem[{{Lusso} {et~al.}(2012){Lusso}, {Comastri}, {Simmons}, {Mignoli},
  {Zamorani}, {Vignali}, {Brusa}, {Shankar}, {Lutz}, {Trump}, {Maiolino},
  {Gilli}, {Bolzonella}, {Puccetti}, {Salvato}, {Impey}, {Civano}, {Elvis},
  {Mainieri}, {Silverman}, {Koekemoer}, {Bongiorno}, {Merloni}, {Berta}, {Le
  Floc'h}, {Magnelli}, {Pozzi}, \& {Riguccini}}]{Lusso12}
{Lusso}, E., {Comastri}, A., {Simmons}, B.~D., {et~al.} 2012, \mnras, 425, 623

\bibitem[{{Lusso} {et~al.}(2010){Lusso}, {Comastri}, {Vignali}, {Zamorani},
  {Brusa}, {Gilli}, {Iwasawa}, {Salvato}, {Civano}, {Elvis}, {Merloni},
  {Bongiorno}, {Trump}, {Koekemoer}, {Schinnerer}, {Le Floc'h}, {Cappelluti},
  {Jahnke}, {Sargent}, {Silverman}, {Mainieri}, {Fiore}, {Bolzonella}, {Le
  F{\`e}vre}, {Garilli}, {Iovino}, {Kneib}, {Lamareille}, {Lilly}, {Mignoli},
  {Scodeggio}, \& {Vergani}}]{Lusso10}
{Lusso}, E., {Comastri}, A., {Vignali}, C., {et~al.} 2010, \aap, 512, A34

\bibitem[{{Lusso} {et~al.}(2019){Lusso}, {Piedipalumbo}, {Risaliti},
  {Paolillo}, {Bisogni}, {Nardini}, \& {Amati}}]{Lusso19}
{Lusso}, E., {Piedipalumbo}, E., {Risaliti}, G., {et~al.} 2019, arXiv e-prints,
  arXiv:1907.07692

\bibitem[{Lusso \& Risaliti(2016)}]{lusso-risaliti16}
Lusso, E. \& Risaliti, G. 2016, The Astrophysical Journal, 819, 154

\bibitem[{{Lusso} \& {Risaliti}(2017)}]{LussoRisaliti17}
{Lusso}, E. \& {Risaliti}, G. 2017, \aap, 602, A79

\bibitem[{{Lusso} {et~al.}(2015){Lusso}, {Worseck}, {Hennawi}, {Prochaska},
  {Vignali}, {Stern}, \& {O'Meara}}]{Lusso15}
{Lusso}, E., {Worseck}, G., {Hennawi}, J.~F., {et~al.} 2015, \mnras, 449, 4204

\bibitem[{{Merloni}(2003)}]{Merloni03}
{Merloni}, A. 2003, \mnras, 341, 1051

\bibitem[{{Merloni} {et~al.}(2014){Merloni}, {Bongiorno}, {Brusa}, {Iwasawa},
  {Mainieri}, {Magnelli}, {Salvato}, {Berta}, {Cappelluti}, {Comastri},
  {Fiore}, {Gilli}, {Koekemoer}, {Le Floc'h}, {Lusso}, {Lutz}, {Miyaji},
  {Pozzi}, {Riguccini}, {Rosario}, {Silverman}, {Symeonidis}, {Treister},
  {Vignali}, \& {Zamorani}}]{Merloni14}
{Merloni}, A., {Bongiorno}, A., {Brusa}, M., {et~al.} 2014, \mnras, 437, 3550

\bibitem[{{Mortlock} {et~al.}(2011){Mortlock}, {Warren}, {Venemans}, {Patel},
  {Hewett}, {McMahon}, {Simpson}, {Theuns}, {Gonz{\'a}les-Solares}, {Adamson},
  {Dye}, {Hambly}, {Hirst}, {Irwin}, {Kuiper}, {Lawrence}, \&
  {R{\"o}ttgering}}]{Mortlock}
{Mortlock}, D.~J., {Warren}, S.~J., {Venemans}, B.~P., {et~al.} 2011, \nat,
  474, 616

\bibitem[{{Nanni} {et~al.}(2018){Nanni}, {Gilli}, {Vignali}, {Mignoli},
  {Comastri}, {Vanzella}, {Zamorani}, {Calura}, {Lanzuisi}, {Brusa}, {Tozzi},
  {Iwasawa}, {Cappi}, {Vito}, {Balmaverde}, {Costa}, {Risaliti}, {Paolillo},
  {Prandoni}, {Liuzzo}, {Rosati}, {Chiaberge}, {Caminha}, {Sani}, {Cappelluti},
  \& {Norman}}]{Nanni18}
{Nanni}, R., {Gilli}, R., {Vignali}, C., {et~al.} 2018, \aap, 614, A121

\bibitem[{{Nanni} {et~al.}(2017){Nanni}, {Vignali}, {Gilli}, {Moretti}, \&
  {Brandt}}]{Nanni17}
{Nanni}, R., {Vignali}, C., {Gilli}, R., {Moretti}, A., \& {Brandt}, W.~N.
  2017, \aap, 603, A128

\bibitem[{{P{\^a}ris} {et~al.}(2017){P{\^a}ris}, {Petitjean}, {Ross}, {Myers},
  {Aubourg}, {Streblyanska}, {Bailey}, {Armengaud}, {Palanque-Delabrouille},
  {Y{\`e}che}, {Hamann}, {Strauss}, {Albareti}, {Bovy}, {Bizyaev}, {Niel
  Brandt}, {Brusa}, {Buchner}, {Comparat}, {Croft}, {Dwelly}, {Fan},
  {Font-Ribera}, {Ge}, {Georgakakis}, {Hall}, {Jiang}, {Kinemuchi},
  {Malanushenko}, {Malanushenko}, {McMahon}, {Menzel}, {Merloni}, {Nandra},
  {Noterdaeme}, {Oravetz}, {Pan}, {Pieri}, {Prada}, {Salvato}, {Schlegel},
  {Schneider}, {Simmons}, {Viel}, {Weinberg}, \& {Zhu}}]{Paris17}
{P{\^a}ris}, I., {Petitjean}, P., {Ross}, N.~P., {et~al.} 2017, \aap, 597, A79

\bibitem[{{Read} {et~al.}(2014){Read}, {Guainazzi}, \&
  {Sembay}}]{read-guainazzi}
{Read}, A.~M., {Guainazzi}, M., \& {Sembay}, S. 2014, \aap, 564, A75

\bibitem[{{Risaliti} {et~al.}(2007){Risaliti}, {Elvis}, {Fabbiano}, {Baldi},
  {Zezas}, \& {Salvati}}]{Risaliti07}
{Risaliti}, G., {Elvis}, M., {Fabbiano}, G., {et~al.} 2007, \apjl, 659, L111

\bibitem[{{Risaliti} \& {Lusso}(2015)}]{RisalitiLusso15}
{Risaliti}, G. \& {Lusso}, E. 2015, \apj, 815, 33

\bibitem[{{Risaliti} \& {Lusso}(2019)}]{RisalitiLusso18}
{Risaliti}, G. \& {Lusso}, E. 2019, Nature Astronomy, 3, 272

\bibitem[{{Shakura} \& {Sunyaev}(1973)}]{SS73}
{Shakura}, N.~I. \& {Sunyaev}, R.~A. 1973, \aap, 24, 337

\bibitem[{{Shemmer} {et~al.}(2017){Shemmer}, {Brandt}, {Paolillo}, {Kaspi},
  {Vignali}, {Lira}, \& {Schneider}}]{Shemmer17}
{Shemmer}, O., {Brandt}, W.~N., {Paolillo}, M., {et~al.} 2017, \apj, 848, 46

\bibitem[{{Shemmer} {et~al.}(2006){Shemmer}, {Brandt}, {Schneider}, {Fan},
  {Strauss}, {Diamond-Stanic}, {Richards}, {Anderson}, {Gunn}, \&
  {Brinkmann}}]{Shemmer06}
{Shemmer}, O., {Brandt}, W.~N., {Schneider}, D.~P., {et~al.} 2006, \apj, 644,
  86

\bibitem[{{Shemmer} {et~al.}(2005){Shemmer}, {Brandt}, {Vignali}, {Schneider},
  {Fan}, {Richards}, \& {Strauss}}]{Shemmer05}
{Shemmer}, O., {Brandt}, W.~N., {Vignali}, C., {et~al.} 2005, \apj, 630, 729

\bibitem[{{Shen} {et~al.}(2011){Shen}, {Richards}, {Strauss}, {Hall},
  {Schneider}, {Snedden}, {Bizyaev}, {Brewington}, {Malanushenko},
  {Malanushenko}, {Oravetz}, {Pan}, \& {Simmons}}]{Shen}
{Shen}, Y., {Richards}, G.~T., {Strauss}, M.~A., {et~al.} 2011, \apjs, 194, 45

\bibitem[{{Steffen} {et~al.}(2006){Steffen}, {Strateva}, {Brandt}, {Alexander},
  {Koekemoer}, {Lehmer}, {Schneider}, \& {Vignali}}]{Steffen06}
{Steffen}, A.~T., {Strateva}, I., {Brandt}, W.~N., {et~al.} 2006, \aj, 131,
  2826

\bibitem[{{Strateva} {et~al.}(2005){Strateva}, {Brandt}, {Schneider}, {Vanden
  Berk}, \& {Vignali}}]{Strateva05}
{Strateva}, I.~V., {Brandt}, W.~N., {Schneider}, D.~P., {Vanden Berk}, D.~G.,
  \& {Vignali}, C. 2005, \aj, 130, 387

\bibitem[{{Svensson} \& {Zdziarski}(1994)}]{SZ94}
{Svensson}, R. \& {Zdziarski}, A.~A. 1994, \apj, 436, 599

\bibitem[{{Vagnetti} {et~al.}(2013){Vagnetti}, {Antonucci}, \&
  {Trevese}}]{Vagnetti13}
{Vagnetti}, F., {Antonucci}, M., \& {Trevese}, D. 2013, \aap, 550, A71

\bibitem[{{Vagnetti} {et~al.}(2010){Vagnetti}, {Turriziani}, {Trevese}, \&
  {Antonucci}}]{Vagnetti10}
{Vagnetti}, F., {Turriziani}, S., {Trevese}, D., \& {Antonucci}, M. 2010, \aap,
  519, A17

\bibitem[{{Vanden Berk} {et~al.}(2001){Vanden Berk}, {Richards}, {Bauer},
  {Strauss}, {Schneider}, {Heckman}, {York}, {Hall}, {Fan}, {Knapp},
  {Anderson}, {Annis}, {Bahcall}, {Bernardi}, {Briggs}, {Brinkmann}, {Brunner},
  {Burles}, {Carey}, {Castander}, {Connolly}, {Crocker}, {Csabai}, {Doi},
  {Finkbeiner}, {Friedman}, {Frieman}, {Fukugita}, {Gunn}, {Hennessy},
  {Ivezi{\'c}}, {Kent}, {Kunszt}, {Lamb}, {Leger}, {Long}, {Loveday}, {Lupton},
  {Meiksin}, {Merelli}, {Munn}, {Newberg}, {Newcomb}, {Nichol}, {Owen}, {Pier},
  {Pope}, {Rockosi}, {Schlegel}, {Siegmund}, {Smee}, {Snir}, {Stoughton},
  {Stubbs}, {SubbaRao}, {Szalay}, {Szokoly}, {Tremonti}, {Uomoto}, {Waddell},
  {Yanny}, \& {Zheng}}]{VandenBerk2001}
{Vanden Berk}, D.~E., {Richards}, G.~T., {Bauer}, A., {et~al.} 2001, \aj, 122,
  549

\bibitem[{{Vignali} {et~al.}(2001){Vignali}, {Brandt}, {Fan}, {Gunn}, {Kaspi},
  {Schneider}, \& {Strauss}}]{vignali01}
{Vignali}, C., {Brandt}, W.~N., {Fan}, X., {et~al.} 2001, \aj, 122, 2143

\bibitem[{{Vignali} {et~al.}(2003{\natexlab{a}}){Vignali}, {Brandt}, \&
  {Schneider}}]{Vignali03c}
{Vignali}, C., {Brandt}, W.~N., \& {Schneider}, D.~P. 2003{\natexlab{a}}, \aj,
  125, 433

\bibitem[{{Vignali} {et~al.}(2003{\natexlab{b}}){Vignali}, {Brandt},
  {Schneider}, {Anderson}, {Fan}, {Gunn}, {Kaspi}, {Richards}, \&
  {Strauss}}]{Vignali03d}
{Vignali}, C., {Brandt}, W.~N., {Schneider}, D.~P., {et~al.}
  2003{\natexlab{b}}, \aj, 125, 2876

\bibitem[{{Vignali} {et~al.}(2005){Vignali}, {Brandt}, {Schneider}, \&
  {Kaspi}}]{Vignali05}
{Vignali}, C., {Brandt}, W.~N., {Schneider}, D.~P., \& {Kaspi}, S. 2005, \aj,
  129, 2519

\bibitem[{{Wu} {et~al.}(2015){Wu}, {Wang}, {Fan}, {Yi}, {Zuo}, {Bian}, {Jiang},
  {McGreer}, {Wang}, {Yang}, {Yang}, {Thompson}, \& {Beletsky}}]{Wu15}
{Wu}, X.-B., {Wang}, F., {Fan}, X., {et~al.} 2015, \nat, 518, 512

\bibitem[{{Young} {et~al.}(2010){Young}, {Elvis}, \& {Risaliti}}]{Young10}
{Young}, M., {Elvis}, M., \& {Risaliti}, G. 2010, \apj, 708, 1388

\end{thebibliography}
%
%
%
%
\onecolumn
\begin{landscape}
{ 
\renewcommand{\arraystretch}{1.3}
\begin{longtable}{llllcccccccccc}
\multicolumn{1}{l}{\textbf{Table 1}}\\
\multicolumn{1}{l}{\Large{Quasar properties}}\\\hline
Name					&	RA					&	DEC					&	z	&	$f_{\rm 2500\AA}$			&	$\Gamma_{\rm X}$		&	$L_{\rm 0.5-2\ keV}$	&	$L_{\rm 2-10\ keV}$	&	$f_{\rm 2keV}$		&	$\chi_{\rm dof}^2$	&	$\#_{\rm dof}$	&	$t_{\rm exp}$	&	C/X &	Ref.\\
(1)						&	(2)					&	(3)					&	(4)	&	(5)					&	(6)				&	(7)				&	(8)				&	(9)		&	(10)		&	(11)	&	(12)	&	(13)&	(14)\\\hline\hline
ULAS J1120+0641			&		11h20m01.48s	&	+06d41m24.3s		&	7.08	&	$5.7^{1}$		&	$2.0^{+0.4}_{-0.4}$	&	12$^{+4}_{-7}$	 	&	12$^{+3}_{-3}$ 		&	$1.9^{+1.1}_{-0.8}$	&	1.16		&	36		&	2.3		&	X	&	1\\
SDSS J114816.7+525150.4		&		11h48m16.65s	&	+52d51m50.2s		&	6.43	&	$0.84^{2}$		&	$1.5^{+0.5}_{-0.5}$	&	3.$^{+8}_{-2}$ 		&	8$^{+5}_{-3}$ 		&	$1.0^{+1.5}_{-0.6}$	&	0.47		&	32		&	5		&	X	&	2\\
SDSS 1030+0524			&		10h30m27.10s	&	+05d24m55.0s		&	6.31	&	$0.42^{3}$		&	$2.1^{+0.2}_{-0.2}$	&	5$^{+7}_{-3}$ 		&	5$^{+3}_{-2}$ 		&	$7.6^{+2}_{-1.5}$	&	1.07		&	156		&	103.8	&	X	&	3\\
SDSS J010013.0+280225.9		&		01h00m13.02s	&	+28d02m55.8s		&	6.30	&	$6.9^{4}$		&	$2.4^{+0.2}_{-0.2}$	&	100$^{+70}_{-40}$ 	&	59$^{+15}_{-12}$ 	&	$14^{+5}_{-4}$	&	1.10		&	182		&	65.4		&	X	&	4\\
SDSS 1623+3112			&		16h23m31.81s	&	+31d12m00.0s		&	6.22	&	$0.40^{5}$		&	$1.9^{\dagger}$	&	$<5$			 	&	$<4$ 			&	$<1.8$			&	16.50		&	7		&	17.2		&	C	&	5\\
SDSS 1602+4228			&		16h02m53.98s	&	+42d28m24.9s		&	6.07	&	$1.37^{5}$		&	$2.0^{+0.5}_{-0.5}$	&	23$^{+47}_{-16}$ 	&	28$^{+15}_{-10}$ 	&	$6^{+7}_{-3}$	&	1.00		&	26		&	13.2		&	C	&	5\\
SDSS 1630+4012			&		16h30m33.90s	&	+40d12m09.6s		&	6.05	&	$0.28^{2}$		&	$ 2.0^{+0.7}_{-0.7}$	&	9$^{+30}_{-7}$ 		&	10$^{+7}_{-4}$ 		&	$2.3^{+4}_{-1.5}$	&	1.36		&	16		&	27.3		&	C	&	2\\
SDSS 1306+0356			&		13h06m08.26s	&	+03d56m26.3s		&	5.99	&	$0.32^{6}$		&	$ 1.9^{+0.2}_{-0.2}$	&	8$^{+6}_{-3}$ 		&	13$^{+3}_{-2}$ 		&	$2.8^{+1.2}_{-0.8}$	&	0.79		&	91		&	119.7		&	C	&	6\\
SDSS 1411+1217			&		14h11m11.29s	&	+12d17m37.4s		&	5.93	&	$0.37^{6}$		&	$ 2.3^{+0.8}_{-0.8}$	&	18$^{+90}_{-15.0}$ 	&	16$^{+-8}_{-8}$ 	&	$5^{+8}_{-3}$	&	1.34		&	13		&	14.4		&	C	&	6\\
SDSS 0836+0054			&		08h36m43.85s	&	+00d54m53.3s		&	5.82	&	$1.10^{6}$		&	$ 1.9^{+0.5}_{-0.5}$	&	30$^{+60}_{-20}$ 	&	48$^{+25}_{-18}$ 	&	$10^{+11}_{-6}$	&	0.96		&	24		&	5.7		&	C	&	6\\
SDSS 0231-0728			&		02h31m37.65s	&	-07d28m54.5s		&	5.41	&	$1.24^{\star}$	&	$2.1^{+0.6}_{-0.5}$	&	1.80$^{+2.0}_{-1.0}$ &	70$^{+30}_{-20}$ 	&	$20^{+20}_{-11}$	&	1.08		&	20		&	4.2		&	C	&	7\\
SDSS 1053+5804			&		10h53m22.99s	&	+58d04m12.1s		&	5.21	&	$1.51^{\star}$	&	$1.9^{\dagger}$	&	$<0.8$		 	&	$<20	$		 	&	$<8.4$			&	0.98		&	4		&	4.7		&	C$^{{*}{\vartriangle}}$	&	7\\
SDSS 2228-0757			&		22h28m45.15s	&	-07d57m55.4s		&	5.14	&	$0.68^{\star}$		&	$1.9^{\dagger}$	&	$<0.1$	 		&	$<4$ 			&	$<2.3$			&	2.10		&	3		&	7.0		&	C	&	7\\
SDSS 0756+4104			&		07h56m18.14s	&	+41d04m08.6s		&	5.09	&	$0.74^{\star}$		&	$1.8^{+0.7}_{-0.6}$	&	1.0$^{+1.0}_{-0.6}$ 	&	22$^{+13}_{-9}$ 	&	$5^{+7}_{-3}$	&	2.36		&	18		&	7.4		&	C	&	7\\
SDSS 1204-0021			&		12h04m41.73s	&	-00d21m49.6s		&	5.03	&	$1.94^{\star}$	&	$1.7^{+1.1}_{-1.2}$	&	0.6$^{+1.0}_{-0.5}$ 	&	13$^{+8}_{-6}$ 		&	$3^{+6}_{-2}$	&	0.33		&	9		&	6.3		&	C	&	7\\
SDSS J133422.63+475033.6	&		13h34m22.64s	&	+47d50m33.6s		&	5.00	&	2.4			&	$1.9^{+0.6}_{-0.6}$	&	0.8$^{+1.0}_{-0.5}$ 	&	14$^{+8}_{-6}$ 		&	$4^{+5}_{-2}$	&	1.31		&	18		&	11.6		&	C	&	7\\
SDSS 2216+0013			&		22h16m44.00s	&	+00d13m48.3s		&	4.99	&	$0.67^{\star}$		&	$2.3^{+1.1}_{-1.0}$	&	0.8$^{+2.0}_{-0.6}$ 	&	10$^{+10}_{-6}$ 	&	$5^{+10}_{-3}$	&	0.64		&	7		&	7.5		&	C	&	8\\
SDSS 0040-0915			&		00h40m54.65s	&	-09d15m26.8s		&	4.98	&	$2.5^{\star}$		&	$1.6^{+1.2}_{-0.7}$	&	6$^{+30}_{-5}$ 		&	13$^{+10}_{-7}$ 	&	$3^{+6}_{-1.9}$	&	1.07		&	139		&	$^{\ddagger}$	&	X	&	7\\
SDSS 1026+4719			&		10h26m22.87s	&	+47d19m07.2s		&	4.94	&	2.9			&	$1.3^{+1.1}_{-1.1}$	&	0.9$^{+5}_{-0.8}$ 	&	25$^{+39}_{-17}$ 	&	$4^{+20}_{-4}$		&	1.12		&	5		&	2.3		&	C$^{\vartriangle}$	&	7\\
SDSS 1536+5008			&		15h36m50.26s	&	+50d08m10.3s		&	4.93	&	3.0			&	$1.3^{+0.6}_{-0.6}$	&	1.8$^{+1.0}_{-0.7}$ 	&	27$^{+13}_{-10}$ 	&	$3^{+5}_{-2}$	&	1.77		&	14		&	4.6		&	C	&	7\\
SDSS J105123.04+354534.3	&		10h51m23.04s	&	+35d45m34.3s		&	4.91	&	3.7			&	$1.3^{+0.3}_{-0.3}$	&	14$^{+20}_{-8}$ 	&	46$^{+24}_{-17}$ 	&	$8^{+8}_{-4}$	&	1.34		&	62		&	$^{\ddagger}$	&	X	&	7\\
SDSS J142103.83+343332.0	&		14h21m03.83s	&	+34d33m32.0s		&	4.90	&	2.0			&	$1.9^{\dagger}$	&	$<0.1$ 			&	$<3$ 			&	$<1.3$			&	5.68		&	4		&	12.9		&	C	&	7\\
SDSS J173744.88+582829.62	&		17h37m44.88s	&	+58d28m29.6s		&	4.89	&	2.9			&	$1.9^{\dagger}$	&	$<0.1$			&	$<3$ 			&	$<1.7$			&	4.16		&	1		&	4.6		&	C	&	7\\
SDSS 2225-0014			&		22h25m09.20s	&	-00d14m06.9s		&	4.89	&	2.3			&	$1.9^{\dagger}$	&	$<0.7$ 			&	$<7$				&	$<3.5$			&	2.65		&	2		&	3.4		&	C	&	7\\
SDSS 1023+6335			&		10h23m32.07s	&	+63d35m08.0s		&	4.88	&	0.75			&	$1.6^{+1.2}_{-1.1}$	&	0.4$^{+1.0}_{-0.3}$ 	&	7$^{+8}_{-4}$ 		&	$1.5^{+5}_{-1.3}$	&	0.66		&	4		&	4.7		&	C	&	7\\
SDSS 0211-0009			&		02h11m02.70s	&	-00d09m10.3s		&	4.87	&	$0.17^{\star}$	&	$1.7^{+1.2}_{-1.1}$	&	0.4$^{+1.0}_{-0.3}$ 	&	6$^{+7}_{-4}$ 		&	$1.4^{+4}_{-1.2}$	&	2.76		&	4		&	4.9		&	C	&	8\\
SDSS 0951+5945			&		09h51m51.19s	&	+59d45m56.2s		&	4.86	&	0.83			&	$1.9^{\dagger}$	&	$<0.8$	 		&	$<10$ 			&	$<3.5$			&	5.45		&	3		&	5.1		&	C	&	7\\
SDSS J075652.07+450258.86	&		07h56m52.08s	&	+45d02m58.9s		&	4.83	&	0.68			&	$1.9^{\dagger}$	&	$<5$ 			&	$<40$		 	&	$<2.1$	&	9.20		&	3		&	7.0		&	C	&	7\\
SDSS 0941+5947			&		09h41m08.36s	&	+59d47m25.7s		&	4.82	&	1.78			&	$1.3^{+1.3}_{-1.1}$	&	0.8$^{+5.0}_{-0.7}$	&	14$^{+30}_{-9}$ 	&	$2.0^{+6}_{-1.7}$	&	1.65		&	11		&	4.2		&	C	&	7\\
SDSS J142705.86+330817.9	&		14h27m05.86s	&	+33d08m18.0s		&	4.71	&	2.8			&	$1.6^{+1.0}_{-0.9}$	&	1.5$^{+5.}_{-1.2}$ 	&	21$^{+25}_{-13}$ 	&	$5^{+16}_{-4}$	&	3.98		&	8		&	5.0		&	C$^{\vartriangle}$		&	7\\
PSS 1347+4956			&		13h47m43.29s	&	+49d56m21.3s		&	4.51	&	5.8			&	$2.14^{+0.5}_{-0.5}$	&	5$^{+4}_{-2}$ 		&	34$^{+12}_{-9}$ 	&	$13^{+10}_{-6}$	&	0.77		&	28		&	5.9		&	C	&	7\\
SDSS 1302+0030			&		13h02m16.13s	&	+00d30m32.1s		&	4.50	&	0.87			&	$1.9^{\dagger}$	&	$<0.8$ 			&	$<7$		 		&	$<1.5$	&	3.91		&	2		&	10.8		&	C	&	7\\
\hline
\label{table:data}
\end{longtable}}
{ 
\renewcommand{\arraystretch}{1.3}
\begin{longtable}{llllcccccccccc}
\hline
Name					&	RA					&	DEC			&	z	&	$f_{\rm 2500\AA}$		&	$\Gamma_{\rm X}$		&	$L_{\rm 0.5-2\ keV}$	&	$L_{\rm 2-10\ keV}$	&	$f_{\rm 2keV}$		&	$\chi_{\rm dof}^2$	&	$\#_{\rm dof}$	&	$t_{\rm exp}$	&	C/X\\
(1)						&	(2)					&	(3)			&	(4)	&	(5)				&	(6)				&	(7)				&	(8)				&	(9)				&	(10)		&	(11)	&	(12)	&	(13)\\ \hline \hline
PSS 0808+5215			&		08h08m49.43s	&	+52d15m15.3s		&	4.44	&	3.8		&	$1.5^{+1.0}_{-1.0}$	&	0.7$^{+2.0}_{-0.6}$ 	&	9$^{+8}_{-5}$ 		&	$2.1^{+5}_{-1.7}$	&	1.70		&	7		&	4.9		&	C	&	7\\
BRI 0103+0032			&		01h06m19.24s	&	+00d48m23.3s		&	4.44	&	3.7		&	$2.0^{+0.7}_{-0.6}$	&	5$^{+6}_{-3}$ 		&	43$^{+18}_{-14}$ 	&	$15^{+16}_{-8}$	&	3.44		&	20		&	3.7		&	C	&	7\\
SDSS 0831+5235			&		08h31m03.01s	&	+52d35m33.5s		&	4.44	&	1.78		&	$2.0^{+0.4}_{-0.5}$	&	9$^{+8}_{-4}$ 		&	9$^{+2}_{-2}$ 		&	$3.8^{+1.5}_{-1.8}$	&	1.02		&	502		&	$^{\ddagger}$	&	X	&	7\\
PSS 1443+2724			&		14h43m31.18s	&	+27d24m36.8s		&	4.44	&	2.2		&	$1.9^{+0.8}_{-0.9}$	&	3$^{+4}_{-2}$ 		&	23$^{+15}_{-10}$ 	&	$7^{+11}_{-5}$	&	4.23		&	8		&	2.2		&	C	&	7\\
SDSS 1401+0244			&		14h01m46.53s	&	+02d44m34.7s		&	4.44	&	3.7		&	$1.6^{+0.2}_{-0.2}$	&	12$^{+4}_{-3}$ 		&	26$^{+3}_{-3}$ 		&	$7.3^{+1.5}_{-1.4}$	&	1.19		&	613		&	$^{\ddagger}$	&	X	&	7\\
PSS 0747+4434			&		07h47m49.74s	&	+44d34m17.0s		&	4.43	&	1.31		&	$3.0^{+1.3}_{-1.1}$	&	1.3$^{+1.0}_{-0.6}$ 	&	10$^{+8}_{-5}$ 		&	$9^{+16}_{-7}$	&	1.35		&	6		&	4.5		&	C	&	7\\
BR 1600+0724			&		16h03m20.89s	&	+07d21m04.5s		&	4.38	&	5.9		&	$2.6^{+1.1}_{-1.0}$	&	3$^{+5}_{-23}$ 		&	12$^{+10}_{-6}$ 	&	$7^{+12}_{-5}$	&	3.78		&	6		&	4.6		&	C	&	7\\
SDSS 2357+0043			&		23h57m18.36s	&	+00d43m50.4s		&	4.36	&	1.22		&	$1.4^{+0.6}_{-0.6}$	&	0.7$^{+1.0}_{-0.4}$ 	&	9$^{+5}_{-3}$ 		&	$2.0^{+3}_{-1.2}$	&	2.43		&	18		&	12.8		&	C	&	7\\
PSS 1317+3531			&		13h17m43.13s	&	+35d31m31.8s		&	4.34	&	0.79		&	$2.9^{+2.5}_{-1.8}$	&	2.0$^{+8.0}_{-1.8}$ 	&	7$^{+8}_{-5}$ 		&	$5.0^{+14}_{-4}$	&	0.39		&	2		&	2.8		&	C	&	7\\
PSS 0955+5940			&		09h55m11.33s	&	+59d40m30.6s		&	4.34	&	3.5		&	$1.8^{+0.9}_{-0.9}$	&	5$^{+3}_{-2}$	 	&	31$^{+7}_{-6}$ 		&	$5^{+6}_{-3}$	&	1.53		&	11		&	43.6		&	C	&	7\\
SDSS 0050-0053			&		00h50m06.35s	&	-00d53m19.3s		&	4.33	&	3.2		&	$1.4^{+0.7}_{-0.7}$	&	0.8$^{+1.0}_{-0.6}$ 	&	12$^{+5}_{-4}$ 		&	$2.7^{+3}_{-1.7}$	&	2.39		&	21		&	13.1		&	C	&	7\\
PSS 0957+3308			&		09h57m44.46s	&	+33d08m20.8s		&	4.20	&	6.0		&	$1.3^{+0.9}_{-0.9}$	&	1.6$^{+3.0}_{-1.1}$ 	&	20$^{+8}_{-7}$ 		&	$4^{+6}_{-3}$	&	0.59		&	16		&	6.1		&	C	&	7\\
PSS 0059+0003			&		00h59m22.66s	&	+00d03m01.4s		&	4.18	&	1.77		&	$1.3^{+0.7}_{-0.7}$	&	1.4$^{+3.0}_{-1.0}$ 	&	17$^{+12}_{-8}$ 	&	$4^{+6}_{-3}$	&	3.13		&	10		&	2.7		&	C	&	7\\
SDSS 1444-0123			&		14h44m28.63s	&	-01d23m44.0s		&	4.18	&	1.10		&	$4.1^{+2.2}_{-1.5}$	&	3$^{+16}_{-2}$ 		&	7$^{+6}_{-4}$ 		&	$11^{+19}_{-7}$	&	3.87		&	29		&	10.1		&	C	&	7\\
PSS 0133+0400			&		01h33m40.30s	&	+04d00m59.9s		&	4.15	&	$2.3^{\star}$	&	$1.9^{+0.2}_{-0.2}$	&	4.2$^{+1.0}_{-1.1}$ 	&	30$^{+5}_{-4}$ 		&	$10^{+3}_{-3}$	&	1.06		&	136		&	65.0		&	C	&	8\\
PSS 0209+0517			&		02h09m44.59s	&	+05d17m13.3s		&	4.14	&	$3.7^{\star}$	&	$2.4^{+0.6}_{-0.6}$	&	6$^{+5}_{-3}$ 		&	26$^{+10}_{-8}$ 	&	$14^{+11}_{-7}$	&	1.60		&	21		&	5.8		&	C	&	8\\
PSS 1057+4555			&		10h57m56.26s	&	+45d55m53.0s		&	4.14	&	17.5		&	$2.0^{+0.6}_{-0.5}$	&	8$^{+6}_{-3}$ 		&	48$^{+16}_{-13}$ 	&	$18^{+12}_{-8}$	&	3.87		&	29		&	2.8		&	C	&	7\\
PSS 1326+0743			&		13h26m11.85s	&	+07d43m58.4s		&	4.09	&	12.0		&	$1.8^{+0.3}_{-0.3}$	&	9$^{+3}_{-2}$ 		&	55$^{+12}_{-10}$ 	&	$19^{+8}_{-6}$		&	0.89		&	57		&	$^{\diamond}$	&	C	&	7\\
SDSS 1413+0000			&		14h13m15.37s	&	+00d00m32.4s		&	4.08	&	1.32		&	$1.7^{+0.6}_{-0.6}$	&	1.4$^{+2.0}_{-0.9}$ 	&	10$^{+5}_{-4}$ 		&	$2.8^{+3}_{-1.6}$	&	1.48		&	19		&	12.5		&	C	&	7\\
SDSS J111812.91+441122.3	&		11h18m12.91s	&	+44d11m22.3s		&	4.02	&	1.65		&	$1.5^{+0.6}_{-0.6}$	&	5$^{+3}_{-2}$ 		&	12$^{+2}_{-2}$ 		&	$3.3^{+1.1}_{-1.0}$	&	1.13		&	600		&	$^{\ddagger}$	&	X	&	7\\
SDSS 1408+0205			&		14h08m50.91s	&	+02d05m22.7s		&	4.01	&	3.3		&	$2.2^{+0.6}_{-0.6}$	&	9$^{+8}_{-4}$ 		&	41$^{+14}_{-11}$ 	&	$19^{+16}_{-9}$	&	0.76		&	29		&	6.0		&	C	&	7\\
\hline
\caption*{\label{tabnote2}{$\star$}: Extrapolation of the rest-frame 2500$\AA$ flux density through fixed slope on SDSS DR7 or DR12 data.
\\\label{tabnote3}{Numbers in column 5}: Extrapolation trough fixed slope on archival observations: (1) \citealt{Mortlock}, (2) \citealt{iwamuro}, (3) \citealt{Hjiort13}, (4) \citealt{Wu15}, (5) \citealt{Gallerani10}, (6) \citealt{jiang07}, (7) \citealt{Shen}, (8) \citealt{Paris17}.
\\\label{tabnoteupp}{$\dagger$}: Upper limit on X-ray monochromatic flux estimate.
\\\label{tabnote-off}*: Off-axis sources.
\\\label{tabnote-camera}{$\vartriangle$}: Sources detected using ACIS-I camera. All of the others were observed using ACIS-S. 
\\\label{tabnote-xmm}{$\ddagger$}: Multiple X-ray observations fitted together.\\
The electronic version of this table will be available online. Please contact the author for information.
}
\end{longtable}}
\end{landscape}
\twocolumn
%
%
\onecolumn
\begin{appendix}
\section{The extrapolation of the 2500$\AA$ flux density}
\begin{figure*}[ht!]
\centering
   \includegraphics[width=17cm]{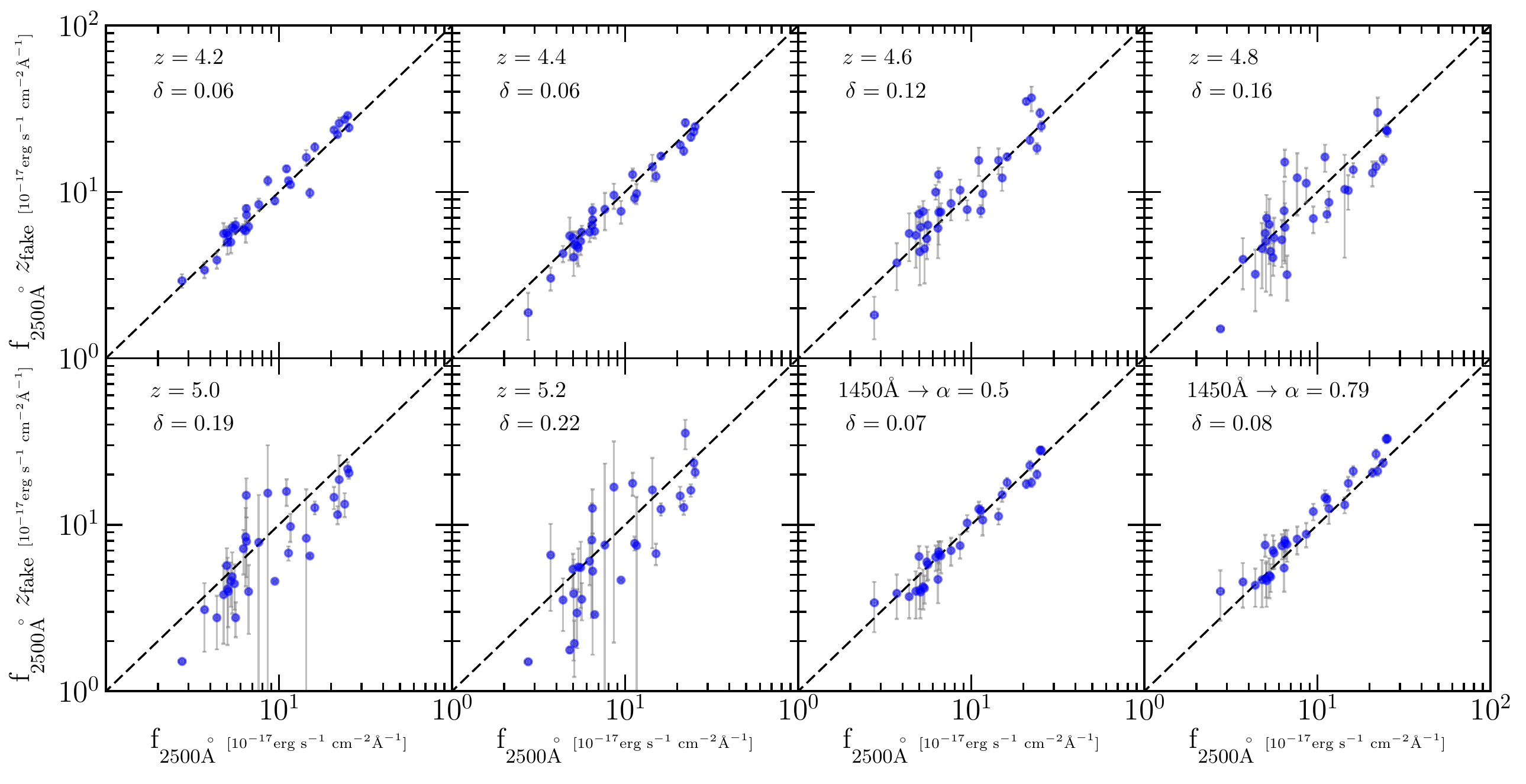}
     \caption{Results of the simulations used to choose between different procedures to estimate the $2500\AA$ flux densities on 30 luminous quasars at $z\simeq$2.
     We compared the flux density observed values with those obtained with each one of the three methods described in the appendix. 
     From the top left panel: the rest-frame $2500\AA$ flux estimated assuming redshift 4.2, 4.4, 4.6, 4.8, 5.0, 5.2 (y-axis), versus the one estimated from the full spectrum (x-axis).
     The last two panels represent the comparison between the true rest-frame $2500\AA$ flux from the spectra, and the one estimated from an extrapolation with a fixed power law, having the slope equal to 0.5 \citep{VandenBerk2001} and 0.79 \citep{vignali01}, respectively.
     The normalization have been assumed in the $1430-1470\AA$ spectral range.}
     \label{fig:susanna}
\end{figure*}

Here we present the procedure adopted to choose the best method to infer the rest-frame 2500$\AA$ flux density for the sources being with the SDSS (DR7 or DR12) but not included in the catalogue by \cite{Shen}.
Given the redshift of the sources in this sample, the rest-frame 2500$\AA$ is outside of the spectral coverage, both for the SDSS DR7 and the DR12.
We compared the following methods to extrapolate it: 1) from the 1450$\AA$ flux density, using a fixed slope $S_{\nu}\propto \nu^{-\alpha}$ with $\alpha=0.5$ (see \citealt{VandenBerk2001}), 2) the same using a slope $\alpha=0.79$ (see \citealt{vignali01}) and 3) the fit to the source spectrum, in order to use the continuum slope extrapolate the 2500$\AA$ flux.
To verify whether the $2500\AA$ flux density values obtained with these methods are reliable, we performed the following test: we selected 30 luminous quasars at redshift $z$=2, fulfilling the selection criteria applied to the main sample, i.e. unobscured (type I) optically-selected quasars, classified as radio quiet sources, and showing no BAL features.
For these sources the rest-frame $2500\AA$ flux density falls well within the SDSS spectral coverage, and we can give a reliable measure of this quantity with a procedure similar to the one performed by \cite{Shen} (i.e. continuum, FeII and emission lines spectral fitting).
We then assumed the redshift of the source to be higher than the true one, i.e. we rest-framed the spectrum and "cut" it at the highest observable wavelength accordingly to the assumed redshift and the SDSS wavelength coverage, perform the fit again and extrapolate the flux density at $2500\AA$.
We carried out this analysis for several values of redshift (z=3, 4, 4.2, 4.4, 4.6, 4.8, 5.0, 5.2) and then compared the extrapolated $2500\AA$ flux with the true value.
This comparison is shown in Fig. \ref{fig:susanna}. 
The extrapolated $2500\AA$ flux densities are in good agreement with the true ones. 
The extrapolation using the constant slope $\alpha=0.50$ turns out to be more accurate than other methods at redshift higher $z$>4.4 (on the basis of the observed dispersion).
We adopted the extrapolation of the flux density with the constant slope $\alpha=0.50$ also for the sources not covered with an SDSS observation, but for which data from other facilities were available in the literature.\\
We also checked the consistency of the UV flux density estimates by \cite{Shen} with the one obtained with the extrapolation with a fixed slope $\alpha$=0.5 (i.e. the first method described above) from the SDSS DR7 spectra for the 33 sources. 
We found that the two different estimates are fully consistent within the uncertainty (assumed of 22\%) for almost the entire sample.
Only a couple of sources showed a significant discrepancy (larger than 2$\sigma$): this can be caused by many reasons, e.g. a wrong fit from the automatic procedure by \cite{Shen} or, equivalently, a significant discrepancy between the assumed slope in the extrapolation (i.e., $\alpha$=0.5) and the intrinsic continuum slope of this particular sources.
However, we tested whether the inclusion or not of these sources could affect the best fit parameters of the $L_{\rm X}-L_{\rm UV}$ relation and we found that they are not statistically significant (i.e. we found fully consistent results in the two cases); then we included them in the sample.
\label{appendix:a}
\section{The X-ray spectral analysis}
Here we present the X-ray spectra for 41 sources in the sample for which we perform the spectral fit (i.e., the sources for which we can only estimate an upper limit to the X-ray fluxes are not included), with the exclusion of the 3 spectra presented in $\S$\ref{xrayanalysis}.
The sources are listed with increasing redshift.
The spectra, binned to a 90\% significance level for presentation purposes, are presented in the top panel along with the best fit model (Galactic absorption and a power law), while the bottom panel shows the residuals, data-model, in units of sigma.
\textbf{The observed-frame energies, in units of keV, are represented in the x-axes.}
\label{appendix:b}

\begin{figure*}[hb!]
\centering
  \includegraphics[width=0.495\textwidth]{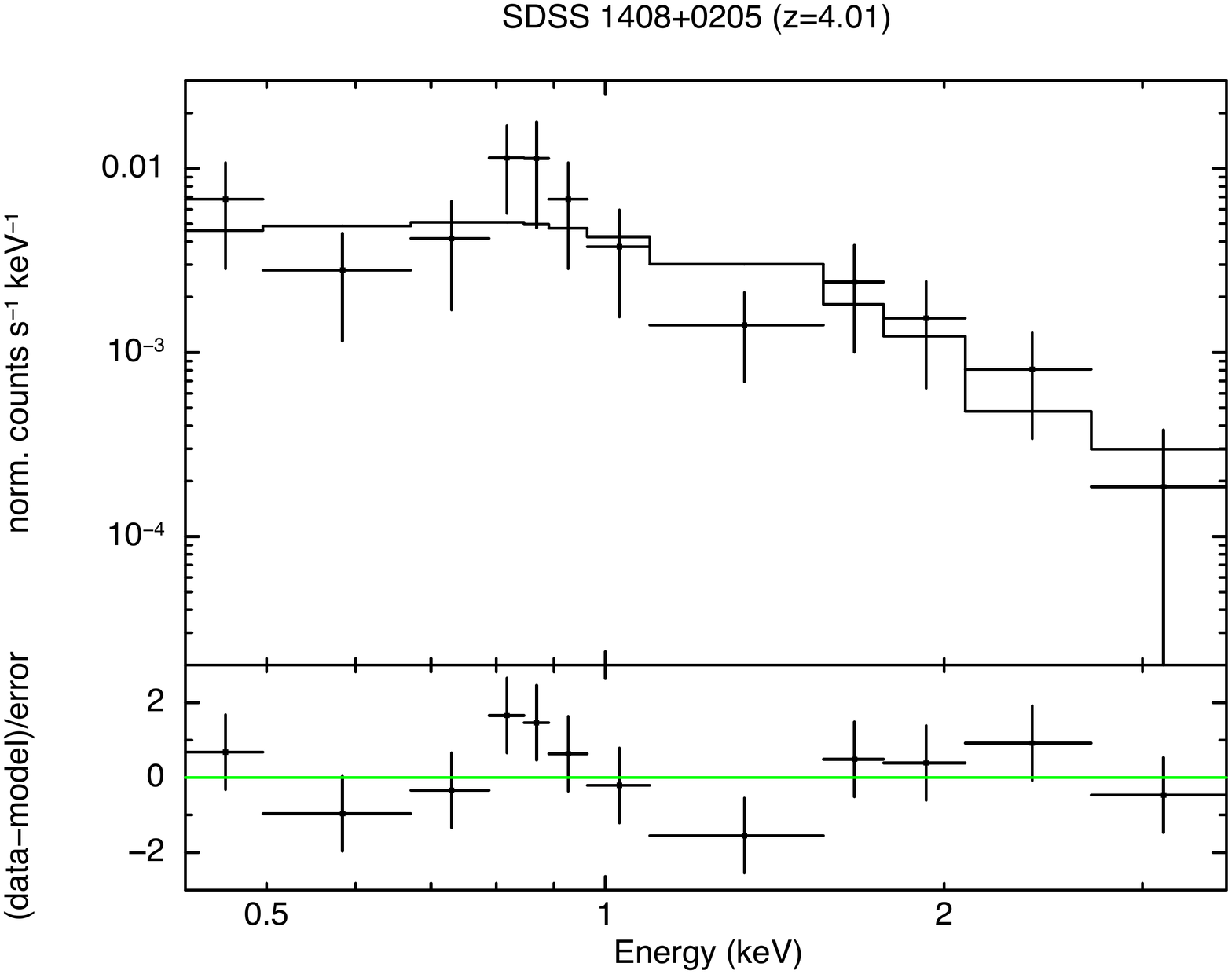}
   \includegraphics[width=0.495\textwidth]{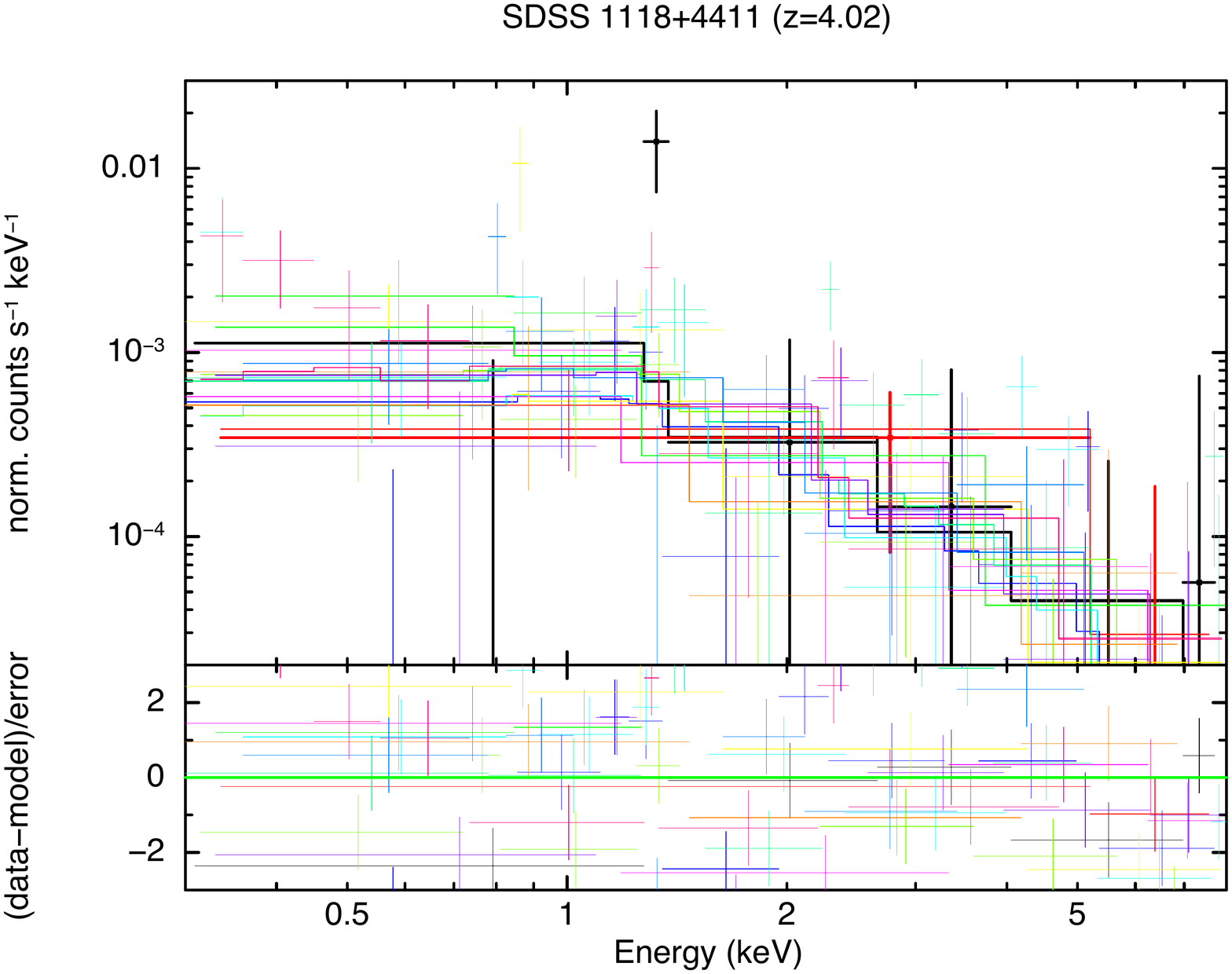}\\
    \vspace{0.4cm}
  \includegraphics[width=0.495\textwidth]{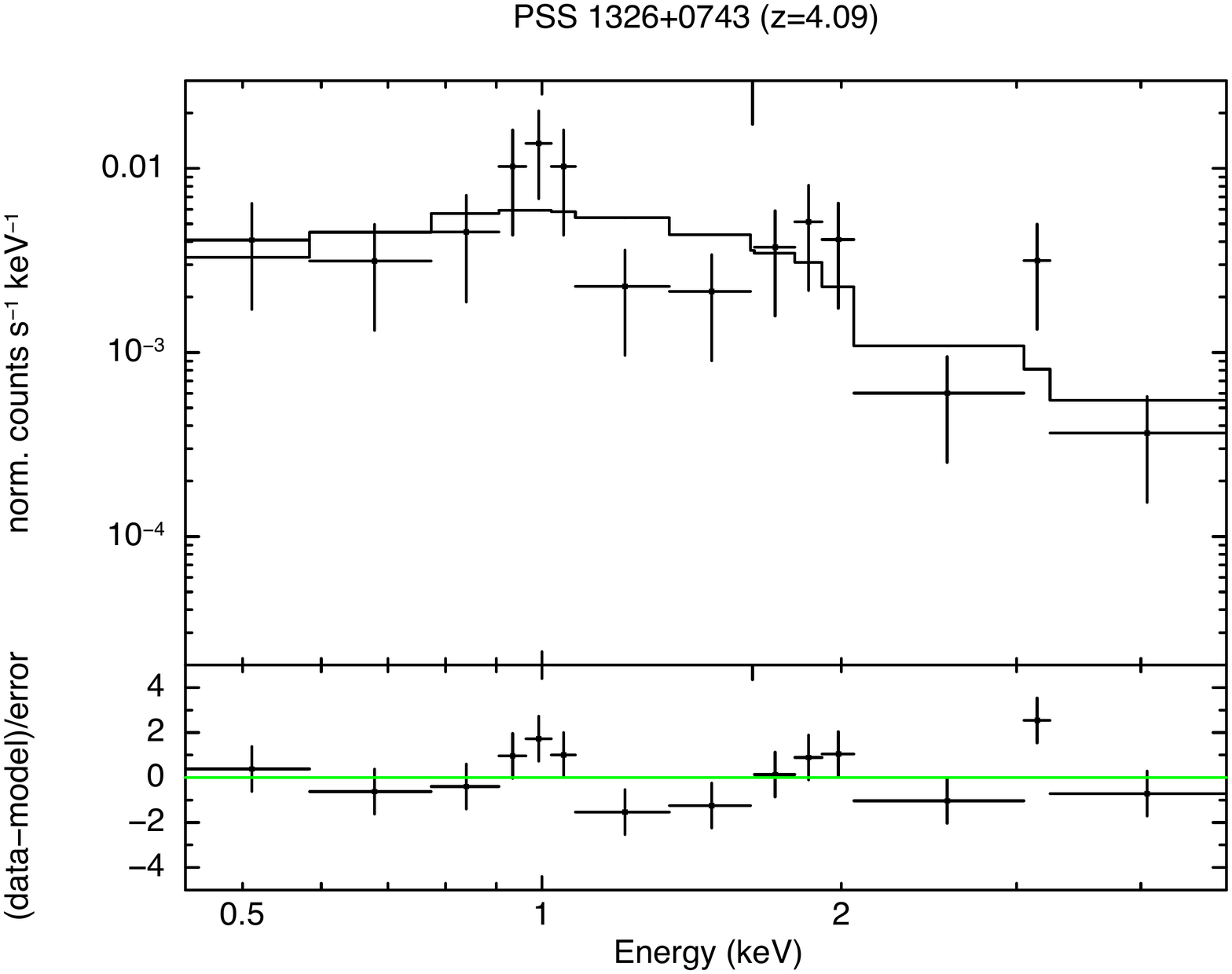}
   \includegraphics[width=0.495\textwidth]{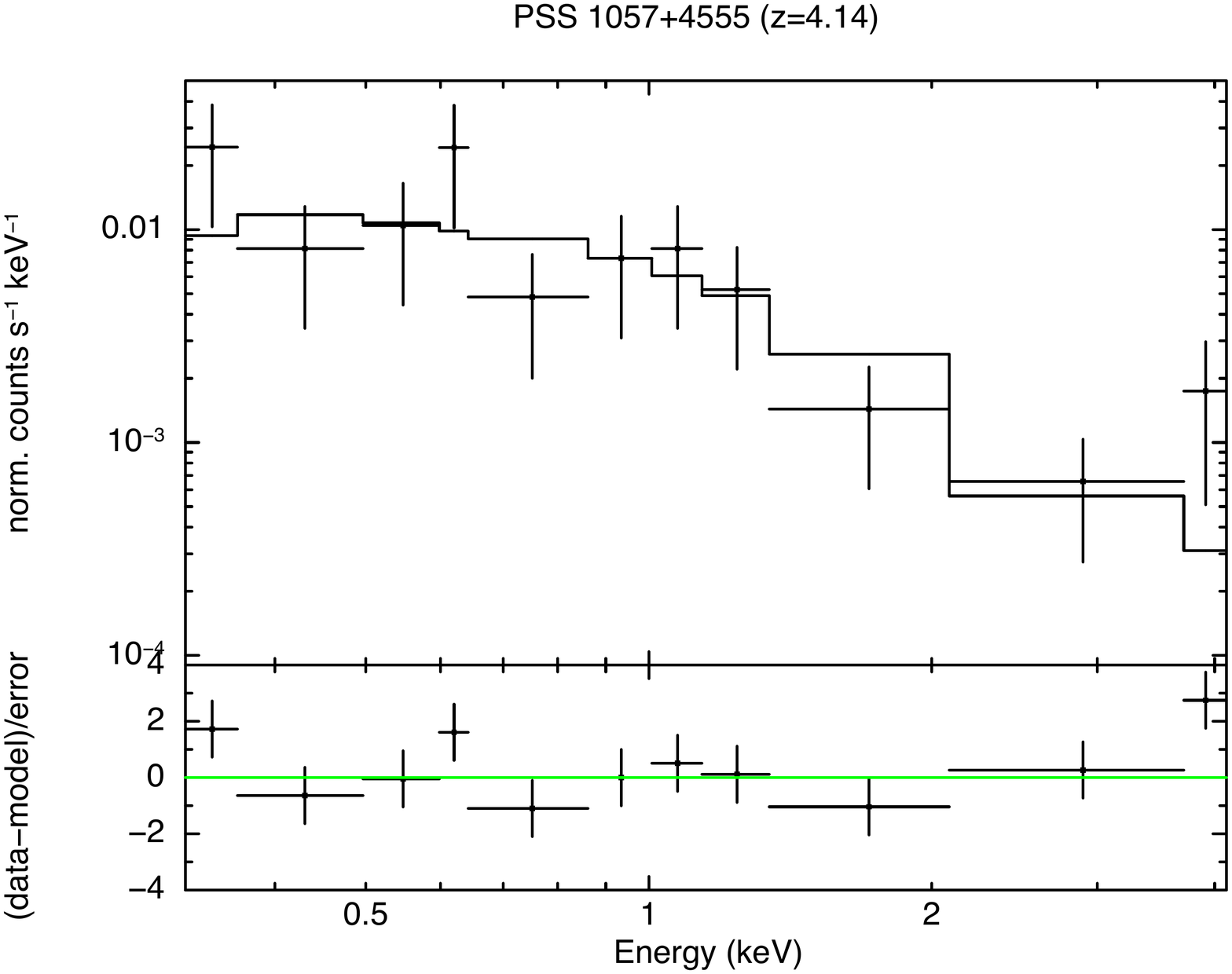}\\
    \vspace{0.4cm}
 \includegraphics[width=0.495\textwidth]{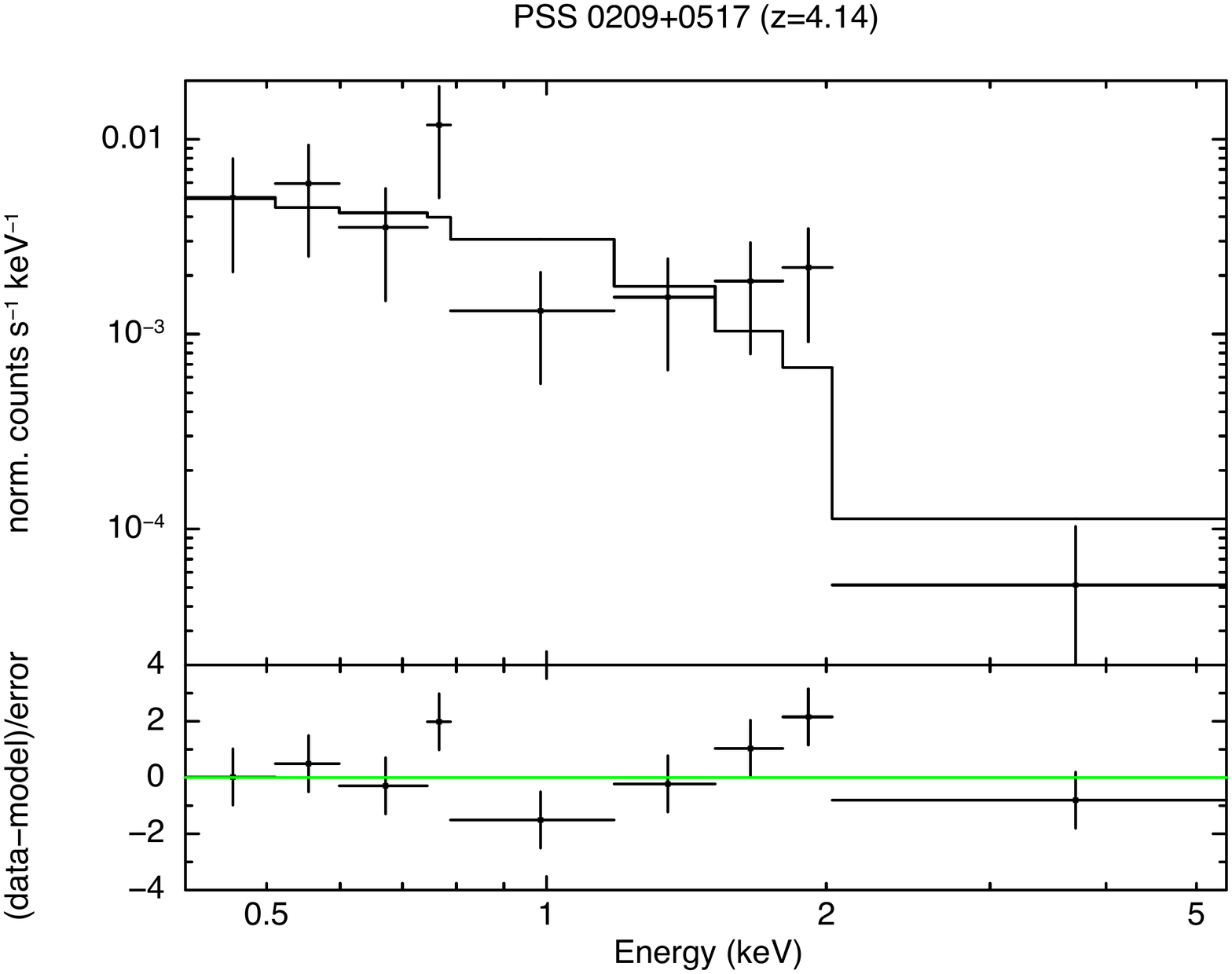}
 \includegraphics[width=0.495\textwidth]{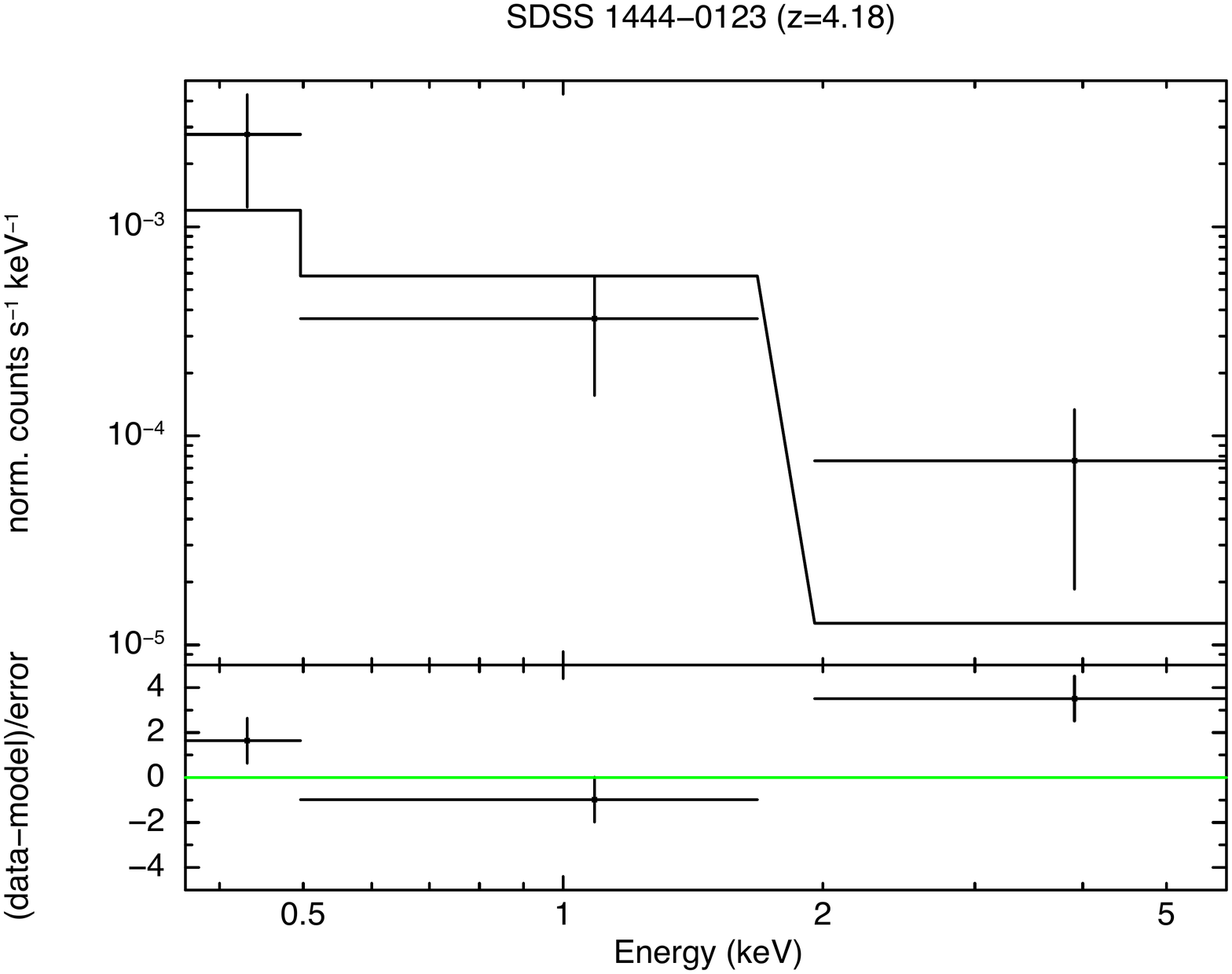}\\
\end{figure*}
\begin{figure*}[b!]
  \centering
    \includegraphics[width=0.495\textwidth]{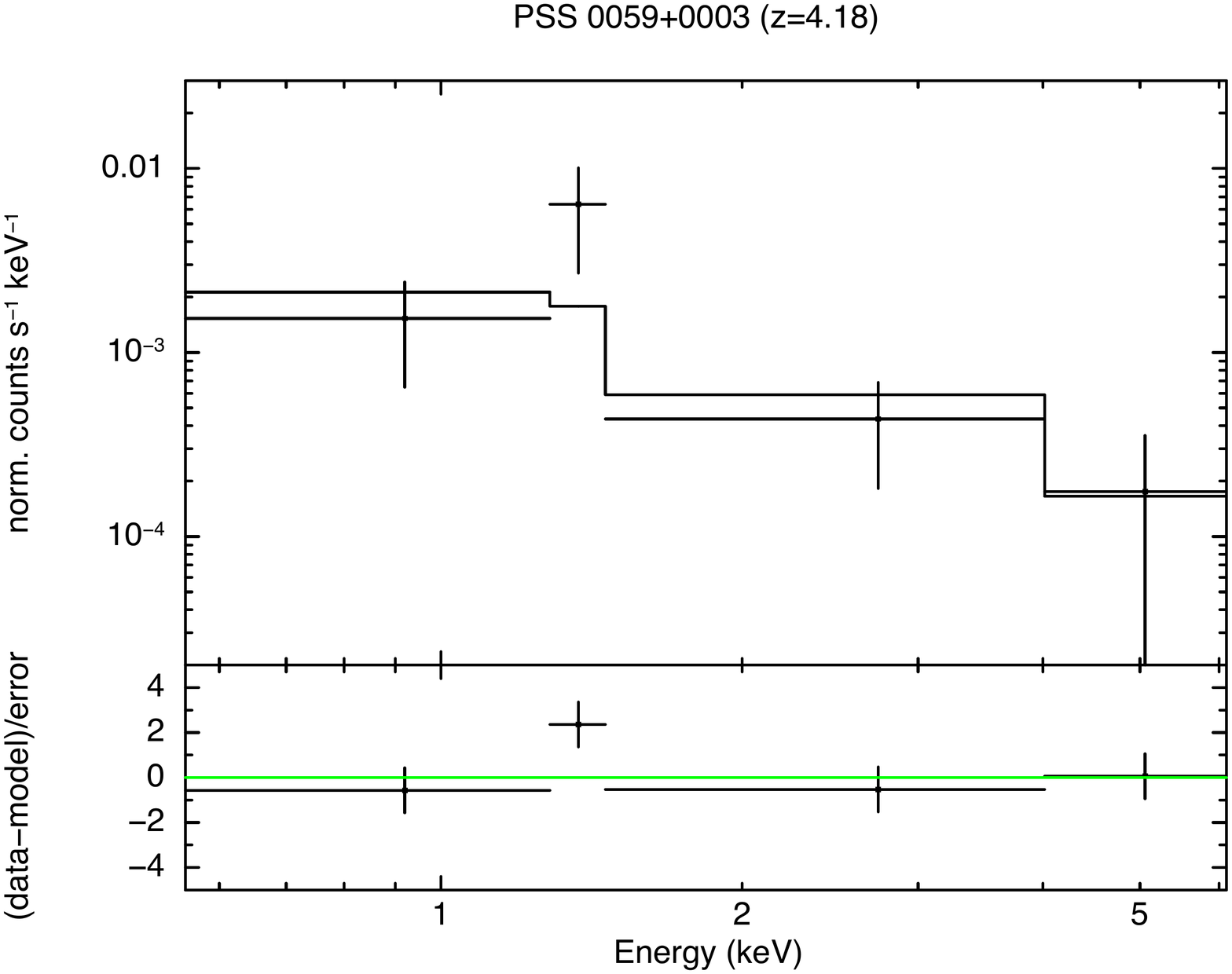}
 \includegraphics[width=0.495\textwidth]{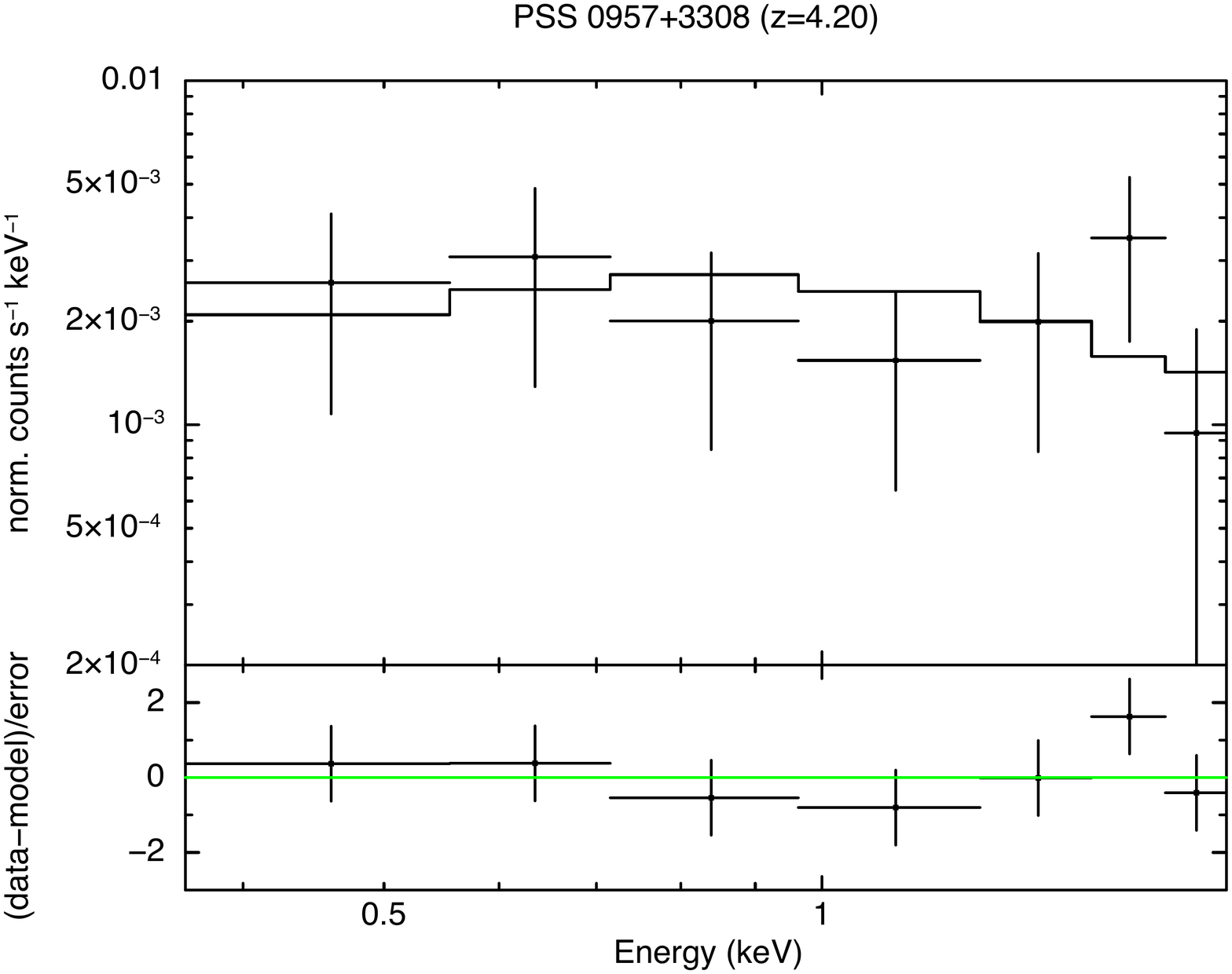}\\
 \vspace{1cm}
   \includegraphics[width=0.495\textwidth]{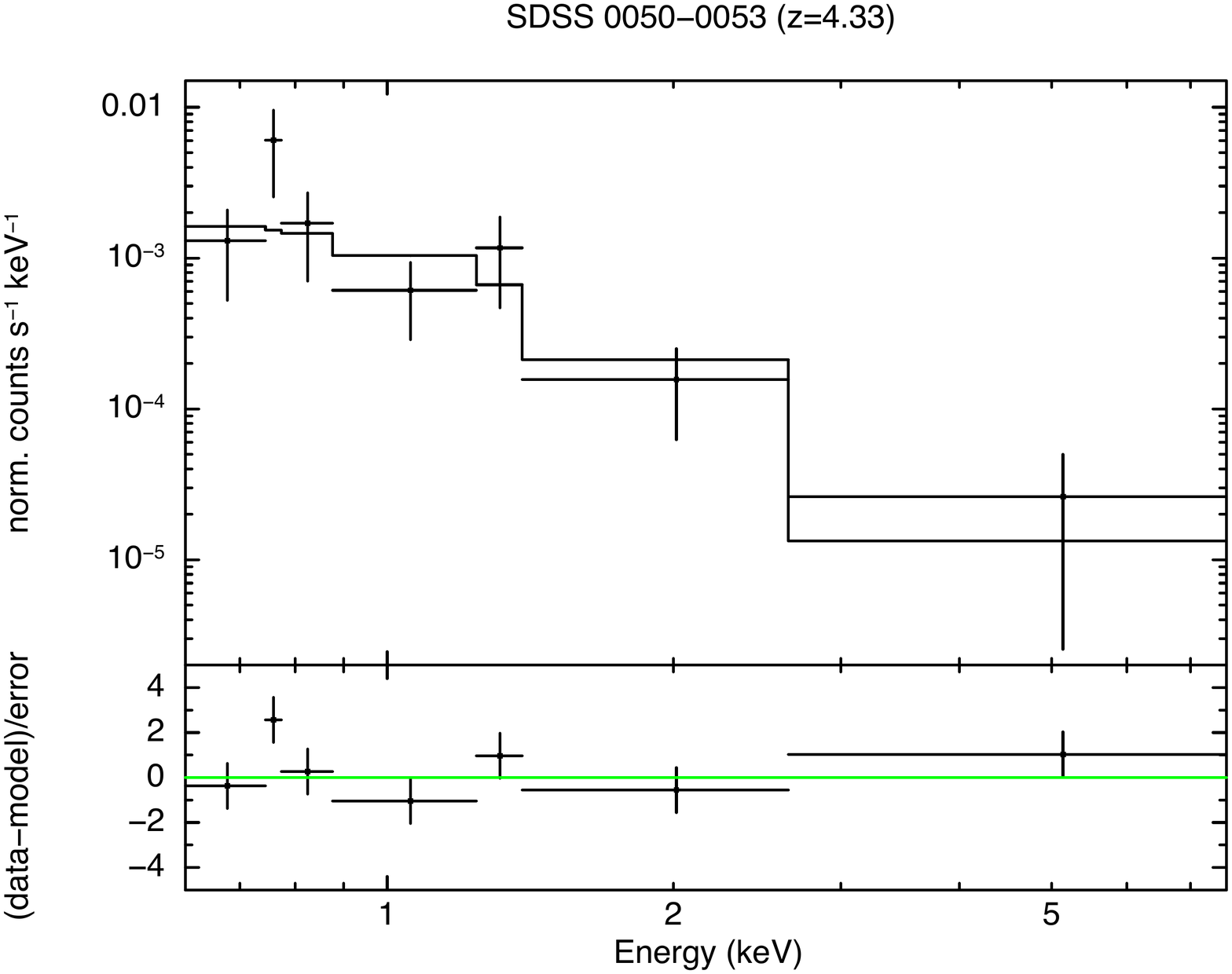}
   \includegraphics[width=0.495\textwidth]{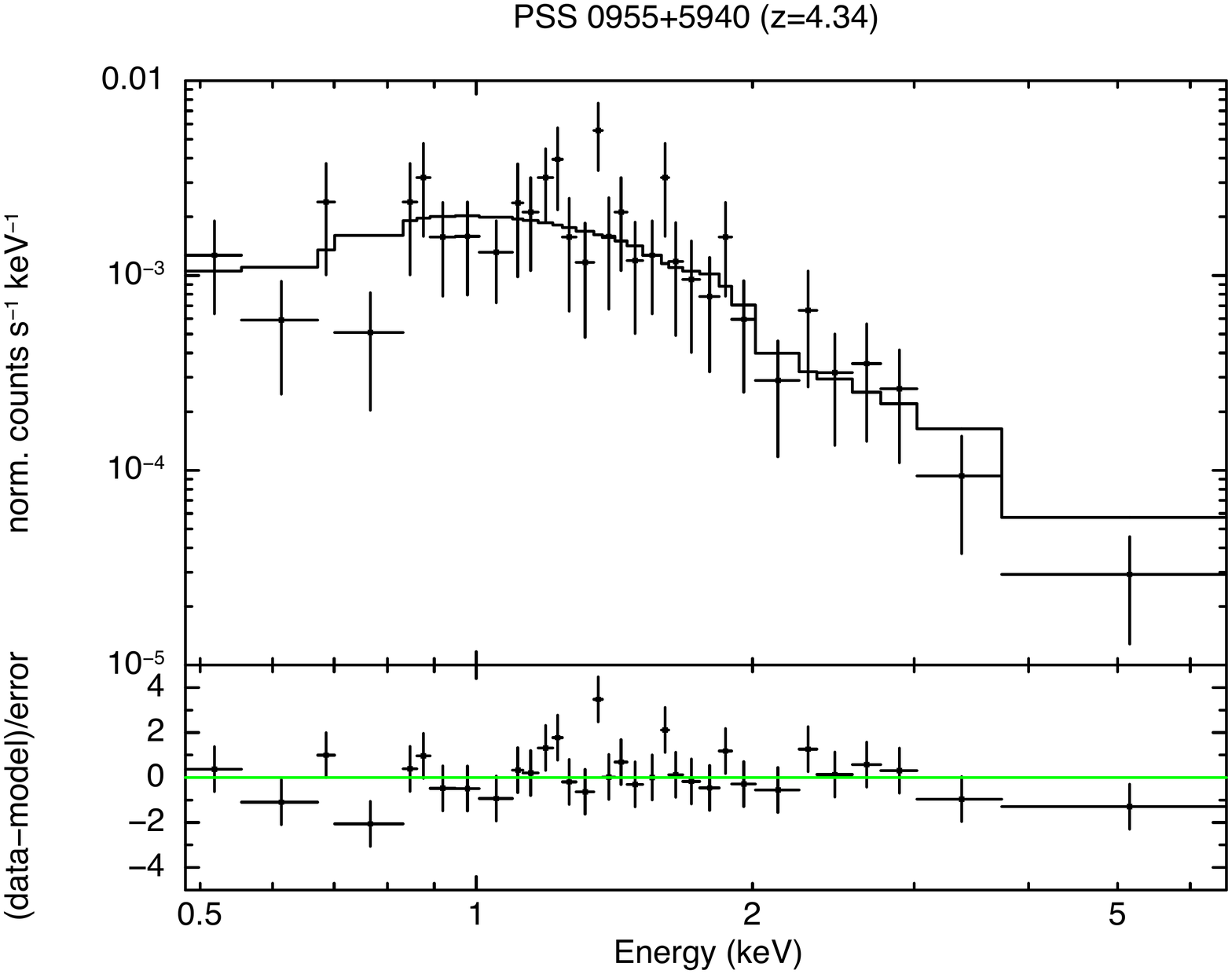}\\
    \vspace{1cm}
 \includegraphics[width=0.495\textwidth]{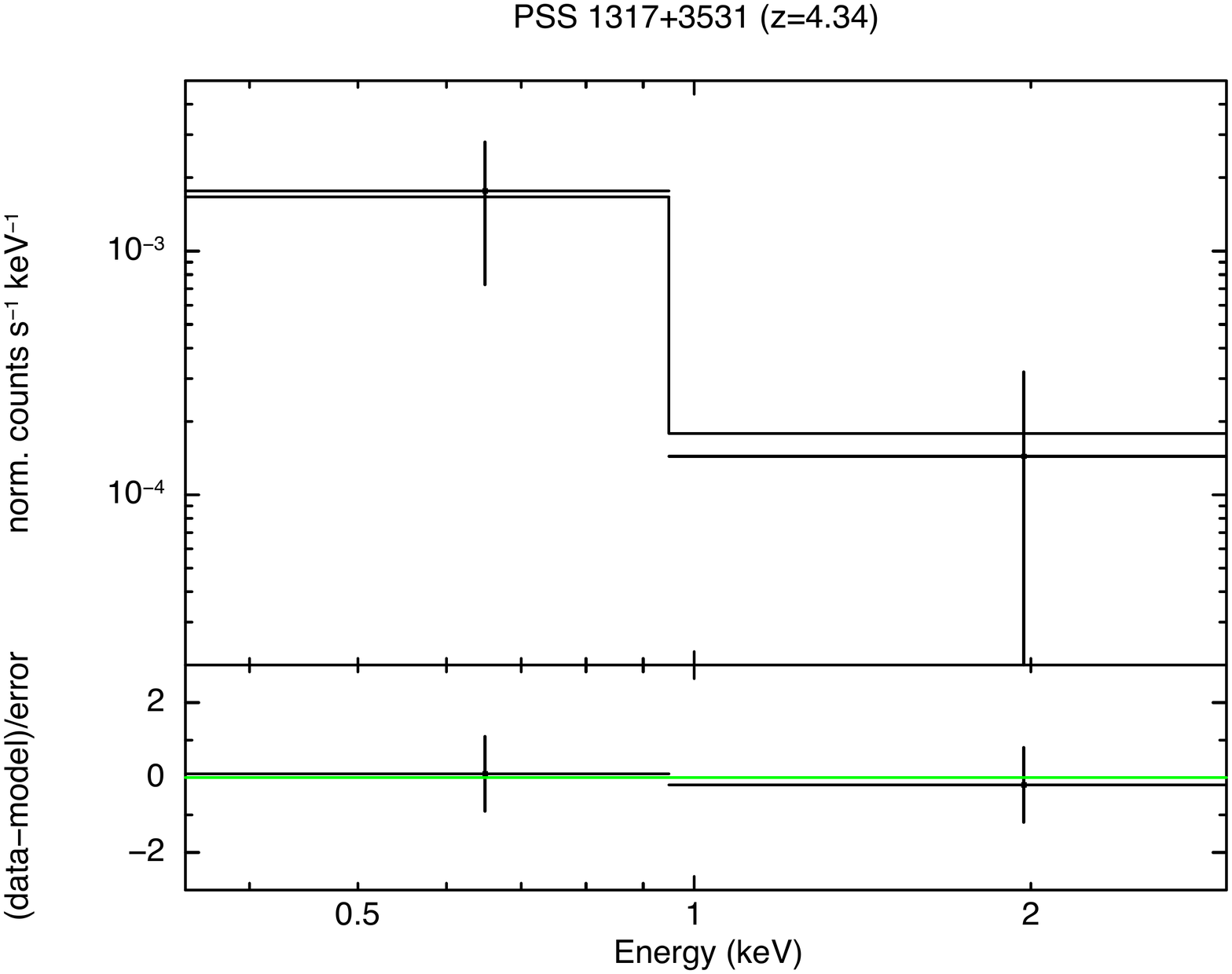}
   \includegraphics[width=0.495\textwidth]{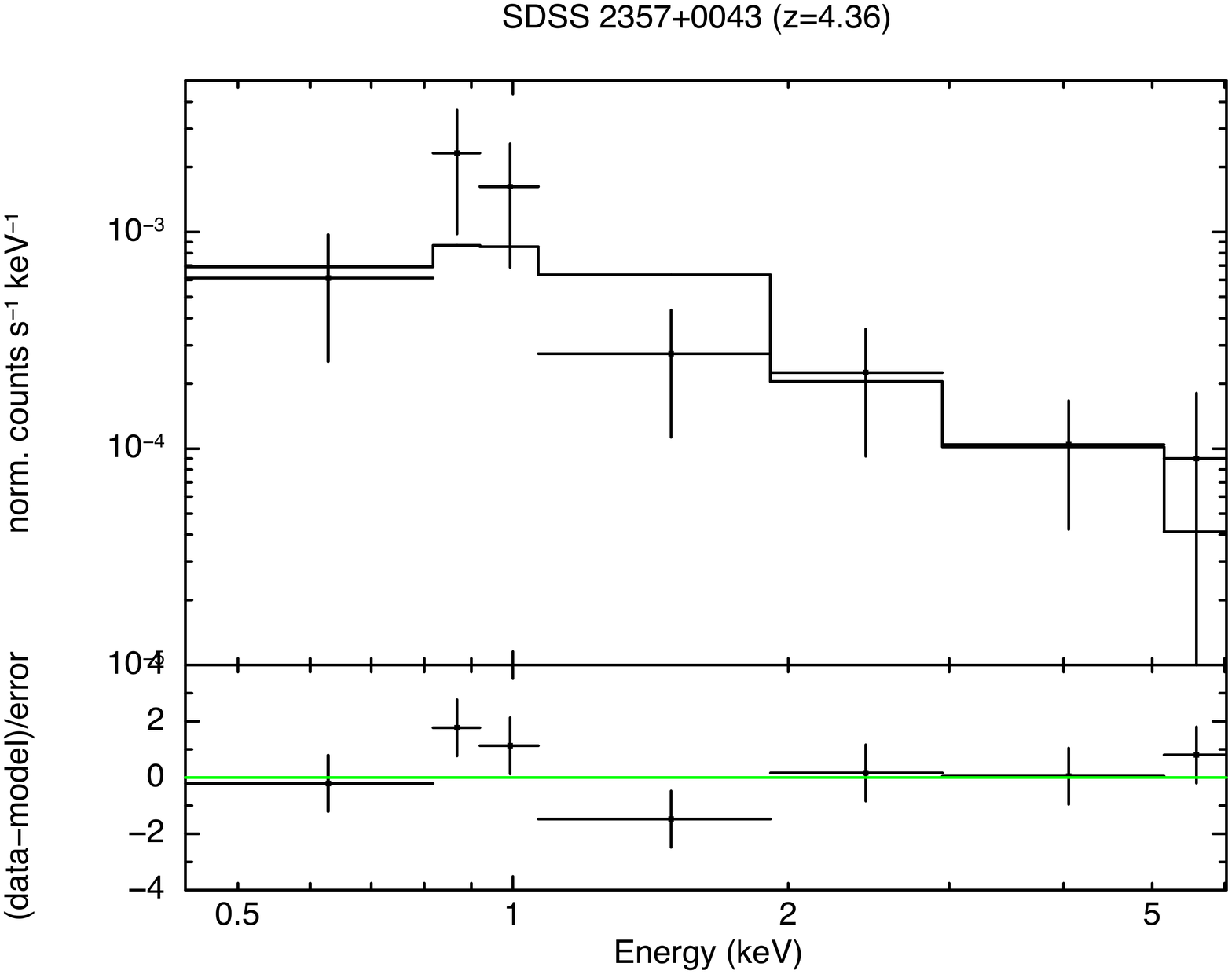}\\
\end{figure*}
\begin{figure*}[b!]
  \centering
 \includegraphics[width=0.495\textwidth]{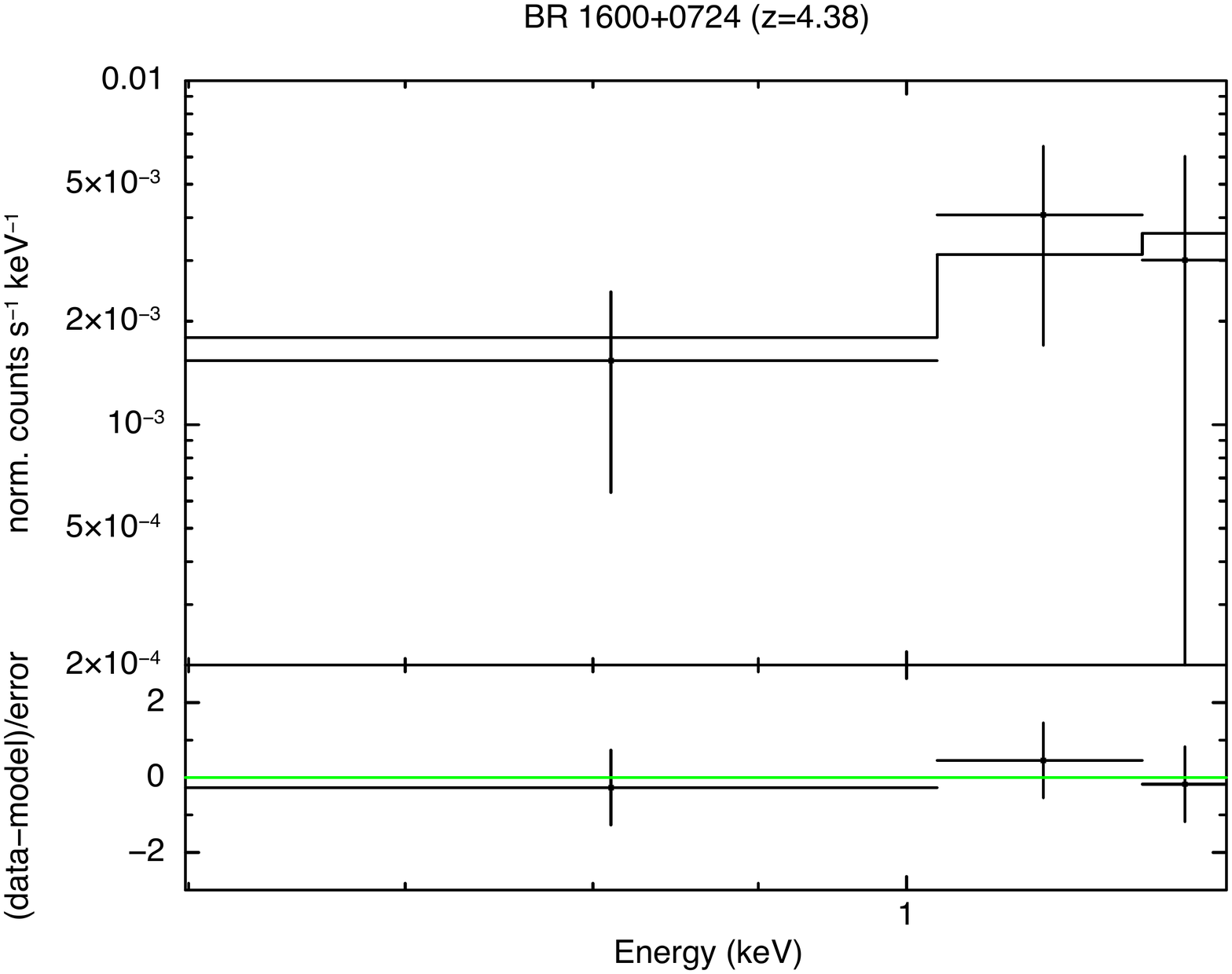}
   \includegraphics[width=0.495\textwidth]{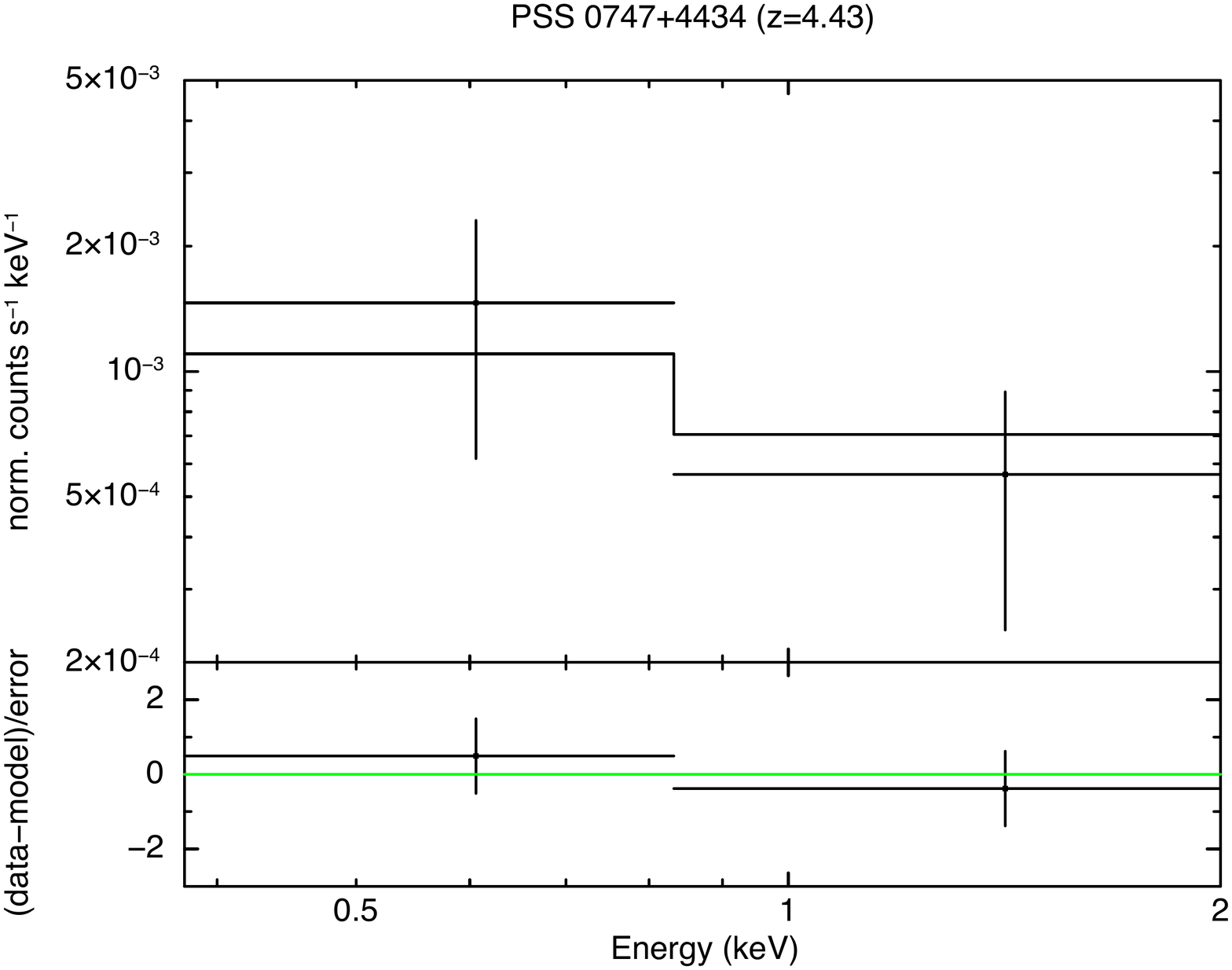}\\ 
    \vspace{1cm}
   \includegraphics[width=0.495\textwidth]{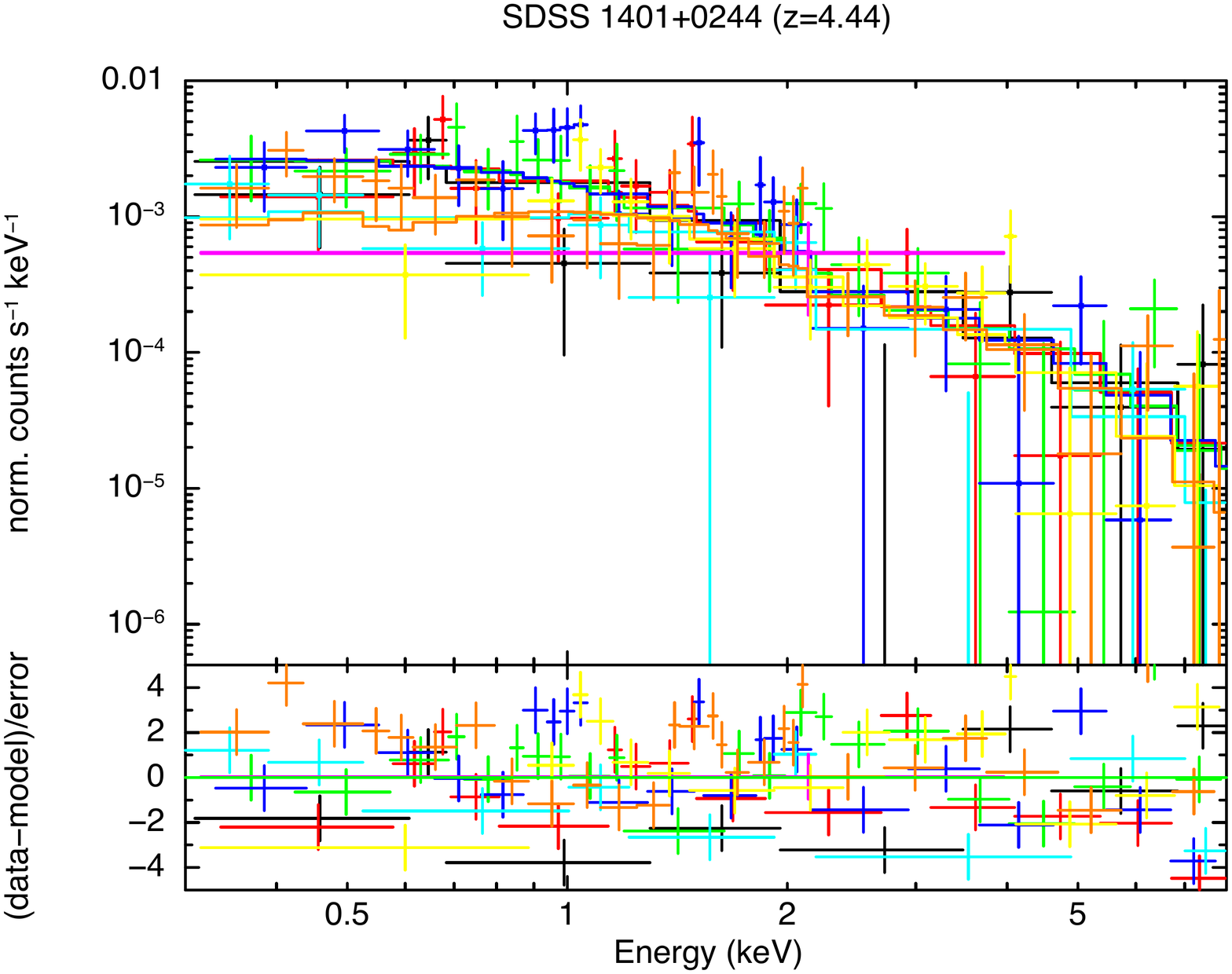}
 \includegraphics[width=0.495\textwidth]{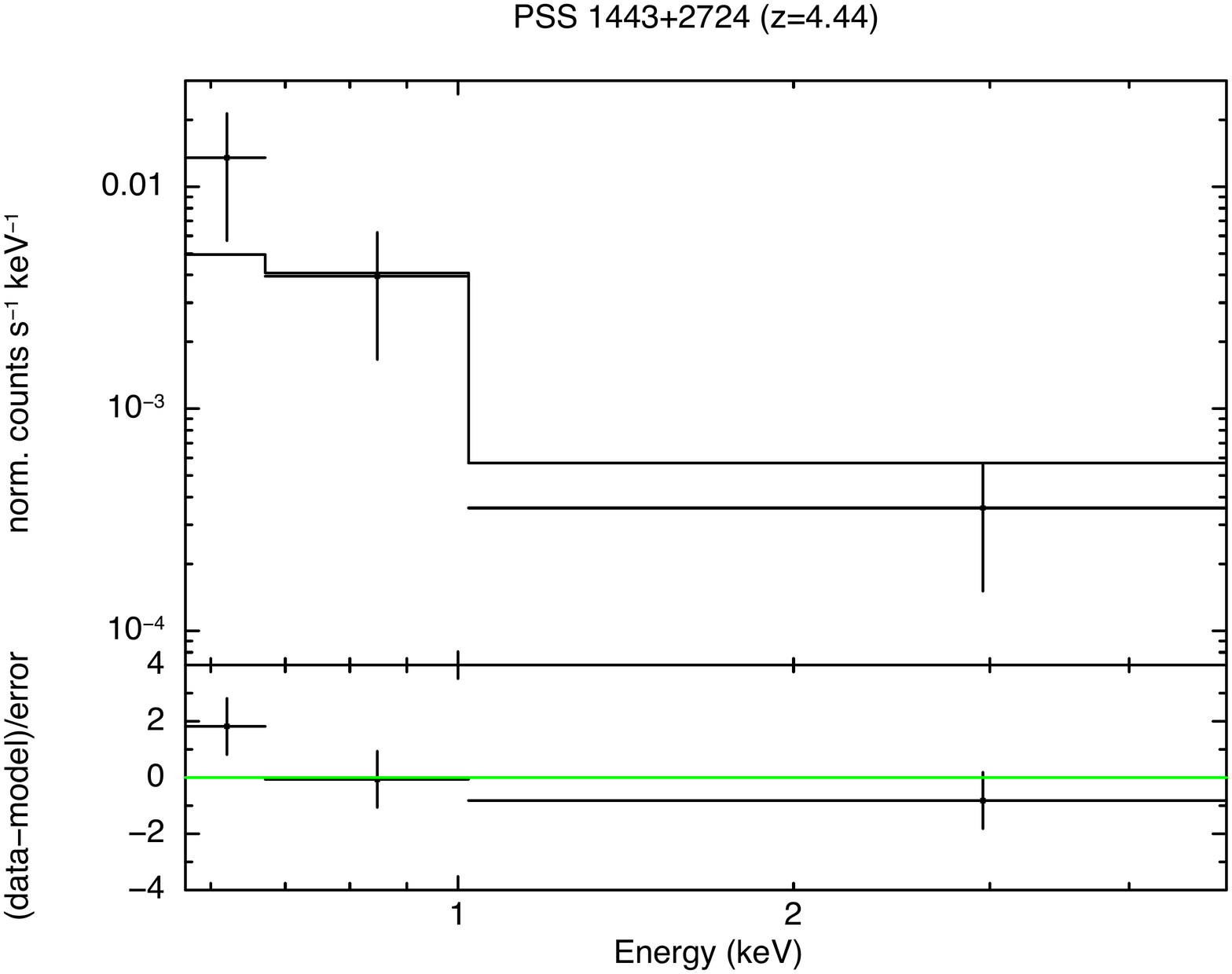}\\
    \vspace{1cm}
   \includegraphics[width=0.495\textwidth]{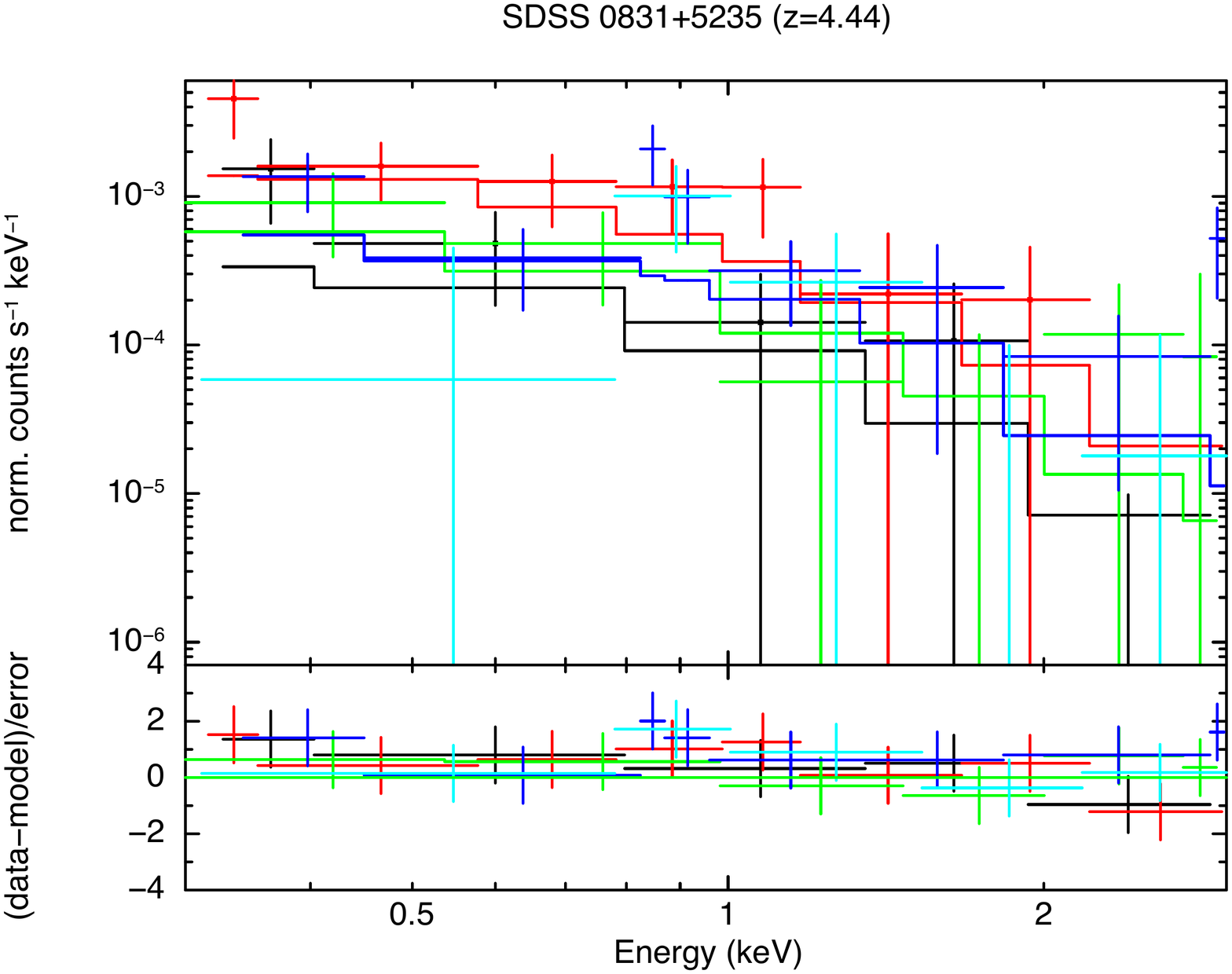}
 \includegraphics[width=0.495\textwidth]{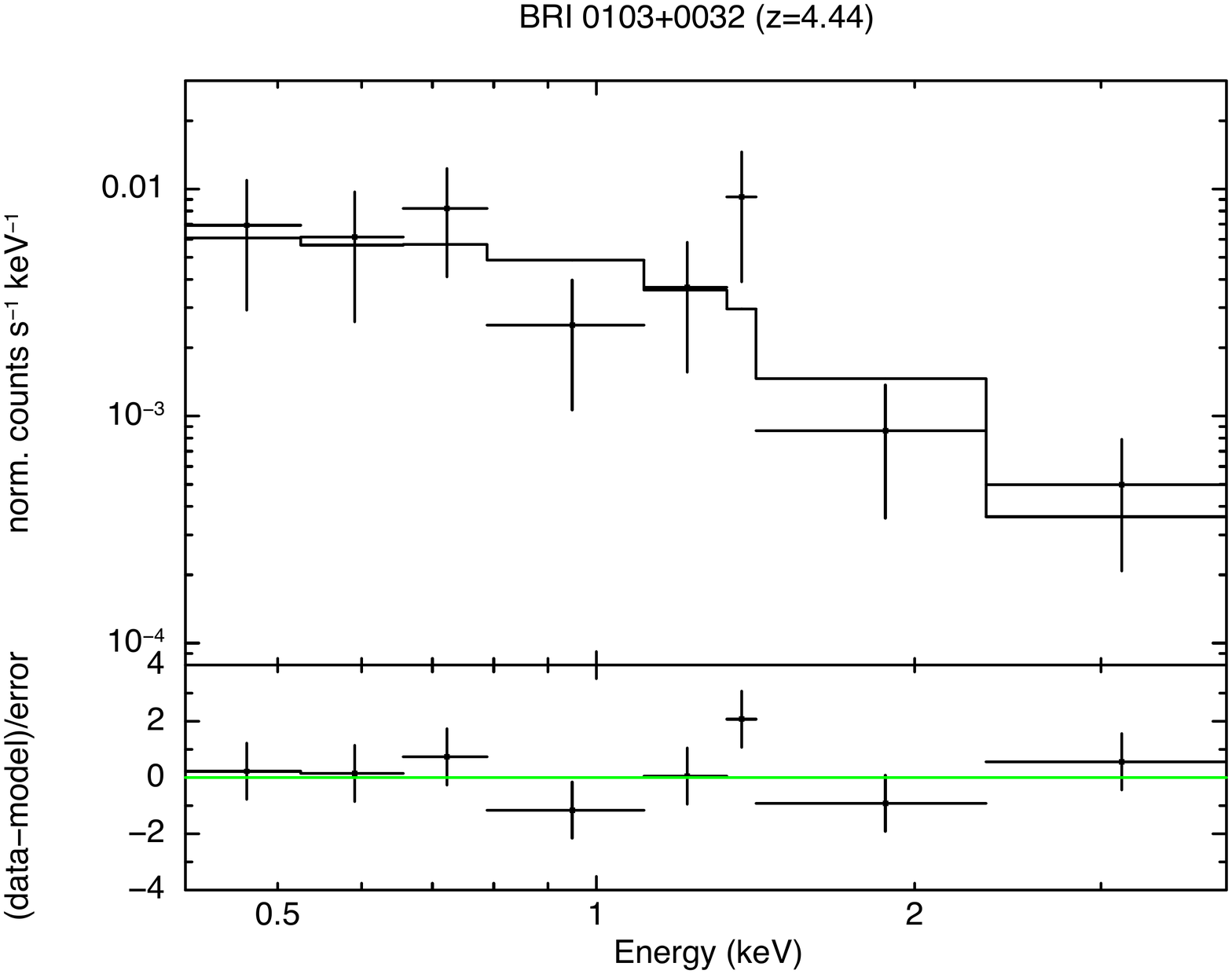}\\
\end{figure*}
\begin{figure*}[b!]
  \centering
   \includegraphics[width=0.495\textwidth]{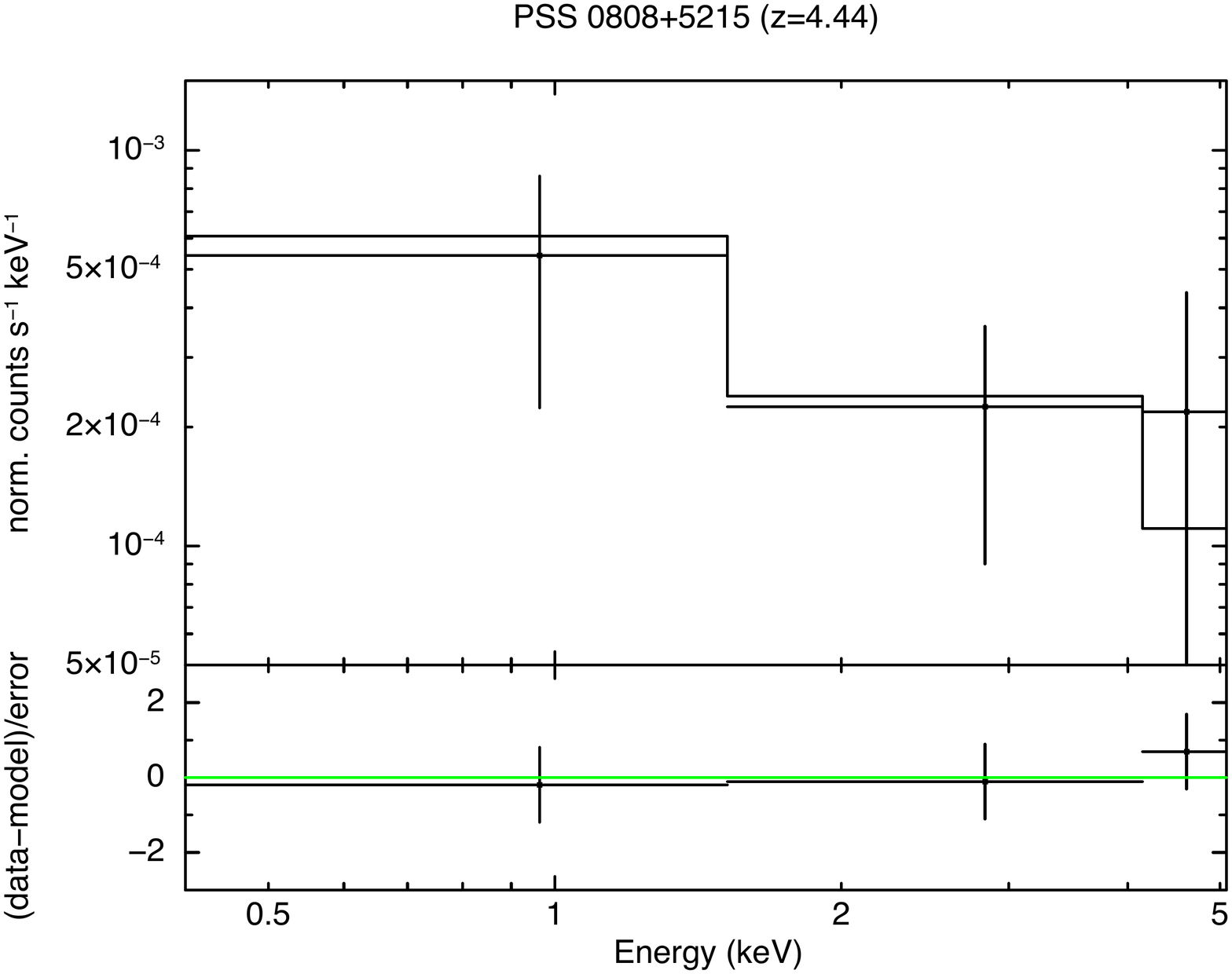}
 \includegraphics[width=0.495\textwidth]{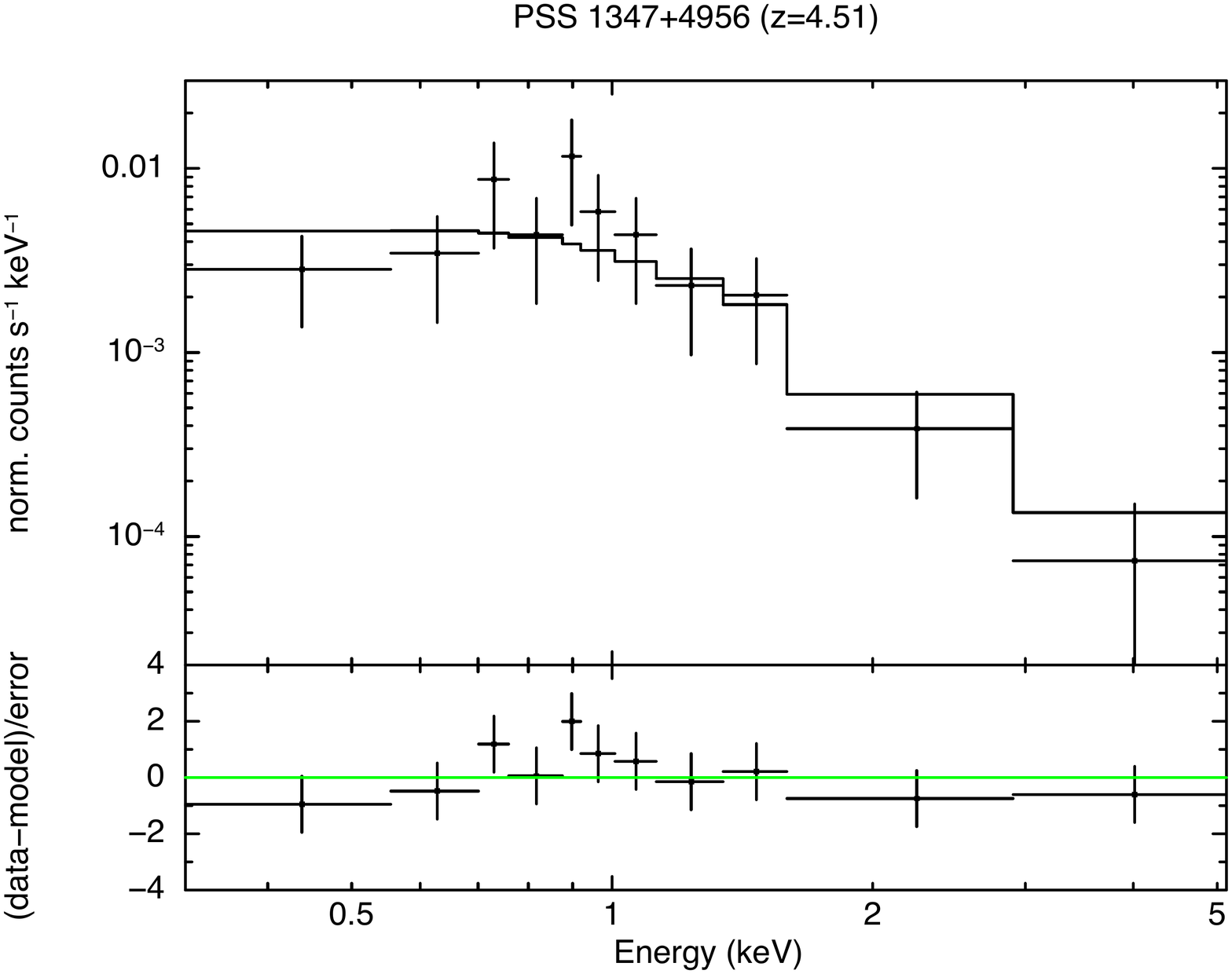}\\
    \vspace{1cm}
   \includegraphics[width=0.495\textwidth]{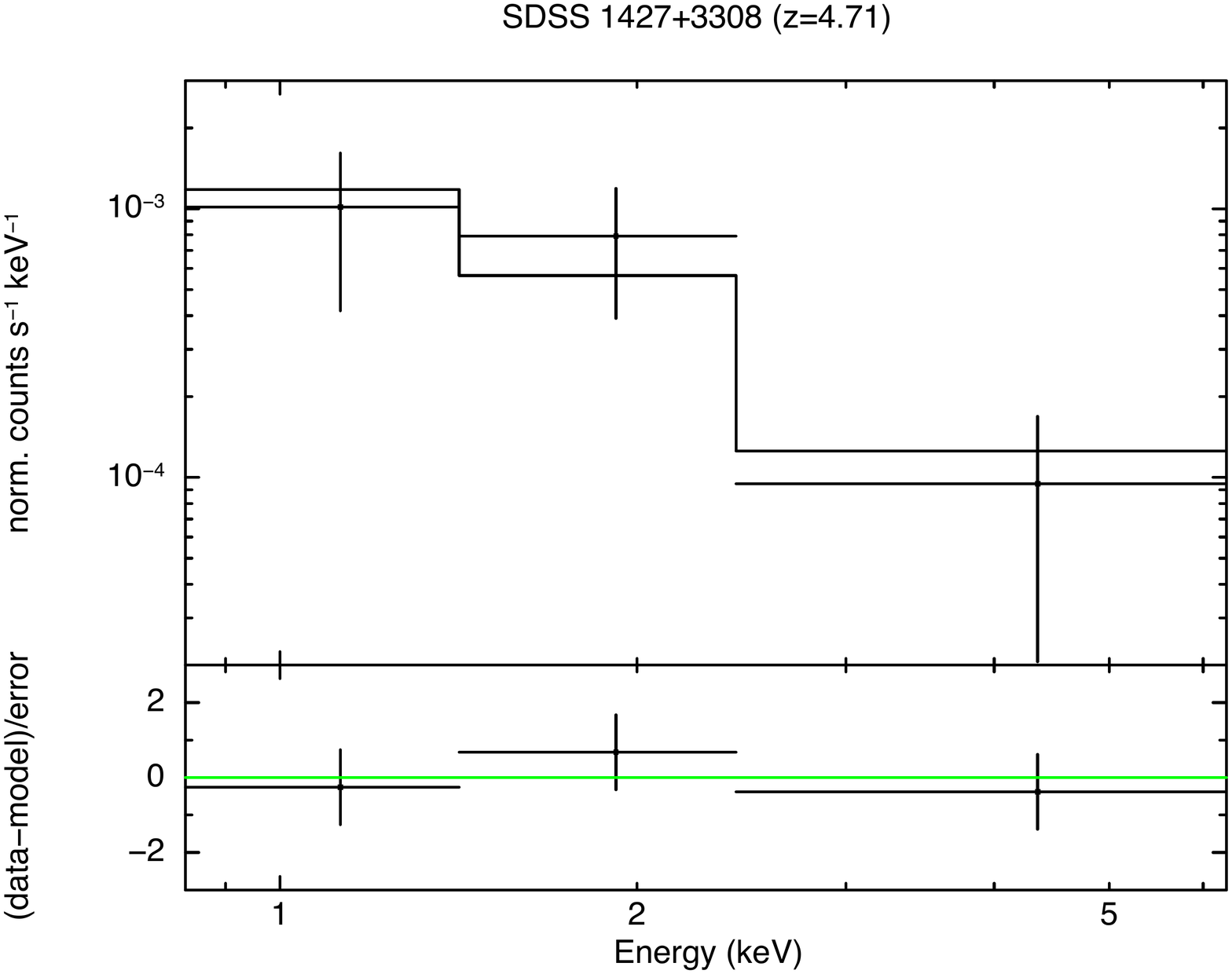}     
 \includegraphics[width=0.495\textwidth]{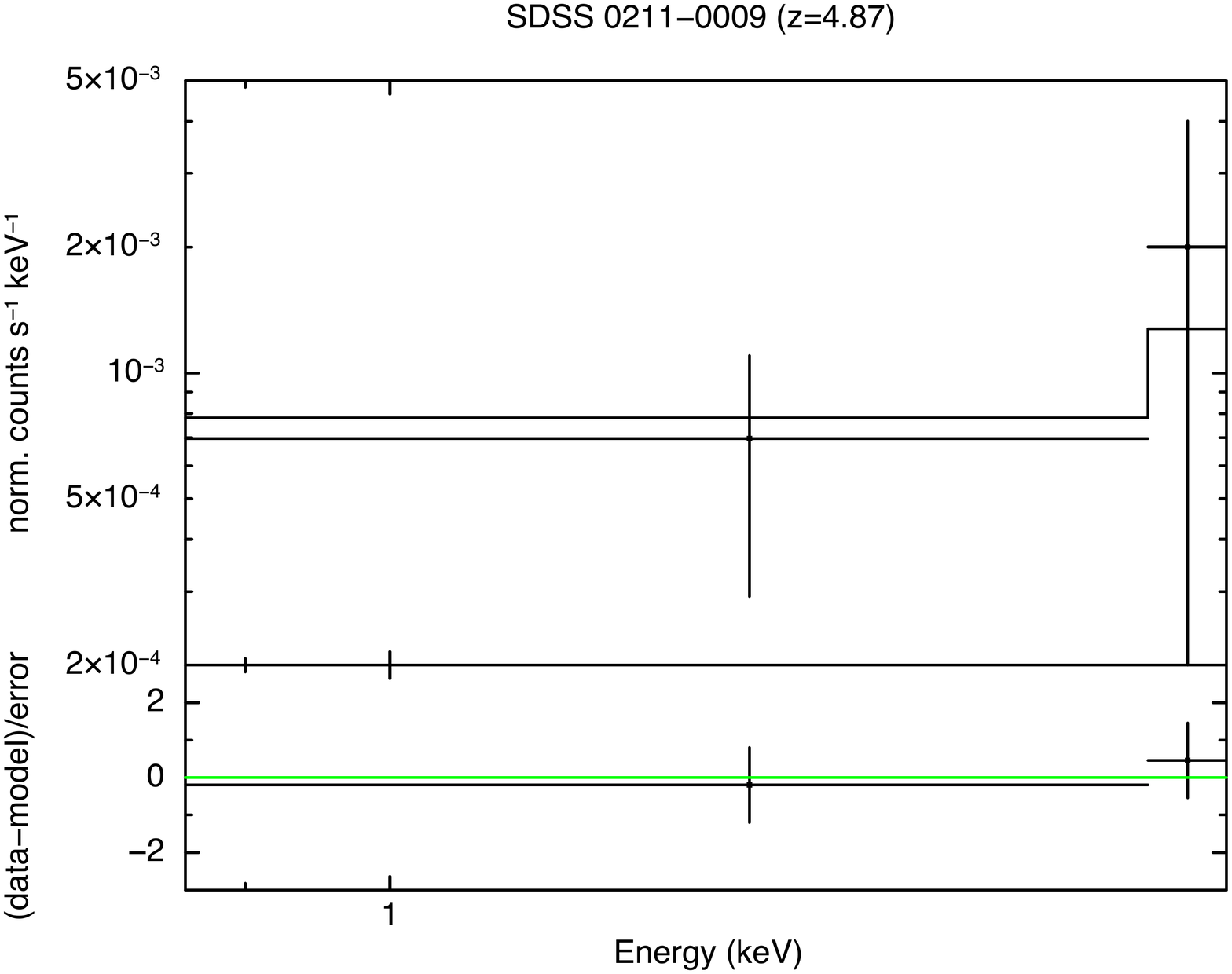}\\
    \vspace{1cm}
   \includegraphics[width=0.495\textwidth]{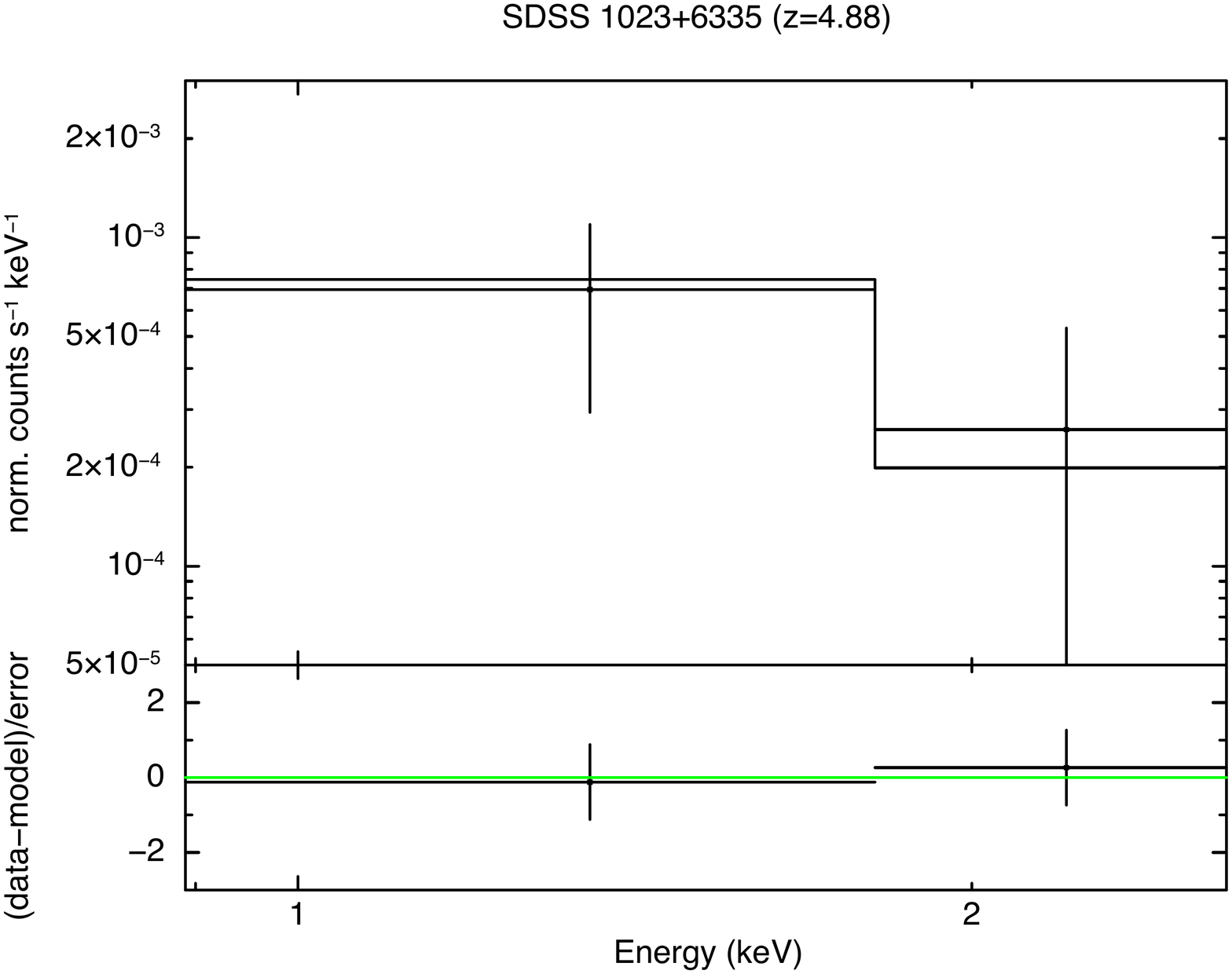}     
 \includegraphics[width=0.495\textwidth]{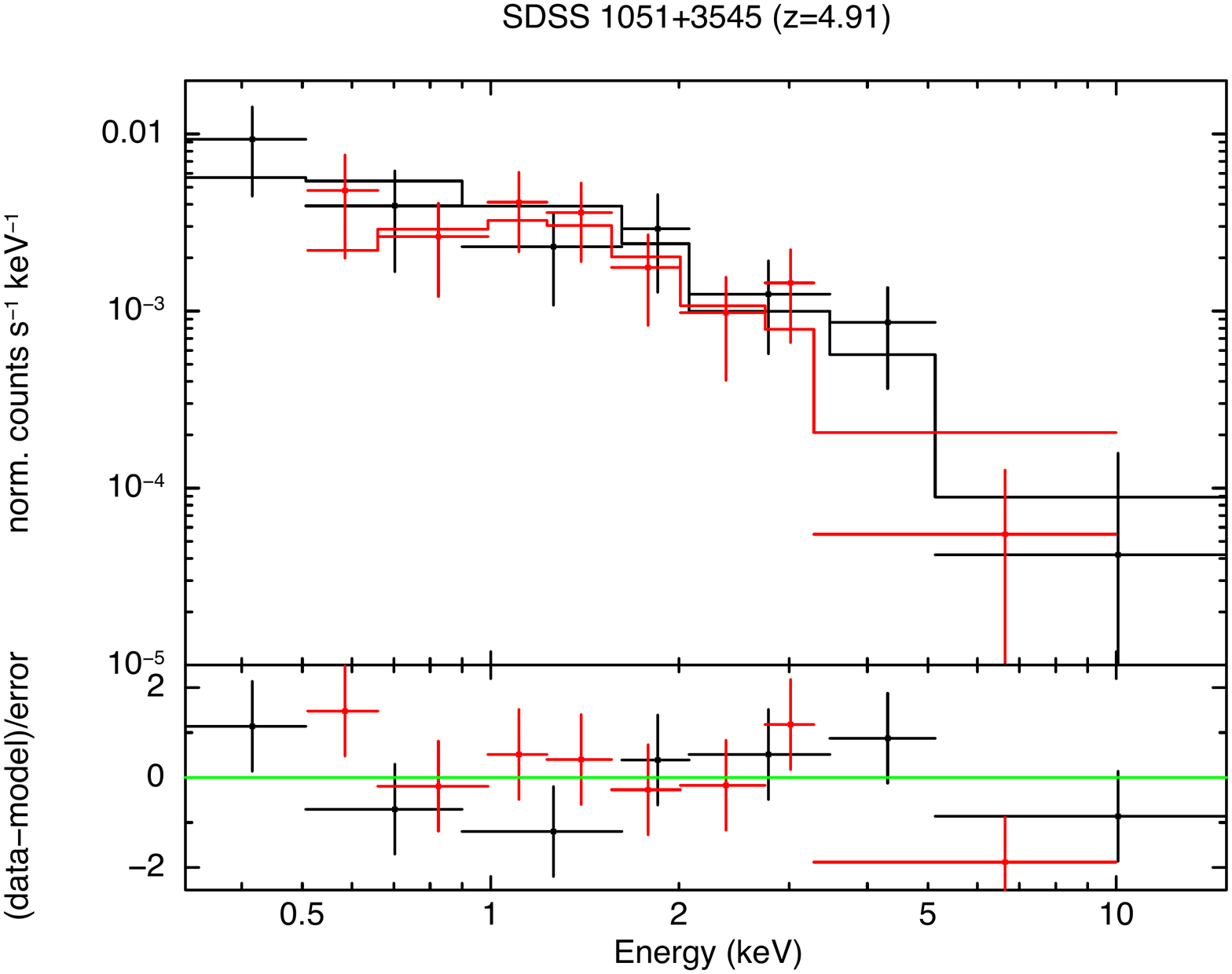}\\
\end{figure*}
\begin{figure*}[b!]
  \centering
   \includegraphics[width=0.495\textwidth]{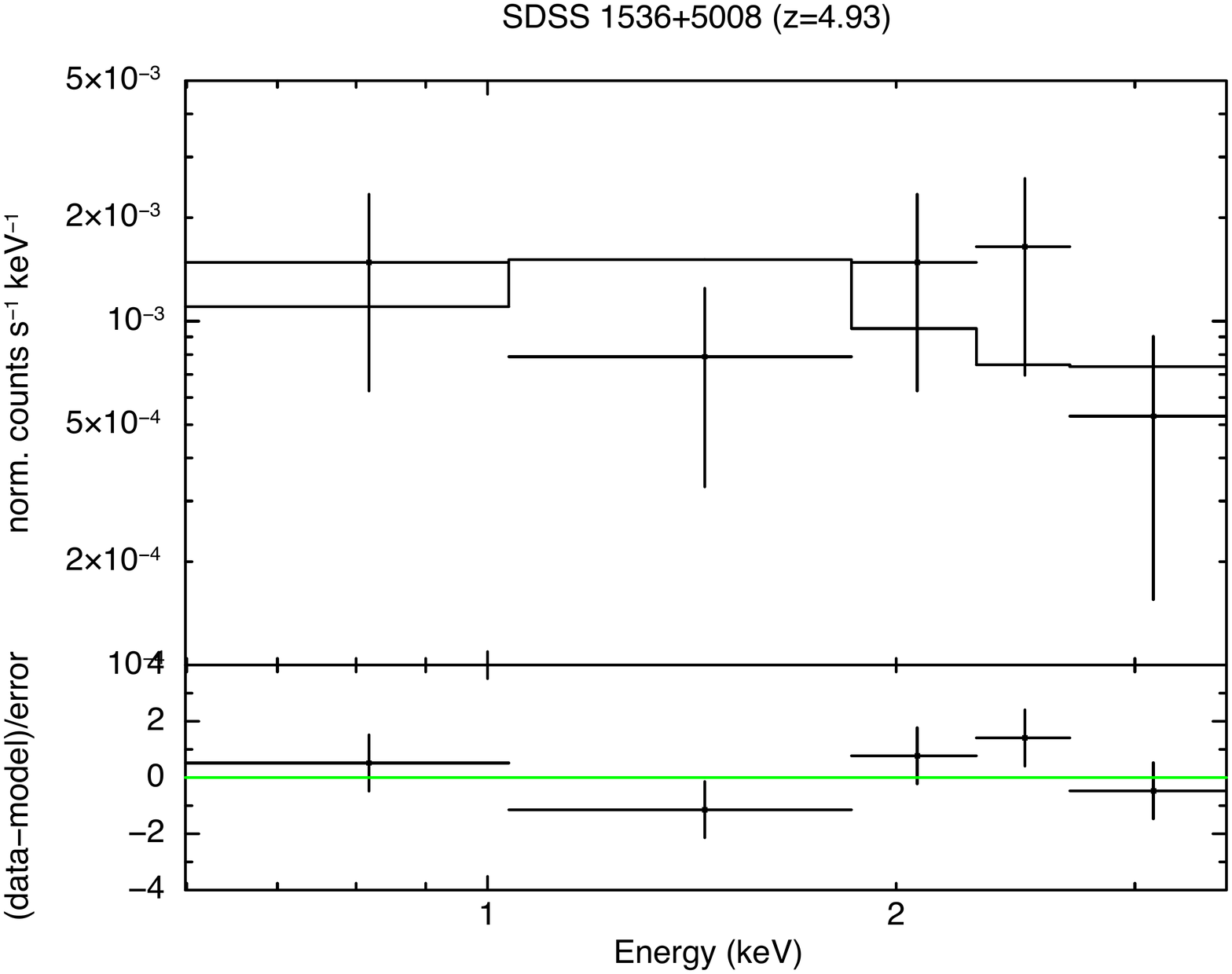}     
 \includegraphics[width=0.495\textwidth]{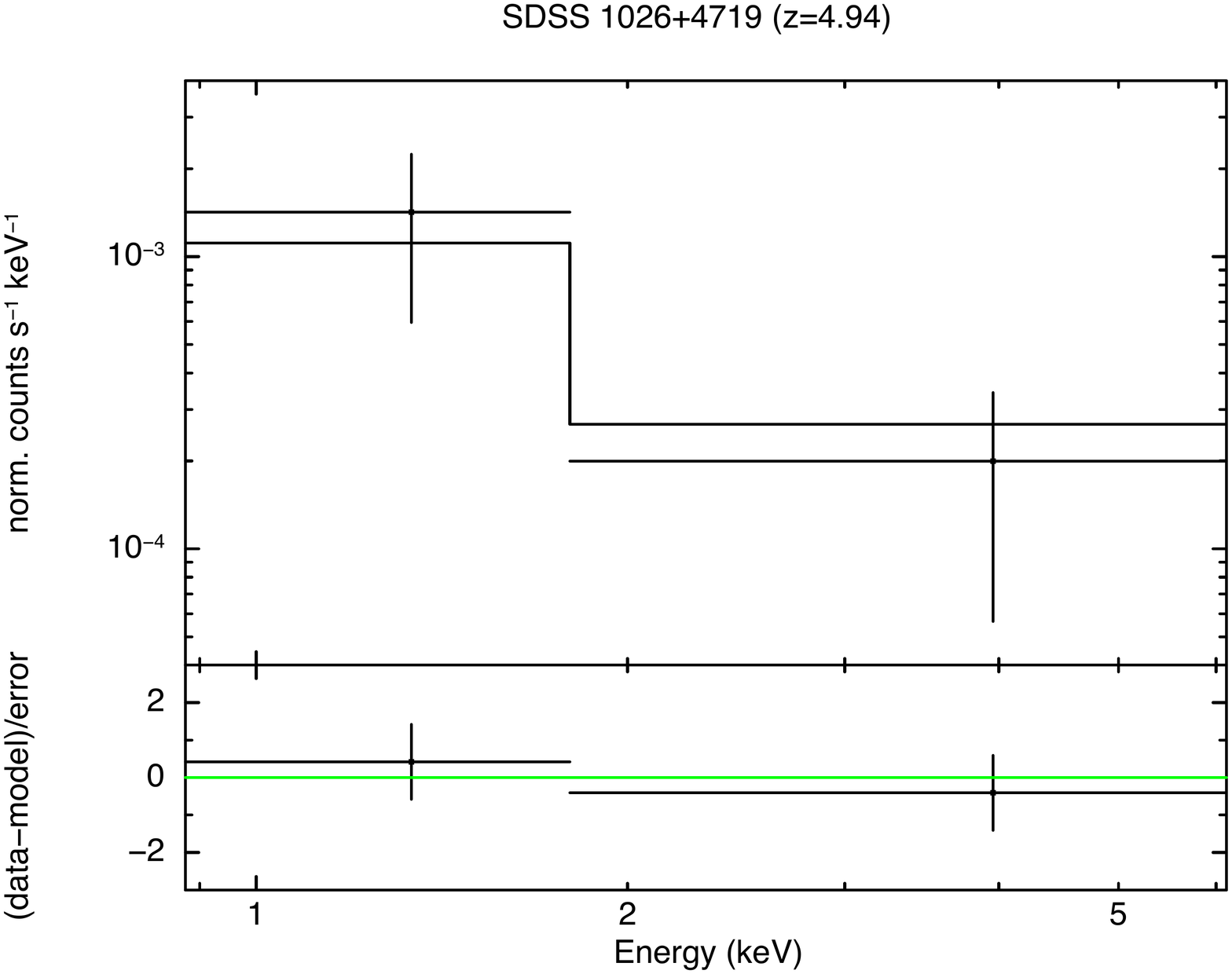}\\
    \vspace{1cm}
   \includegraphics[width=0.495\textwidth]{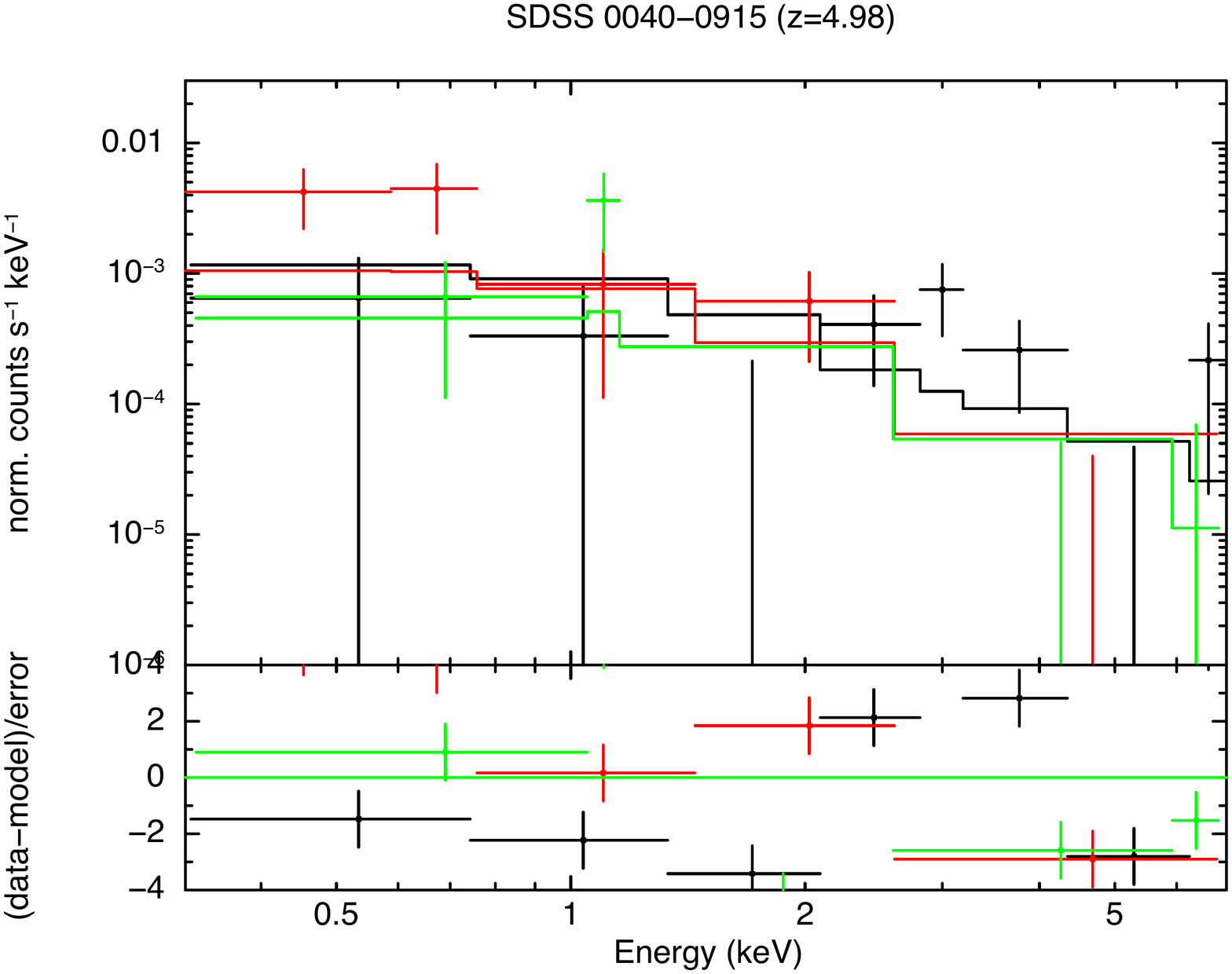}     
 \includegraphics[width=0.495\textwidth]{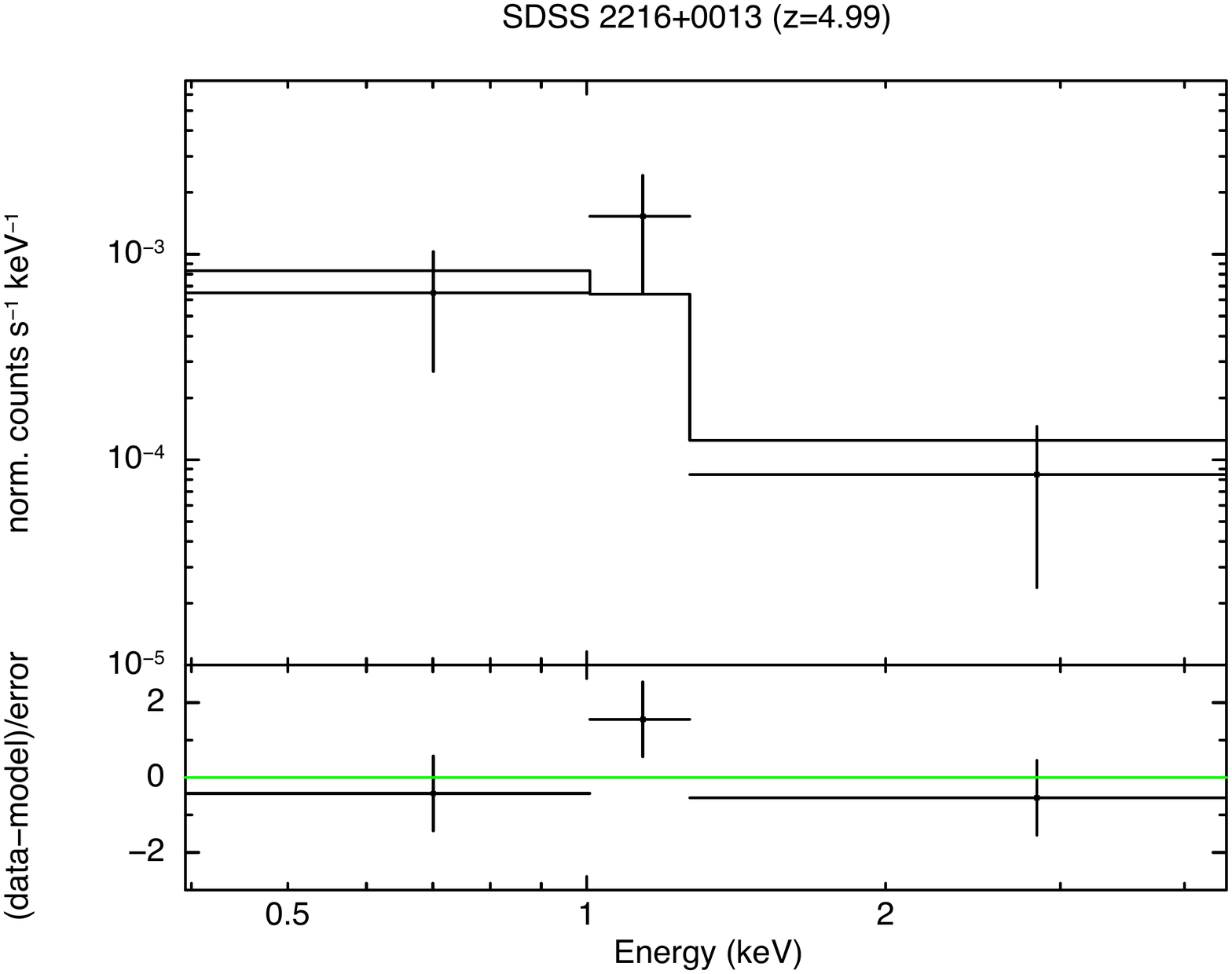}\\
    \vspace{1cm}
   \includegraphics[width=0.495\textwidth]{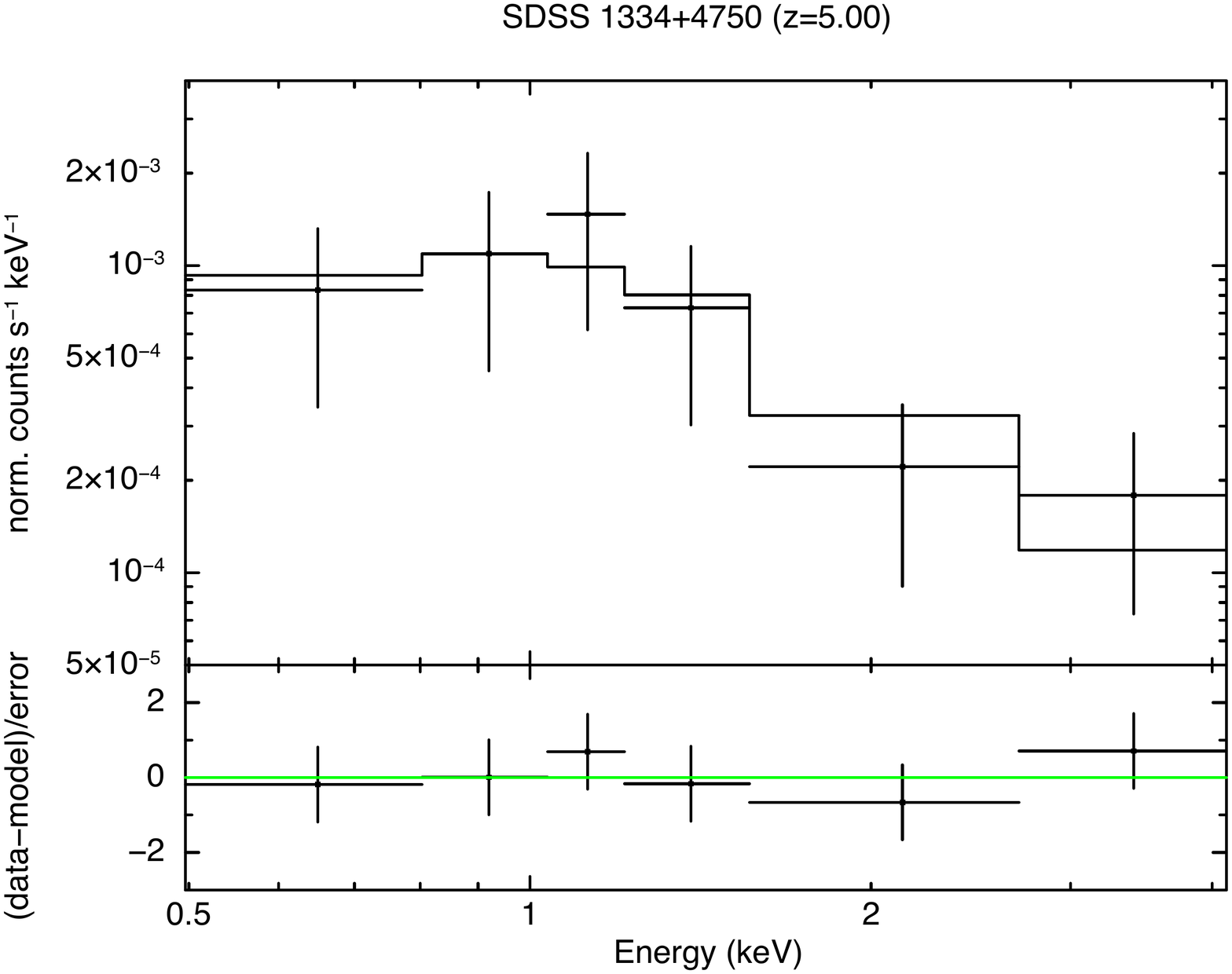}     
 \includegraphics[width=0.495\textwidth]{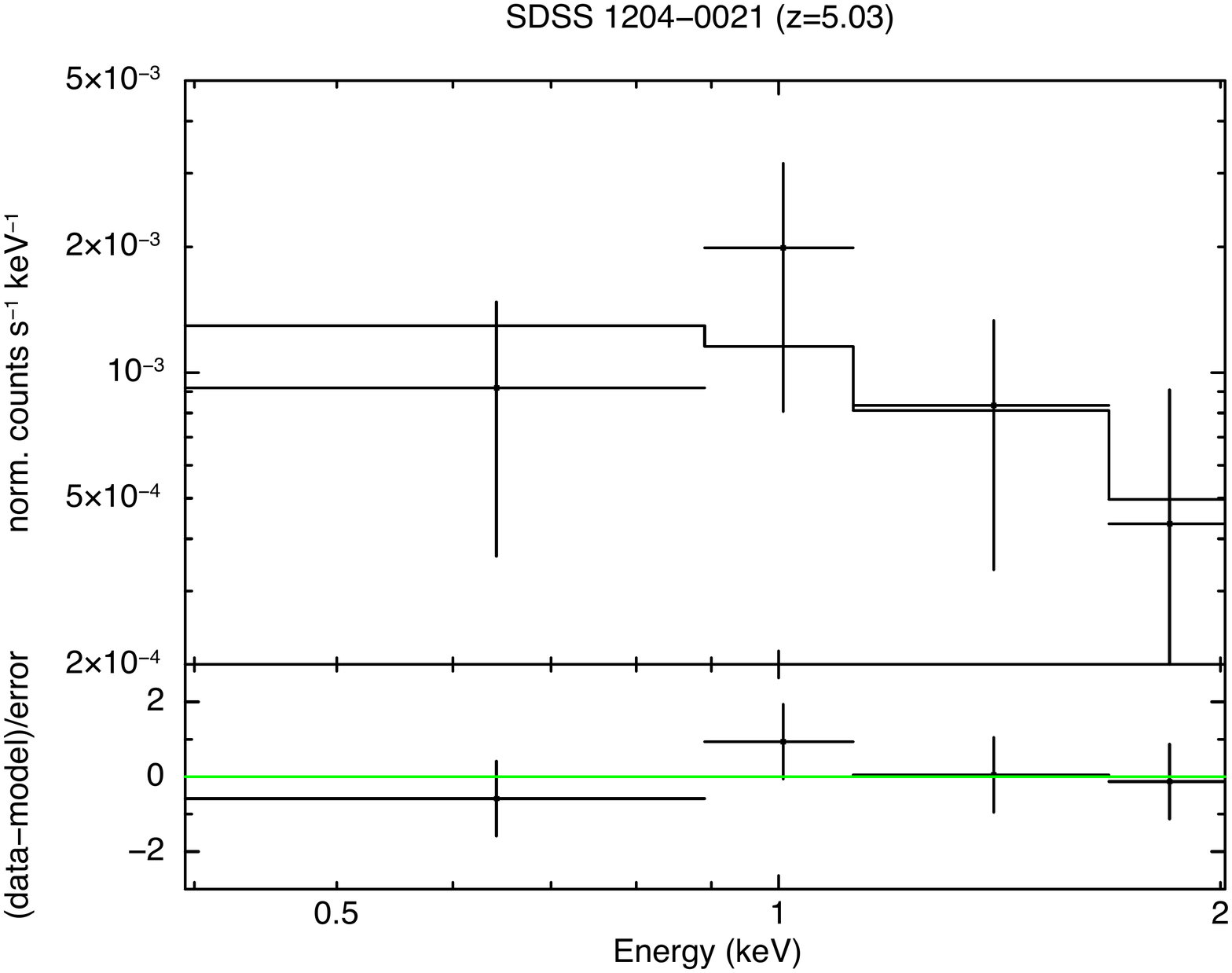}\\
\end{figure*}
\begin{figure*}[hb!]
  \centering
   \includegraphics[width=0.495\textwidth]{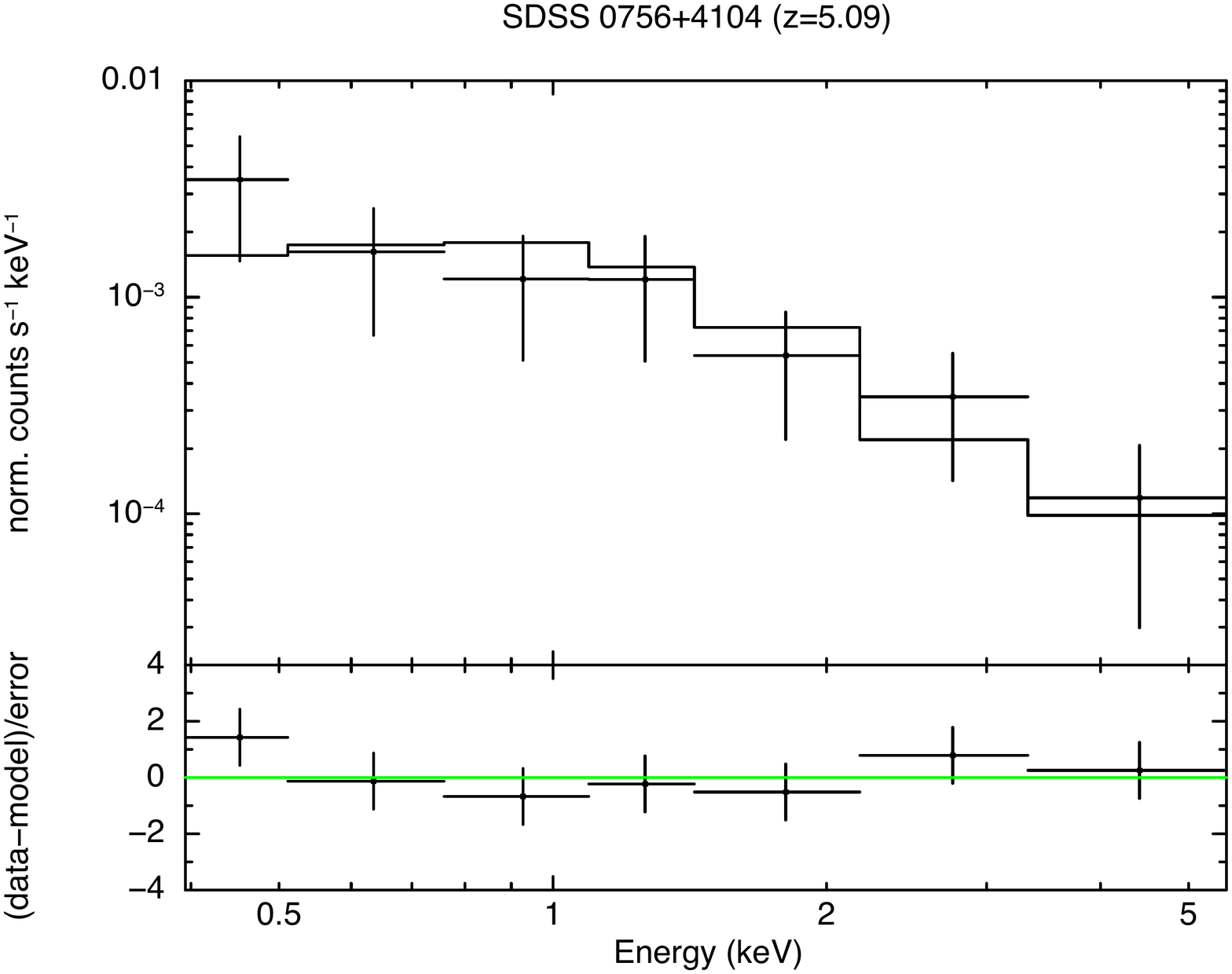}     
 \includegraphics[width=0.495\textwidth]{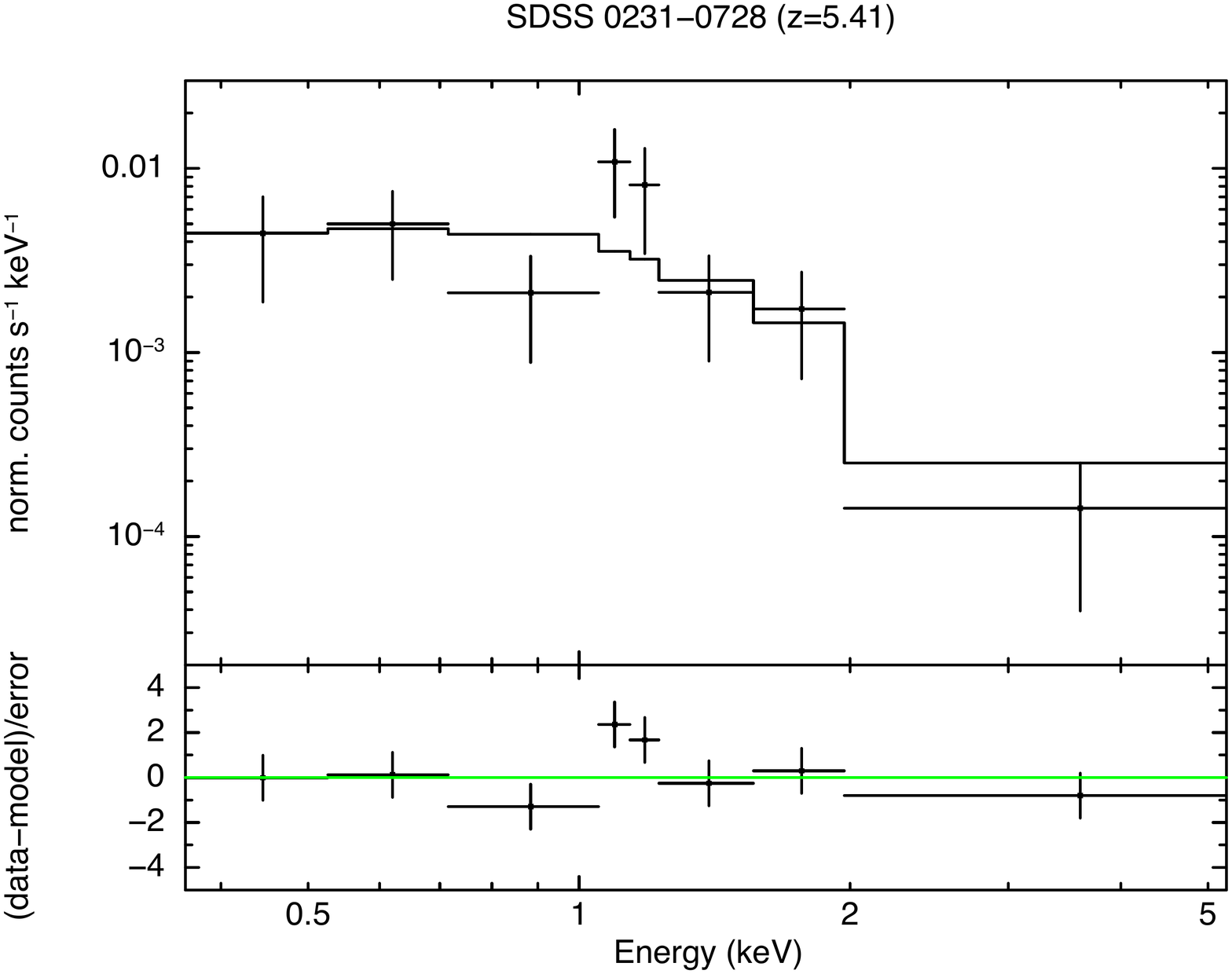}\\
    \vspace{1cm}
   \includegraphics[width=0.495\textwidth]{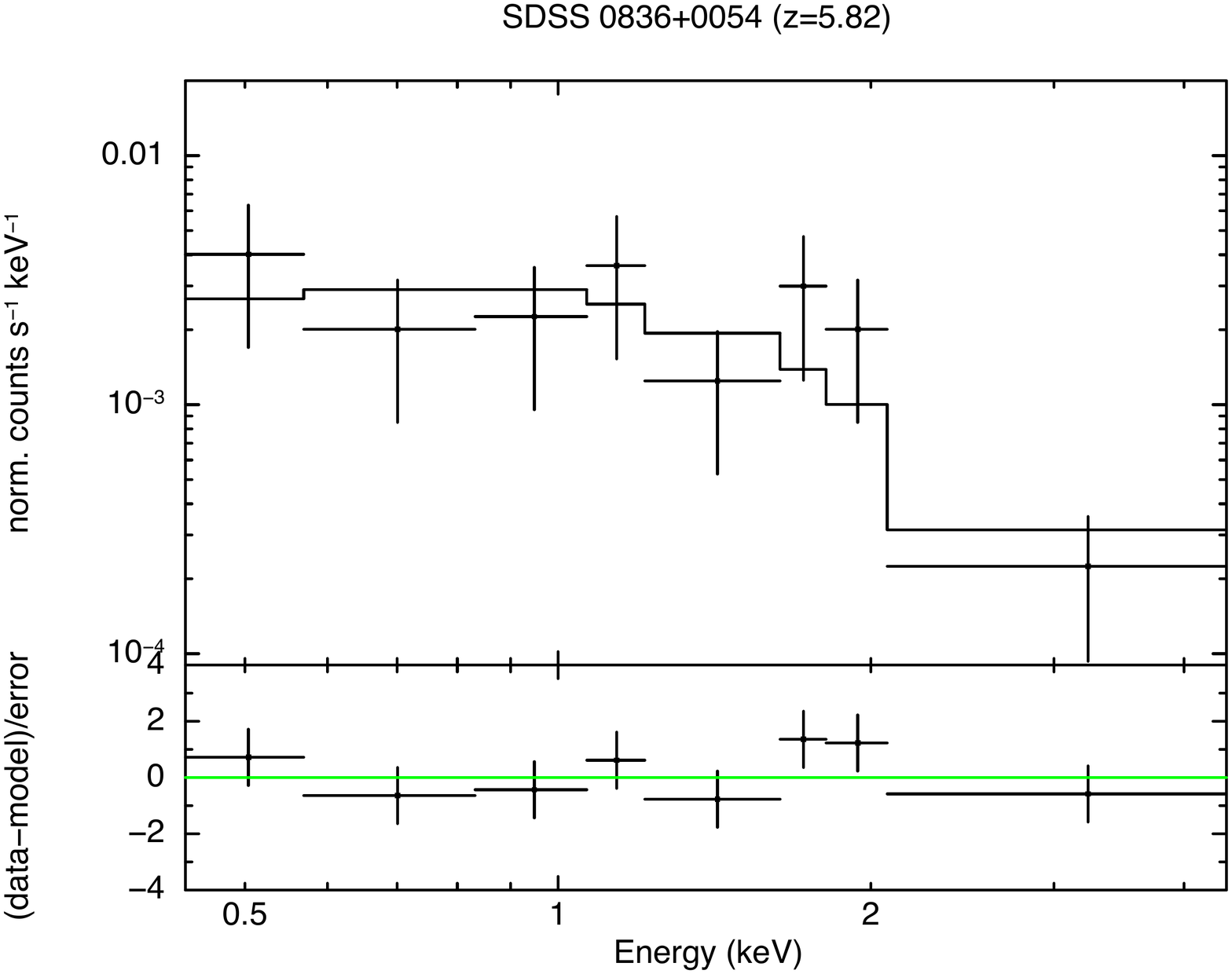}     
 \includegraphics[width=0.495\textwidth]{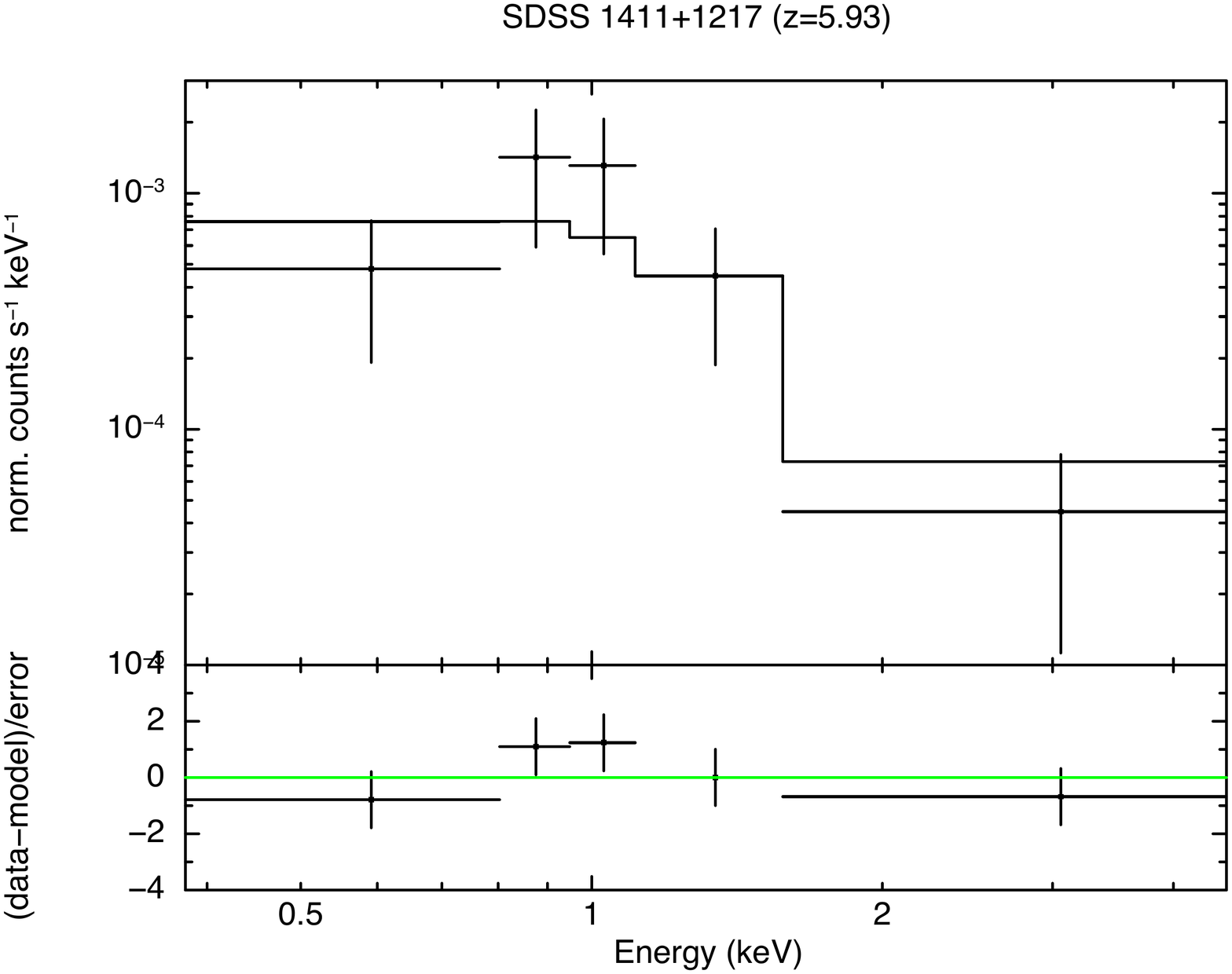}\\
    \vspace{1cm}
   \includegraphics[width=0.495\textwidth]{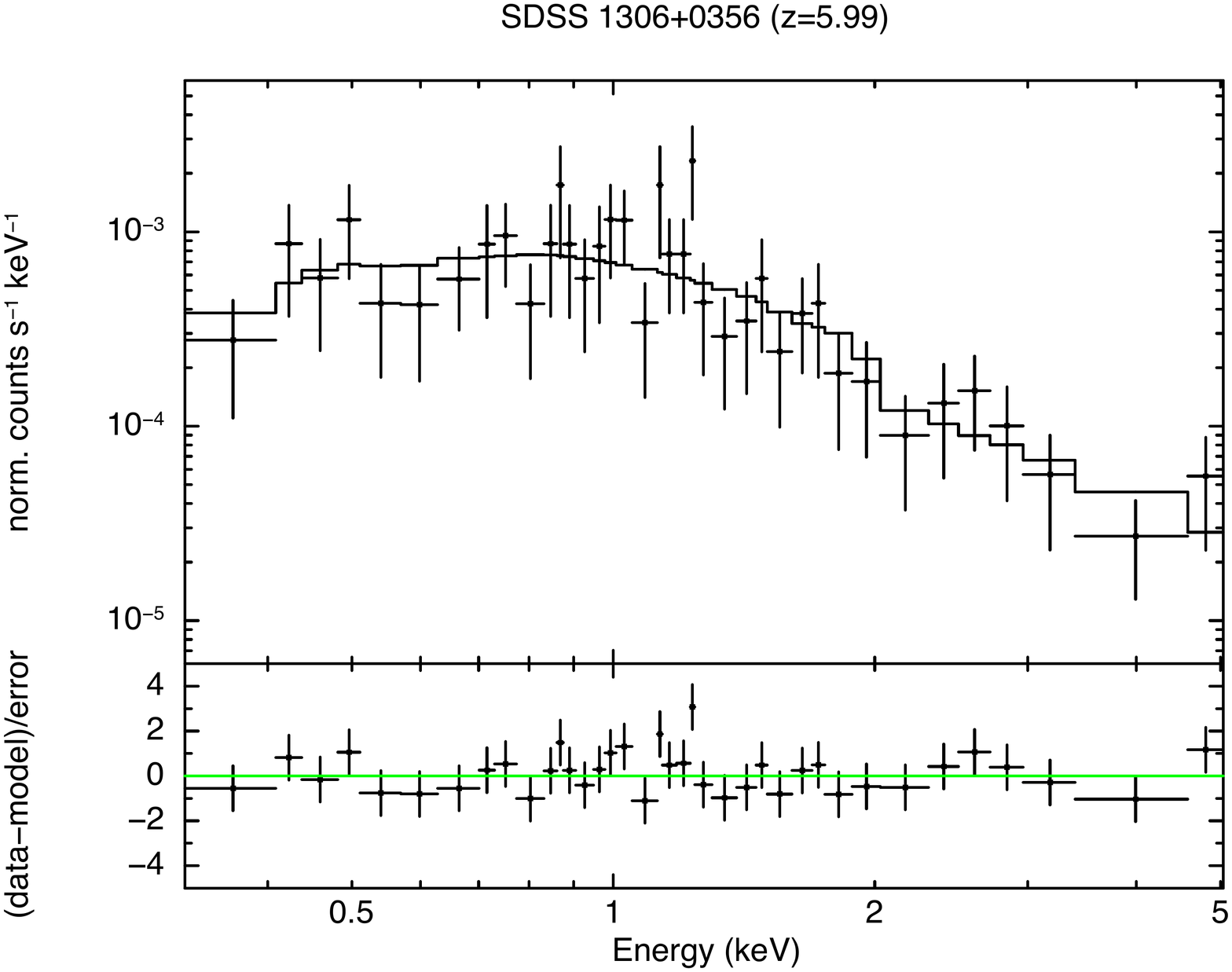}     
 \includegraphics[width=0.495\textwidth]{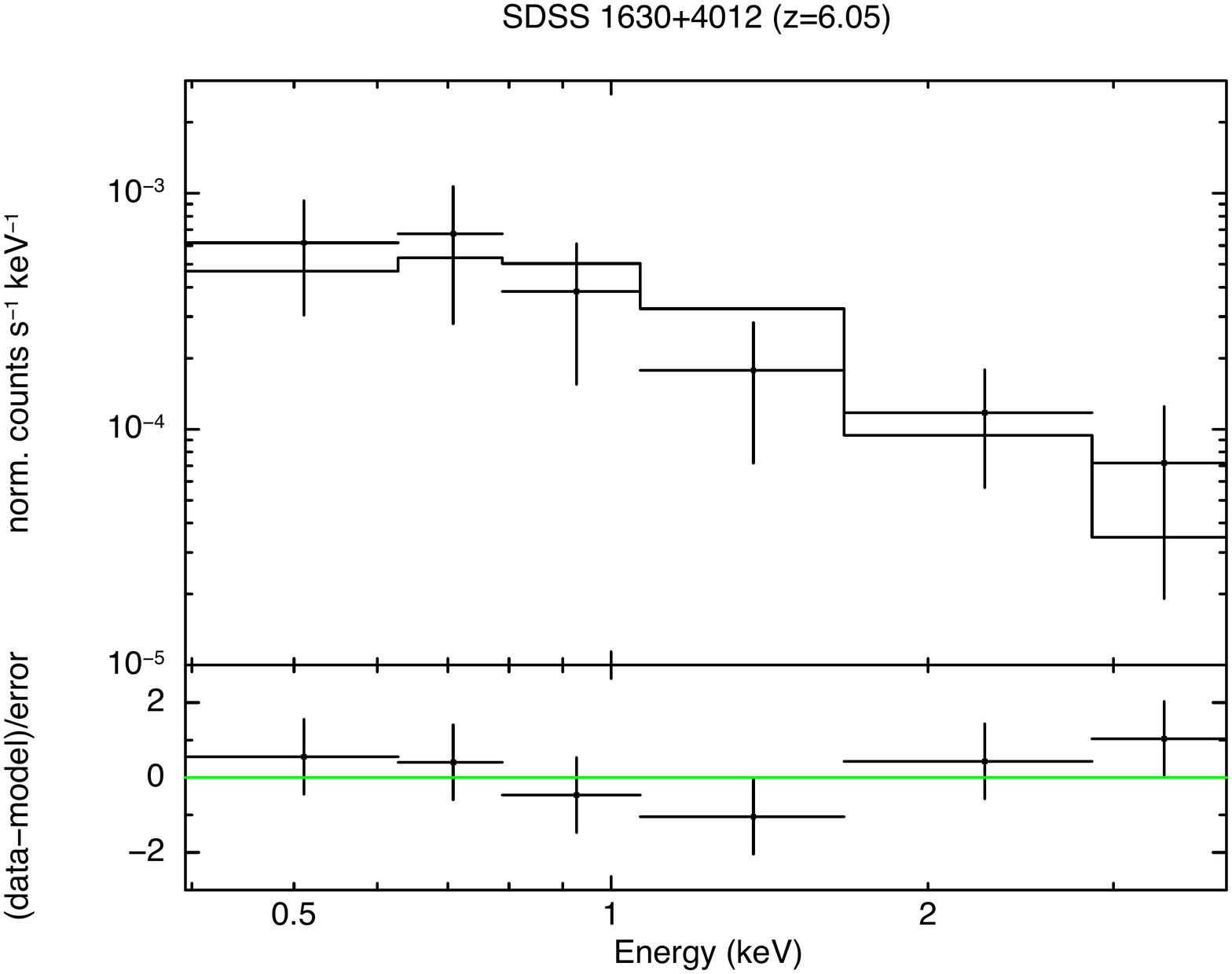}\\
 \end{figure*}
\begin{figure*}[b!]
  \centering
   \includegraphics[width=0.495\textwidth]{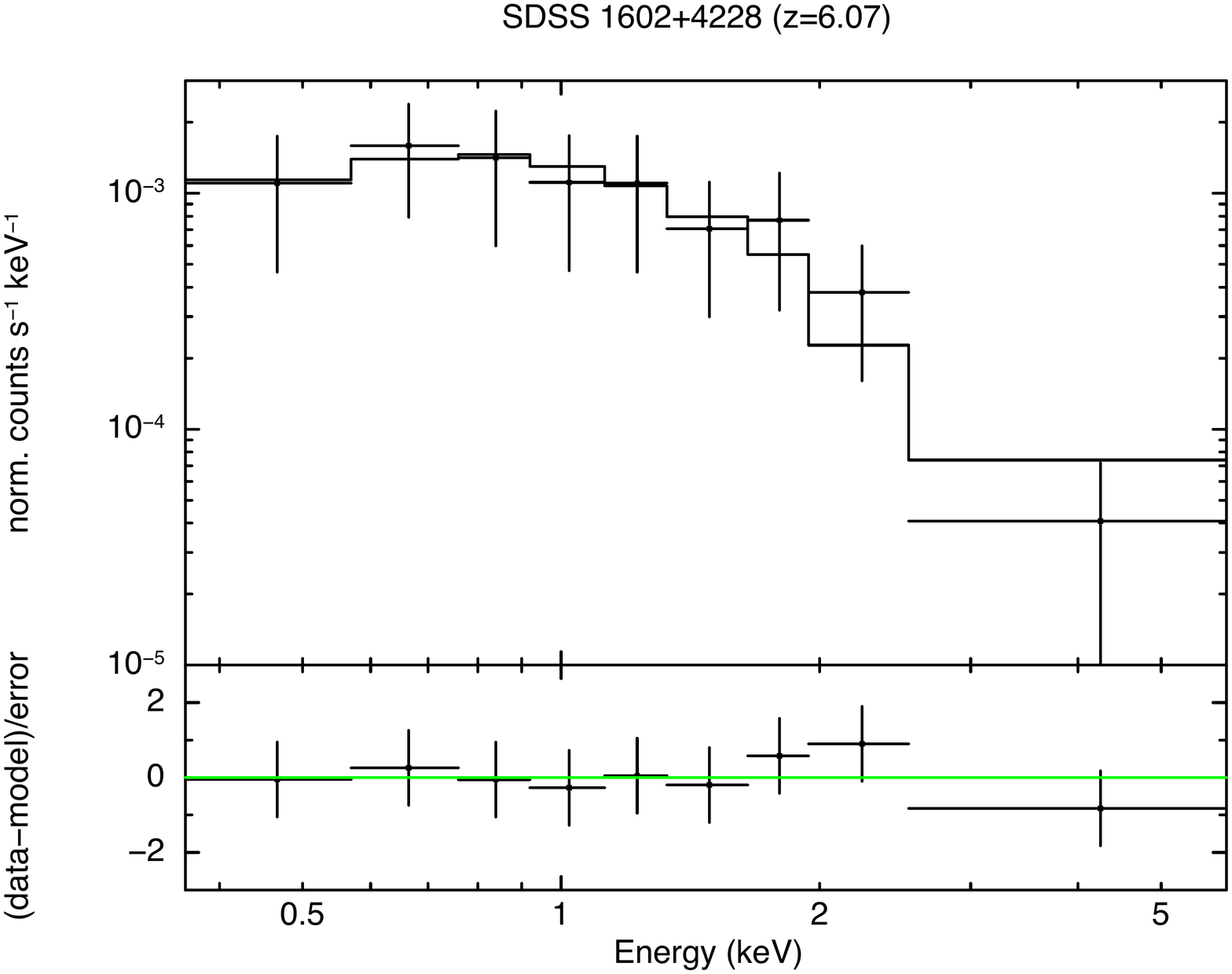}     
 \includegraphics[width=0.495\textwidth]{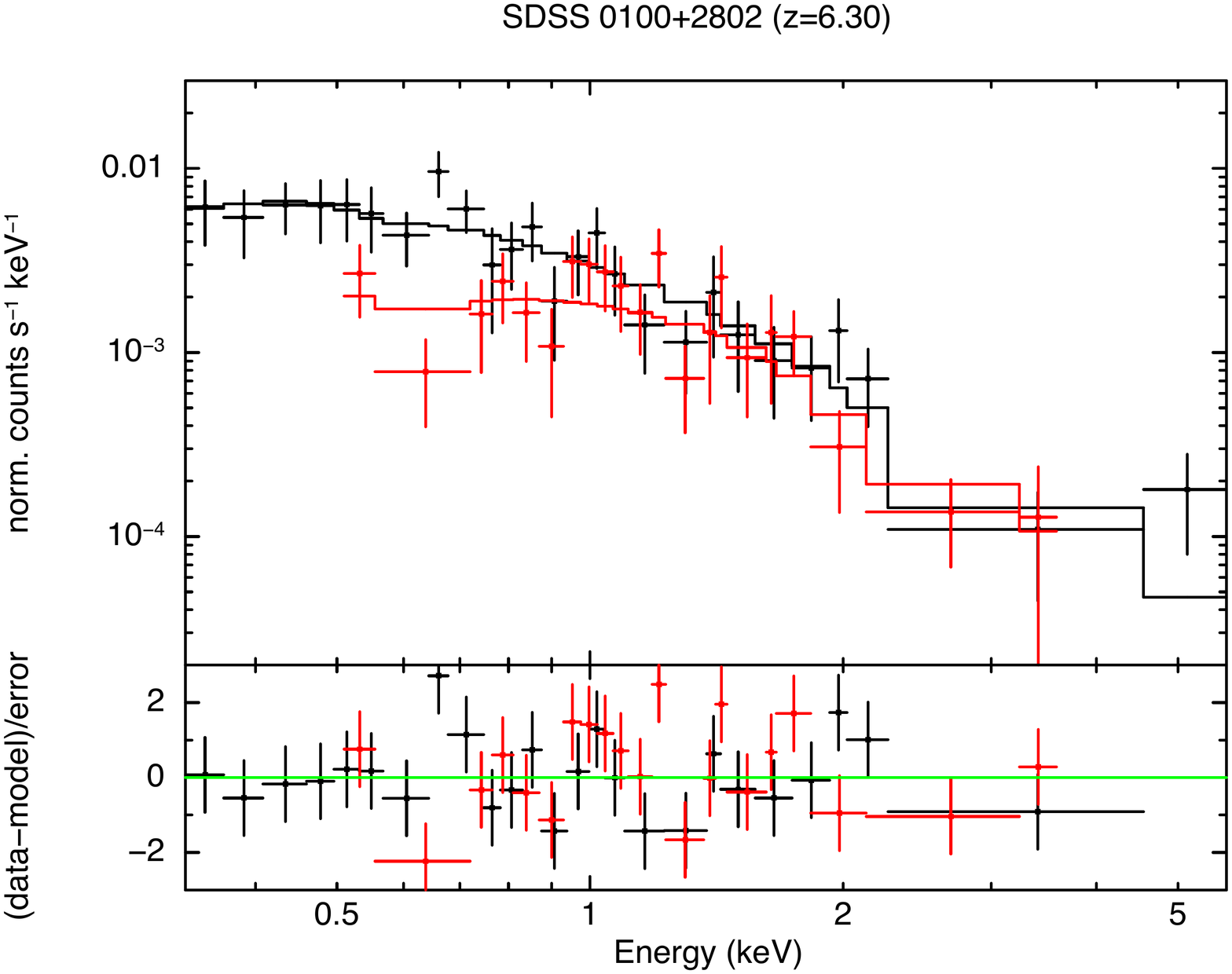}\\
    \vspace{1cm}
   \includegraphics[width=0.495\textwidth]{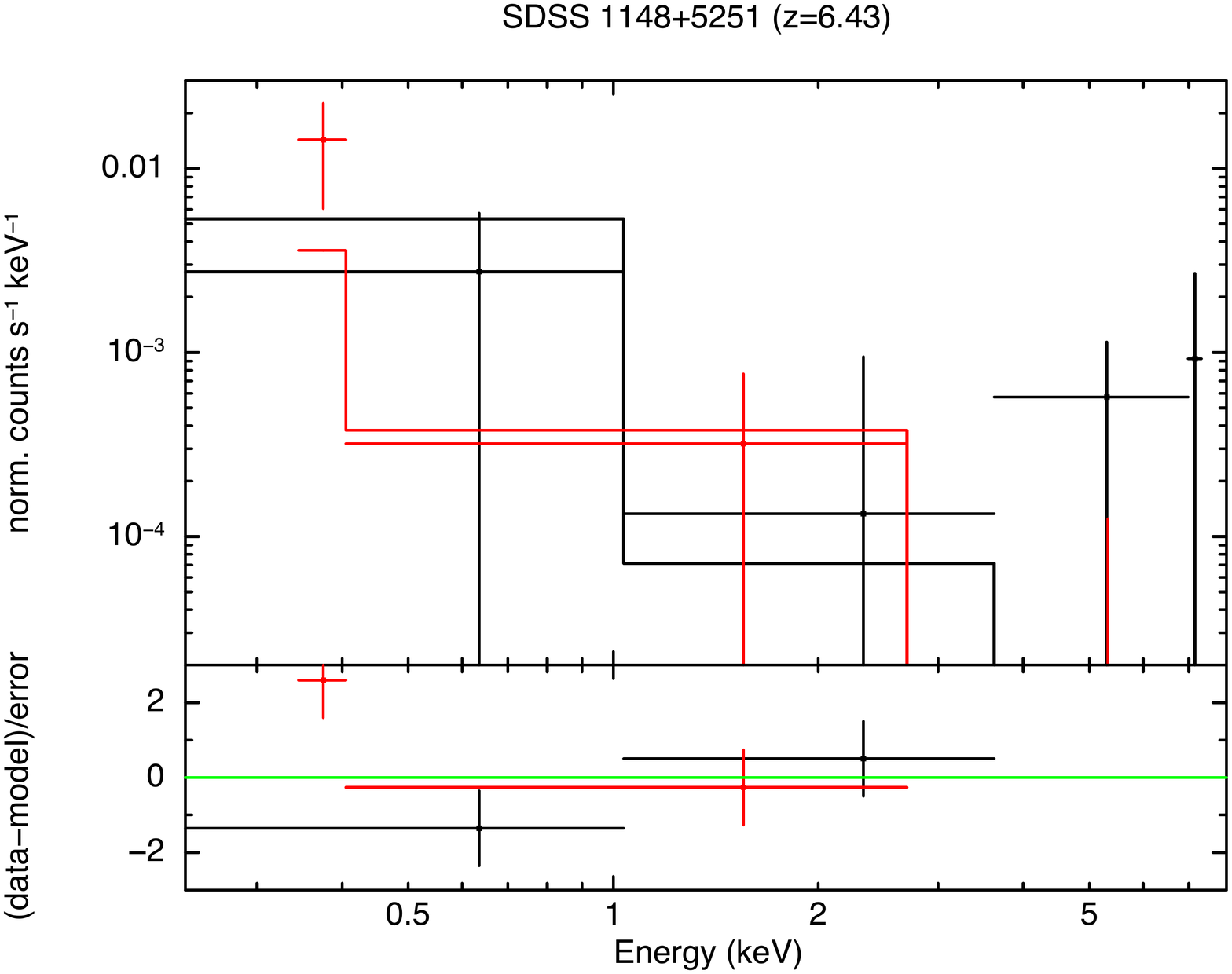}     
 \includegraphics[width=0.495\textwidth]{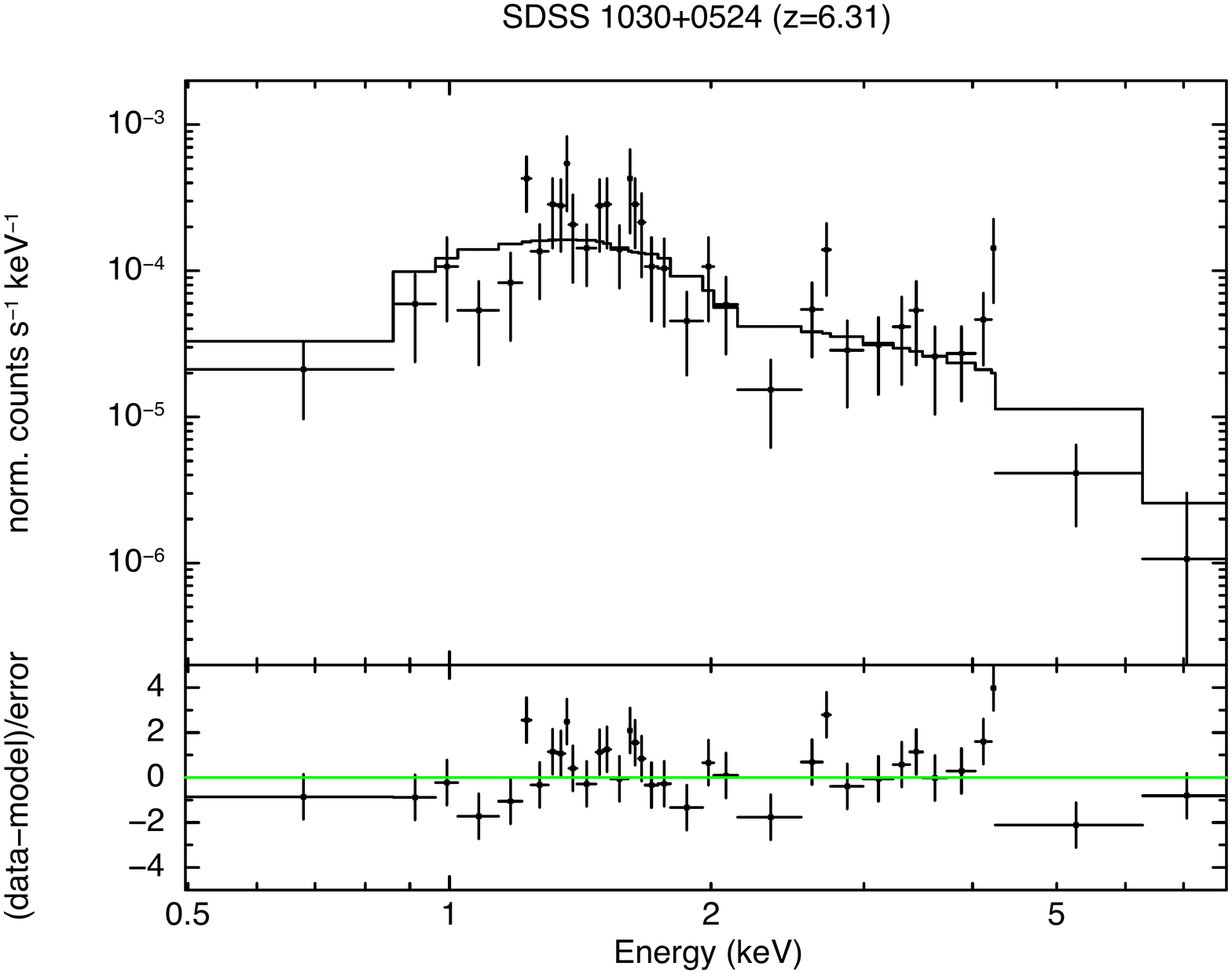}\\
   \includegraphics[width=0.495\textwidth]{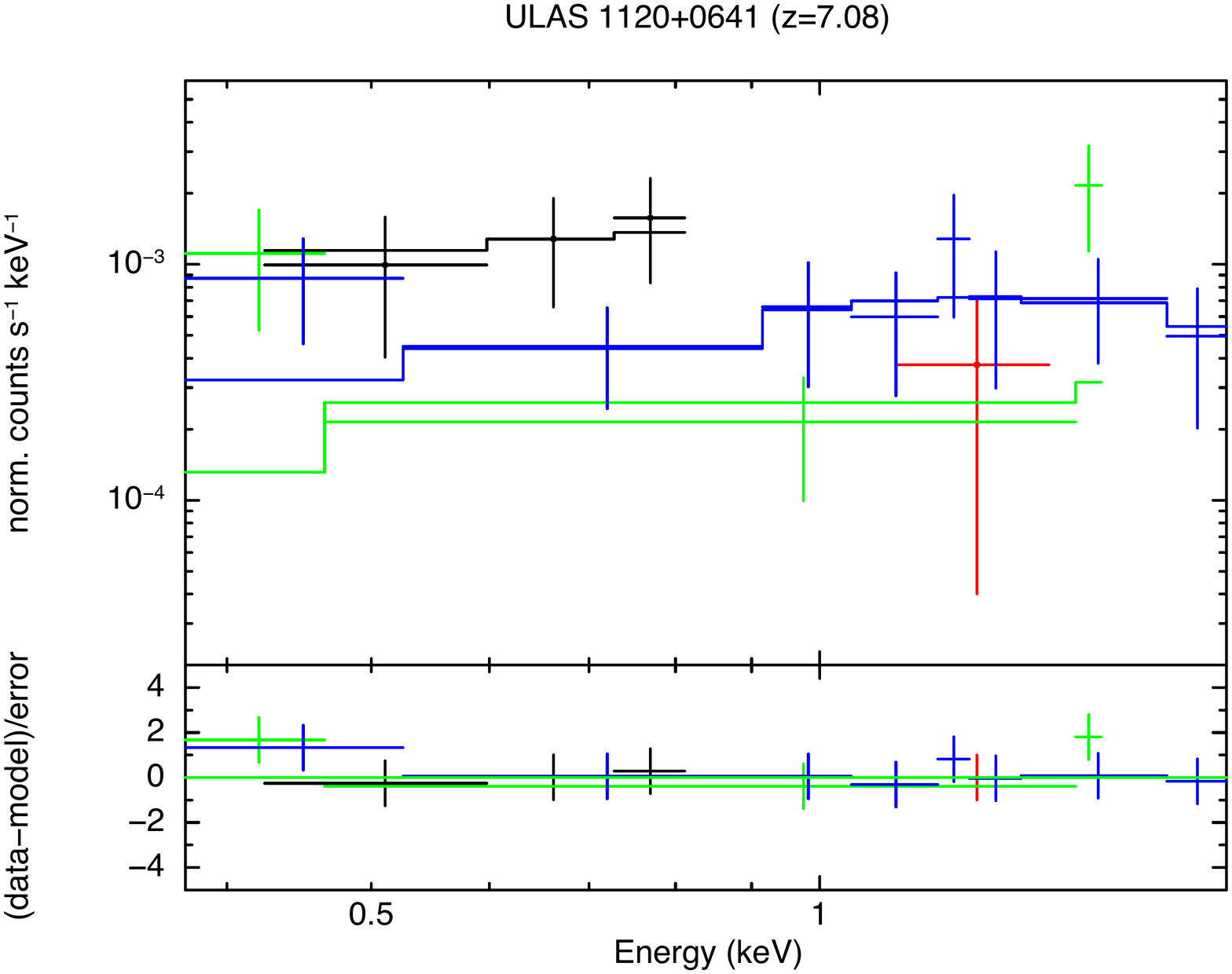}        
      \caption{The best fit models and data for 41 sources are presented in the upper panels of the following plots. Single epoch \emph{Chandra} data are represented with black crossed points, while the best fit model is the black line. In case of multiple \emph{Chandra} data, we chose the more representative observation (e.g., the longest one, or the one for which we have the best fit model with the lowest $\chi^2_{dof}$). In case of single epoch XMM-\emph{Newton} data, PN data are in black, merged MOS in red. Multiple XMM-\emph{Newton} observations have PN and merged MOS in different colours}
     \label{fig:app_b}
\end{figure*}


\end{appendix}
%
%
\end{document}